\documentclass[aps,pra,twocolumn,superscriptaddress,floatfix,nofootinbib]{revtex4-2}

\usepackage{amsmath}
\usepackage{amssymb}
\usepackage{mathtools}
\usepackage{bm}
\usepackage{bbm}
\usepackage{braket}

\usepackage{graphicx}
\usepackage{tikz}
\usetikzlibrary{quantikz2}
\usetikzlibrary{shapes.geometric, matrix, arrows.meta}

\usepackage{array}
\usepackage{multirow}

\usepackage[colorlinks=true, citecolor=blue, linkcolor=blue, urlcolor=magenta]{hyperref}

\usepackage{color}

\newcommand{\Nx}{N_x}          
\newcommand{\Nt}{N_t}          
\newcommand{\Nord}{N_C}          
\newcommand{\Deltat}{\Delta t}        
\newcommand{\NK}{N_K}          
\newcommand{\polylog}{\operatorname{polylog}}

\makeatletter
\renewcommand{\fullwire}[4]{%
  \edef\wiretype{#3}%
  \ifstrequal{#3}{a}{%
    \ifcsundef{wire@type@\the\pgfmatrixcurrentrow}{\edef\wiretype{q}}{%
      \edef\wiretype{\csname wire@type@\the\pgfmatrixcurrentrow\endcsname}%
      \ifdefstring{\wiretype}{u}{\edef\wiretype{q}}{}%
    }%
    \ifcsdef{wire@type@override@\the\pgfmatrixcurrentrow-\the\pgfmatrixcurrentcolumn}{%
      \edef\wiretype{\csname wire@type@override@\the\pgfmatrixcurrentrow-\the\pgfmatrixcurrentcolumn\endcsname}%
    }{}%
  }{}%
  \ifdefstring{\wiretype}{q}{%
    \arrow[#1,arrows,#4]{}%
  }{%
  \ifdefstring{\wiretype}{c}{%
    \arrow[arrows, shift={(0,1.0pt)},  #4]{#1}%
    \arrow[arrows, shift={(0,-1.0pt)}, #4]{#1}%
  }{%
  \ifdefstring{\wiretype}{b}{%
    \arrow[arrows, shift={(0,1.5pt)},  #4]{#1}%
    \arrow[arrows,                     #4]{#1}%
    \arrow[arrows, shift={(0,-1.5pt)}, #4]{#1}%
  }{%
  \ifdefstring{\wiretype}{n}{}{%
  \ifdefstring{\wiretype}{u}{}{%
    \PackageWarning{quantikz}{unknown wire type in cell \the\pgfmatrixcurrentrow-\the\pgfmatrixcurrentcolumn}%
  }}}}}%
}
\makeatother

\newcommand{\hexctrl}[1]{%
  \gate[style={shape=regular polygon, regular polygon sides=6,
               draw, fill=white, inner sep=-2pt}][6pt][6pt]{\phantom{.}}%
  \ifnumcomp{#1}{>}{0}{\wire[d][#1][]{q}}{}%
  \ifnumcomp{#1}{<}{0}{\wire[u][-#1][]{q}}{}%
}
\newcommand{\hexlctrl}[2]{%
  \gate[style={shape=regular polygon, regular polygon sides=6,
               draw, fill=white, inner sep=-2.5pt}][6pt][6pt]{#2}%
  \ifnumcomp{#1}{>}{0}{\wire[d][#1][]{q}}{}%
  \ifnumcomp{#1}{<}{0}{\wire[u][-#1][]{q}}{}%
}

\begin{document}

\title{Explicit Quantum Circuit Simulation of Nonlinear 1-Dimensional Fluid
  with Carleman-linearized Boltzmann Method}

\author{Keita Kanno}
\email{kanno@qunasys.com}
\affiliation{QunaSys Inc., 1-13-7, Hakusan, Bunkyo, Tokyo 113-0001, Japan}

\author{Kazumasa Ueno}
\affiliation{QunaSys Inc., 1-13-7, Hakusan, Bunkyo, Tokyo 113-0001, Japan}

\author{Hayato Higuchi}
\email{higuchi@qunasys.com}
\affiliation{QunaSys Inc., 1-13-7, Hakusan, Bunkyo, Tokyo 113-0001, Japan}

\author{Morimasa Okamoto}
\affiliation{Tokyo Gas Co. Ltd., 1-5-20, Kaigan, Minato, Tokyo 105-8527, Japan}

\author{Yuya Yoshizuru}
\affiliation{Tokyo Gas Co. Ltd., 1-5-20, Kaigan, Minato, Tokyo 105-8527, Japan}
\author{Ryoya Ishimaru}
\affiliation{Tokyo Gas Co. Ltd., 1-5-20, Kaigan, Minato, Tokyo 105-8527, Japan}
\author{Towa Takagi}
\affiliation{Tokyo Gas Co. Ltd., 1-5-20, Kaigan, Minato, Tokyo 105-8527, Japan}
\author{Kentaro Sakamoto}
\affiliation{Tokyo Gas Co. Ltd., 1-5-20, Kaigan, Minato, Tokyo 105-8527, Japan}

\date{\today}

\begin{abstract}
Quantum computation of fluid dynamics has attracted growing
attention as a key application of fault-tolerant quantum computers
anticipated in the coming decade, with lattice Boltzmann methods emerging as a
particularly promising approach.
Explicit and efficient elementary-gate-level
circuit simulations, however, have so far been demonstrated only
in the linear case. Here we include the leading nonlinearity
through second-order Carleman linearization of the
one-dimensional Boltzmann equation, and demonstrate, via
explicit quantum-circuit simulation, the preparation of the
final-time state using a Taylor-expansion-based ODE solver
based on the quantum singular value
transformation.
With this construction, we analyze the gate and qubit complexities,
which scale logarithmically with the grid size, the
nonlinearity captured by the higher-order Carleman
linearization, and the practical utility of higher-order expansions in the Taylor ODE solver.
The construction provides a concrete baseline for computational
cost reduction and further developments such as
extensions to higher dimensions, complex geometries, and the
extraction of physical quantities, towards
industrially useful quantum CFD.
\end{abstract}

\maketitle

\newpage

\section{Introduction}
\label{sec:introduction}

Computational fluid dynamics (CFD) is one of the most resource-intensive tasks
in scientific computing.
High-fidelity simulation of turbulent flows in engineering applications
routinely requires $O(\Nx^D)$ grid points in $D$ spatial dimensions,
where $\Nx$ denotes the number of mesh points along each direction.
Including $\Nt \propto \Nx$ time steps, the total computational cost
scales as $O(\Nx^{D+1})$.
Quantum computing offers a fundamentally different computational paradigm:
$n$ qubits can represent a state vector of dimension $2^n$,
suggesting that quantum algorithms could encode and process
discretized fluid states with exponentially fewer resources
than classical approaches.

This potential has motivated a growing body of research on
quantum algorithms for fluid dynamics~\cite{gaitan2020finding, Budinski2022-us, Itani2022-oz, Li2025-ox, Penuel2024-cl,
Schalkers2024-xh, Jennings2025-theory,Jennings2025-apps,Zamora2025-yj, Ueno2026-linear}.
The lattice Boltzmann method (LBM)~\cite{Qian1992-lbgk, Chen1998-dq, Kruger2016-lj}
is an especially attractive target for quantum implementation~\cite{Budinski2022-us, Itani2022-oz, Li2025-ox, Penuel2024-cl}.
LBM is a method to solve the Boltzmann equation with a set of approximations that is valid under the assumptions such as weak-compressibility.
Under the additional assumption of small Knudsen number (the hydrodynamic limit), LBM reduces to solving the Navier-Stokes equations via the Chapman-Enskog expansion.
LBM admits a built-in explicit time evolution and localized nonlinearity in the collision term acting independently at each lattice site,
making it more amenable to the quantum algorithms based on the Carleman linearization,
although a direct simulation of the Navier-Stokes equations on quantum hardware and explicit comparison with LBM in particular is also an interesting research area~\cite{gaitan2020finding}.
To consider the nonlinear effect of the fluid dynamics while maintaining the exponential advantage in the memory size,
\emph{linearization} techniques~\cite{Carleman1932-zr, 0812.4423, Joseph2020-kv, Liu2021-carleman} that embed the nonlinear dynamics into a linear system are important.

An end-to-end quantum algorithm for fluid dynamics based on the Carleman-linearization
and LBM
is proposed in Ref.~\cite{Li2025-ox}.
They formulate the problem as a Carleman-linearized time-continuous ODE system (unlike the genuine LBM, which is formulated with discretized time), which is solved by the Taylor ODE solver~\cite{Berry2017-vv}.
Ref.~\cite{Penuel2024-cl} applied the framework to
three-dimensional incompressible flows around a sphere, providing
resource estimates for hydrodynamic drag computation, though with
some assumptions on unknown factors and sub-dominant contributions omitted.
Recent works~\cite{Jennings2025-theory, Jennings2025-apps}
take a slightly different approach, directly implementing the fully discrete lattice Boltzmann recurrence
to take advantage of the guaranteed stability and accuracy with larger time steps.

Technical details for implementation, such as the application of various boundary conditions, are given in the literature as well~\cite{Schalkers2024-xh, Penuel2024-cl, Jennings2025-theory,Jennings2025-apps,Zamora2025-yj, Ueno2026-linear}. In particular, Ref.~\cite{Penuel2024-cl} gives an implementation of one-step three-dimensional LBM (D3Q27, i.e., 3 spatial dimensions with 27 representative velocities) with Carleman order 3, including the bounce-back condition around a sphere. Ref.~\cite{Jennings2025-theory} further gives an explicit circuit construction and gate counts for multi-step time evolution with general spatial dimensions and Carleman orders. A quantum circuit implementation is presented in Ref.~\cite{Zamora2025-yj} for D2Q9 with Carleman order 2 for a periodic boundary condition, while an efficient implementation using elementary gates is noted as one of the next challenges. Such an explicit quantum circuit simulation implemented by (multi-controlled) elementary-gates is presented in Ref.~\cite{Ueno2026-linear} for D2Q9, Carleman order 1 (linear) with boundary conditions for inlet, outlet, and bounce-back on obstacles and walls. The multi-step time evolution is implemented based on the QSVT~\cite{Gilyen2019-qsvt} and the ODE solver in Ref.~\cite{Berry2017-vv}, verifying a complete quantum circuit implementation for preparing the final time state of the 2D fluid simulation in the linear case, which will then be used as a subroutine of amplitude estimation algorithms for final result extraction.

In this paper, we take a natural step forward, implementing and simulating a quantum circuit of fluid dynamics with nonlinear effects, using (multi-controlled) elementary-gates. Specifically, we solve the one-dimensional D1Q3 continuous-time Boltzmann equation 
with the second-order Carleman linearization, which captures the leading nonlinearity,
using the Taylor ODE solver~\cite{Berry2017-vv} following the approach of Ref.~\cite{Li2025-ox}.
We construct explicit block encodings at the
elementary-gate level for the second-order Carleman time-evolution
matrix $A$ with bounce-back boundary conditions at both walls, and
for the $L$-matrix (Fig.~\ref{fig:L-matrix}) of the Taylor ODE
solver. The block encoding of $L$ uses that of $A$ as a subroutine, 
and is valid for arbitrary Taylor expansion order $\NK$.
This yields a complete quantum-circuit implementation of state
preparation for the history state, including the final-time state.

We further analyze the effective condition number $\kappa(L)$ of
the $L$-matrix, which governs the computational cost of the Taylor ODE solver.
We confirm that the scaling reported in
Refs.~\cite{Li2025-ox, Jennings2025-theory, Ueno2026-linear},
namely $\kappa(L)$ constant in $\Nx$ and linear in $\Nt$, continues
to hold at $\NK=1,2,3$ and $\Nord =1,2$ in the present
one-dimensional setting with bounce-back walls, identify the
prefactor, and find that both $\NK=1\to 3$ and $\Nord=1\to 2$
inflate $\kappa(L)$ by approximately a factor of~$4$.
Taking the propagation of a one-dimensional pressure wave as a
test problem, we run the QSVT circuit at
$\Nx=32$ for $\Nord=1$ and $\Nx=8$ for $\Nord=2$ 
and simultaneously
verify the quantum-circuit construction and confirm that
$\kappa(L)$ directly controls the achievable accuracy in practice.

The remainder of this paper is organized as follows.
Section~\ref{sec:methods} formulates the problem from the
Boltzmann equation, applies Carleman linearization and the
Taylor ODE solver, and constructs the corresponding block
encodings together with the QSVT-based solver.
Section~\ref{sec:results} reports the numerical analysis gate counts,
and the QSVT-circuit simulations.
Section~\ref{sec:discussion} discusses the cost--accuracy trade-offs and asymptotic scaling of the construction, extensions to higher dimensions and Carleman orders, and 
limitations of the current approach.
Section~\ref{sec:conclusion} summarises our findings future directions.

\section{Methods}
\label{sec:methods}

\subsection{Discretization of Boltzmann equation}
\label{sec:lbm}

Following the approach of Refs.~\cite{Li2025-ox, Penuel2024-cl}, 
we start from the (continuous) Boltzmann equation, which governs the evolution of the one-particle velocity distribution function and describes fluid dynamics at the mesoscopic scale.
In the following, we will first discretize the velocity space, yielding a discrete-velocity Boltzmann equation, and then discretize the
spatial domain, keeping time continuous throughout this section;
the resulting equation is solved, after the Carleman linearization
described in Sec.~\ref{sec:carleman}, using the Taylor ODE solver of Sec.~\ref{sec:qsvt}.

We note that an alternative approach, taken by
Refs.~\cite{Jennings2025-theory, Jennings2025-apps}, is to work
directly with the fully discrete lattice Boltzmann recurrence,
in which the time derivative is replaced by a finite difference
already at the kinetic-equation level, with step size fixed to
$\Deltat = 1$ in lattice units, faithfully following the original
formulation of the lattice Boltzmann method. Although a direct
and rigorous comparison of the two approaches is an interesting
open question for future work, our approach, following
Refs.~\cite{Li2025-ox, Penuel2024-cl}, has the advantage of
keeping the time step $\Deltat$ as a tunable parameter for
managing numerical instability that can occur even in the original
LBM, particularly at high Reynolds numbers. The small time step
required for accuracy, pointed out in
Ref.~\cite{Jennings2025-theory}, can be mitigated by the
higher-order Taylor expansion, as we demonstrate in
Sec.~\ref{sec:results}.

\paragraph{Discrete-velocity Boltzmann equation.}
Following the lattice Boltzmann method~\cite{Qian1992-lbgk},
we discretize the velocity variable of the continuous Boltzmann equation by the D1Q3 velocity set
$\{e_0 = +1,\, e_1 = -1,\, e_2 = 0\}$.
We work in one spatial
dimension throughout this paper, so the position $\alpha$ and the
lattice velocities $e_q$ are scalars.
We further adopt the BGK collision
model~\cite{Bhatnagar1954-bgk}, which replaces the full
Boltzmann collision integral by a single-timescale relaxation
toward a local-equilibrium distribution $f_q^{\mathrm{eq}}$.
The distribution function
$f_q(\alpha, t)$ at velocity direction $q \in \{0,1,2\}$ and
continuous position $\alpha$ obeys the discrete-velocity BGK Boltzmann
equation~\cite{Bhatnagar1954-bgk, Qian1992-lbgk}
\begin{equation}
  \partial_t f_q(\alpha, t)
  + e_q\, \partial_\alpha f_q(\alpha, t)
  = \frac{1}{\tau}\bigl[f_q^{\mathrm{eq}}(\alpha, t)
       - f_q(\alpha, t)\bigr],
  \label{eq:dvbe}
\end{equation}
the left-hand side describes advection, 
while the right-hand side is the BGK collision term, 
relaxing $f_q$ toward its local equilibrium
\begin{equation}
  f_q^{\mathrm{eq}} = w_q \rho
  \left(1 + \frac{e_q u}{c_s^2}
  + \frac{(e_q u)^2}{2 c_s^4} - \frac{u^2}{2 c_s^2}\right),
  \label{eq:feq}
\end{equation}
with lattice sound speed $c_s = 1/\sqrt{3}$, lattice weights
$w_0 = w_1 = 1/6$ and $w_2 = 2/3$, density
$\rho = \sum_q f_q$, and momentum $\rho u = \sum_q f_q e_q$.
$\tau > 0$ is the BGK relaxation time; throughout this work
$\tau$ is fixed per grid size by Eq.~\eqref{eq:tau-nu} in
Sec.~\ref{sec:problem-setup}, where we introduce the
grid-independent label~$\nu$ used to compare runs across
different grid sizes~$\Nx$.

\paragraph{Spatial discretization.}
We next restrict the spatial variable to the lattice
$\alpha \in \{0, 1, \ldots, \Nx-1\}$, working in lattice units where
the site spacing is $\Delta x = 1$, and replace the spatial
derivative $\partial_\alpha f_q$ by the upwind finite difference
aligned with the velocity~$e_q$,
\begin{equation}
  \partial_\alpha f_q(\alpha, t)
  \;\mapsto\; f_q(\alpha, t) - f_q(\alpha - e_q, t),
  \label{eq:upwind}
\end{equation}
so that Eq.~\eqref{eq:dvbe} becomes the system of ordinary
differential equations
\begin{equation}
  \partial_t f_q(\alpha, t)
  + e_q\bigl[f_q(\alpha, t) - f_q(\alpha - e_q, t)\bigr]
  = \frac{1}{\tau}\bigl[f_q^{\mathrm{eq}} - f_q\bigr].
  \label{eq:lbm}
\end{equation}
At the boundaries of the domain we impose bounce-back conditions:
an incoming distribution at a wall is reflected with reversed
velocity, $f_{\bar{q}}(\alpha_\text{wall}) \leftarrow
f_q(\alpha_\text{wall})$, where $\bar{0}=1$, $\bar{1}=0$, and
$\bar{2}=2$. Equation~\eqref{eq:lbm} is the continuous-time,
space-discretized kinetic equation that the rest of this paper
targets with the quantum algorithm.

\paragraph{Linearization of the collision term.}
The equilibrium distribution~\eqref{eq:feq} contains the nonlinear
factor $1/\rho$ in the term proportional to $\rho u^2=(\rho u)^2/\rho$.
Following Ref.~\cite{Li2025-ox} we expand $1/\rho$ around $\rho = 1$:
\begin{equation}
  \frac{1}{\rho} = 2 - \rho + O\bigl((1-\rho)^2\bigr).
  \label{eq:rho-expansion}
\end{equation}
This approximation is valid when $|1-\rho| \ll 1$, which is the target
regime\footnote{We note, though, that
our approach of solving the continuous Boltzmann equation
may have broader application range once the collision term
and discretization along the velocity and the spatial variables are appropriately modified.}
of the LBM formulation~\cite{Li2025-ox}.
For the symmetric initial condition with $|\rho - 1| \leq \Delta\rho/2$ used throughout this paper, the discarded $O((1-\rho)^2)$ terms are bounded by 
$|\rho - 1|^2/(1 - |\rho - 1|) \leq (\Delta\rho/2)^2/(1 - \Delta\rho/2)$. 
This approximation error grows with $\Delta \rho$, which is visible later in Fig.~\ref{fig:carleman-C123}.
With
Eq.~\eqref{eq:rho-expansion}, $f_q^{\mathrm{eq}}$ becomes a polynomial
of degree~3 in the distribution functions~$\{f_q\}$, so that
Eq.~\eqref{eq:lbm} becomes a cubic ODE amenable to Carleman
linearization.

\subsection{Carleman linearization}
\label{sec:carleman}

Quantum linear-system solvers, including the QSVT approach adopted in Sec.~\ref{sec:qsvt}, operate on linear matrix                    equations, so the nonlinear equation~\eqref{eq:lbm} must first be recast in linear form before it can be solved on a quantum computer. Carleman linearization~\cite{Carleman1932-zr, Liu2021-carleman}
transforms a system of polynomial ordinary differential equations
into an infinite-dimensional linear system by introducing
tensor-product variables.
Let $\bm{f}^{(1)} = (f_0(0), f_1(0), f_{Q-1}(0), f_0(1), \ldots,
f_{Q-1}(\Nx-1))^\top \in \mathbb{R}^{Q\Nx}$
denote the vector of all distribution functions across all velocity
directions $q \in \{0,\ldots,Q-1\}$ and lattice sites
$\alpha \in \{0,\ldots,\Nx-1\}$, where $Q=3$.
The second-order Carleman variable is the tensor product
$\bm{f}^{(2)} = \bm{f}^{(1)} \otimes \bm{f}^{(1)} \in \mathbb{R}^{(Q\Nx)^2}$ and higher-order variables 
$\bm{f}^{(k)} = \left(\bm{f}^{(1)}\right)^{\otimes k}$ are defined similarly for $k > 2$.
An exact solution of nonlinear ODEs can be obtained by solving the infinite-dimensional linear system,
while a finite-dimensional approximation is obtained by truncating the system at a finite order~$\Nord$.
Our main focus in this paper is the second-order truncation ($\Nord=2$), 
which captures the leading nonlinear effects of the BGK collision operator, while some $\Nord=3$ 
results are also presented in Sec.~\ref{sec:carleman-error} for reference.

Substituting the $1/\rho$ expansion~\eqref{eq:rho-expansion} into the
collision term of the kinetic equation~\eqref{eq:lbm} makes the
right-hand side a polynomial of degree three in the~$\{f_j\}$. The
resulting system can be written in the Carleman form
\begin{equation}
  \frac{d}{dt}\bm{f}^{(1)}
  = A_{11}\, \bm{f}^{(1)}
  + A_{12}\, \bm{f}^{(2)}
  + A_{13}\, \bm{f}^{(3)},
  \label{eq:ode}
\end{equation}
where $A_{11} \in \mathbb{R}^{Q\Nx \times Q\Nx}$ combines
streaming and the linearized BGK collision, and
$A_{12} \in \mathbb{R}^{Q\Nx \times (Q\Nx)^2}$ encodes the
quadratic contribution of the collision, while $A_{13}$ encodes the cubic contribution, the product of $-\rho$ factor from $1/\rho$ 
and the quadratic term of the local equilibrium~\eqref{eq:feq}.
At $\Nord = 2$, we eliminate all cubic contributions by evaluating the prefactor $1/\rho$  
of the quadratic term in the collision term at $\rho=1$.
This is equivalent to halving $A_{12}$ from the $1/\rho \to 2-\rho$ expression (Eq.~\eqref{eq:A12-kernel}) and dropping $A_{13}$ entirely. See Appendix~\ref{app:scales} for $\Nord = 3$ case.
Rather than treating streaming and collision as simultaneous contributions in Eq.~\eqref{eq:lbm}, we apply collision first
and then stream the result to the neighbor
site for numerical stability.
The matrix elements of $A_{11}$, with column (``input'') index $(\alpha, q)$ and
row (``output'') index $(\alpha', q')$, are
\begin{equation}
  \begin{split}
    (A_{11})_{\alpha' q',\, \alpha q}
    &= \delta_{\alpha',\,\alpha+e_{q'}}\!\Bigl[
        \Bigl(1-\frac{1}{\tau}\Bigr)\delta_{q' q}
        + \frac{w_{q'}}{\tau}(1 + 3\,e_{q'} e_q)\Bigr] \\
    &\quad - \delta_{\alpha'\alpha}\,\delta_{q' q}.
  \end{split}
  \label{eq:A11-elements}
\end{equation}
where the bracketed factor is the linearized BGK collision at
site~$\alpha$, the outer $\delta_{\alpha',\alpha+e_{q'}}$ streams the
result to site~$\alpha + e_{q'}$ (with bounce-back at the walls),
and the subtracted identity accounts for the departure from the
pre-update state.
The nonlinear block $A_{12}$ inherits the same streaming structure
but couples two distribution functions on the same lattice site,
with input index $(\alpha_1, q_1, \alpha_2, q_2)$ and output index
$(\alpha', q')$,
\begin{equation}
  (A_{12})_{\alpha' q',\, \alpha_1 q_1\, \alpha_2 q_2}
  = \delta_{\alpha',\,\alpha_1 + e_{q'}}\,
    \delta_{\alpha_1\alpha_2}\,
    K^{(12)}_{q'\,q_1\,q_2},
  \label{eq:A12-elements}
\end{equation}
with the quadratic collision kernel
\begin{equation}
  K^{(12)}_{q'\,q_1\,q_2}
  = \frac{w_{q'}}{\tau}\left(
    \frac{9}{2}\,(e_{q_1}\!\cdot e_{q'})(e_{q_2}\!\cdot e_{q'})
    - \frac{3}{2}\,(e_{q_1}\!\cdot e_{q_2})\right),
  \label{eq:A12-kernel}
\end{equation}
obtained from the quadratic terms of the local
equilibrium~\eqref{eq:feq} after substituting the $1/\rho$
expansion~\eqref{eq:rho-expansion}.
At first-order Carleman truncation ($\Nord=1$) the $A_{12}$ term is
dropped, yielding the linear ODE
$d\bm{f}^{(1)}/dt = A_{11}\, \bm{f}^{(1)}$.
At second-order truncation ($\Nord=2$), we additionally evolve
$\bm{f}^{(2)}$ using the Leibniz rule
$d(\bm{f}^{(1)} \otimes \bm{f}^{(1)})/dt
= (d\bm{f}^{(1)}/dt) \otimes \bm{f}^{(1)}
+ \bm{f}^{(1)} \otimes (d\bm{f}^{(1)}/dt)$:
\begin{equation}
  \frac{d}{dt}\bm{f}^{(2)}
  = (A_{11} \otimes I + I \otimes A_{11})\, \bm{f}^{(2)},
  \label{eq:carleman-second}
\end{equation}
where higher order terms are truncated, assuming $\Nord=2$.
For the $\Nord=3$ expression, see Appendix~\ref{app:nc3}.
Defining $A_{22} \equiv A_{11} \otimes I
+ I \otimes A_{11}$,
the full second-order Carleman system is
\begin{equation}
  \frac{d}{dt}
  \begin{pmatrix} \bm{f}^{(1)} \\ \bm{f}^{(2)} \end{pmatrix}
  = \begin{pmatrix} A_{11} & A_{12} \\
    0 & A_{22} \end{pmatrix}
  \begin{pmatrix} \bm{f}^{(1)} \\ \bm{f}^{(2)} \end{pmatrix}.
  \label{eq:carleman-system}
\end{equation}
This ODE system is then solved via the Taylor ODE solver
described in Section~\ref{sec:time-discretization}.

\subsection{Taylor ODE solver}
\label{sec:time-discretization}

To integrate the Carleman ODE~\eqref{eq:carleman-system} on a
quantum computer we follow the Taylor ODE solver introduced in Ref.~\cite{Berry2017-vv}: 
a truncated Taylor approximation
of the propagator $e^{\Deltat A}$ over $\Nt$ time steps is recast
as a single linear system
$L\bm{x} = \bm{b}$, whose inverse is then applied via a quantum linear system algorithm such as the one based on the
quantum singular value transformation
(QSVT)~\cite{Gilyen2019-qsvt}.
This subsection constructs $L$ as an explicit matrix, while the quantum circuits that encode
$A$, $L$, and the QSVT procedure that inverts $L$ are given in
the next section Sec.~\ref{sec:circuits}.

Time is discretized into $\Nt$ steps of size $\Deltat = T/\Nt$ for total simulation time $T$.
At each step~\cite{Berry2017-vv}, the propagator $e^{\Deltat A}$ is approximated by
its truncated Taylor series of order~$\NK$,
\begin{equation}
  e^{\Deltat\,A}
  \;\approx\; \sum_{k=0}^{\NK} \frac{(\Deltat\,A)^k}{k!}.
  \label{eq:taylor-evolution}
\end{equation}
The Taylor coefficients are generated by a recurrence:
defining auxiliary vectors $\bm{g}^{(k)}_m$ for time step
$m \in \{0, \ldots, \Nt{-}1\}$ and Taylor index
$k \in \{0, \ldots, \NK\}$, with $\bm{g}^{(0)}_0 = \bm{f}(0)$,
\begin{align}
  \bm{g}^{(k+1)}_m &= \frac{\Deltat\,A}{k+1}\, \bm{g}^{(k)}_m,
  \quad k = 0, \ldots, \NK{-}1,
  \label{eq:taylor-recurrence} \\
  \bm{g}^{(0)}_{m+1} &= \sum_{k=0}^{\NK} \bm{g}^{(k)}_m,
  \label{eq:step-transition}
\end{align}
so that $\bm{g}^{(0)}_{m+1} \approx e^{\Deltat A}\,\bm{g}^{(0)}_m$
and $\bm{g}^{(0)}_{\Nt} \approx e^{TA}\,\bm{f}(0)$.
Rather than applying this recurrence sequentially, 
the Taylor ODE solver~\cite{Berry2017-vv}
encodes all $\Nt$ steps into a single linear system
$L\bm{x} = \bm{b}$ whose solution yields the distribution at
every time step at once.
In addition to the $\Nt$ time-evolution steps, $L$ carries an
\emph{idling block} of $p$ extra rows beyond the final step that
simply repeat the final-time state
$\bm{g}^{(0)}_{\Nt}$~\cite{Berry2017-vv}; this boosts the weight
of the final-time subspace in $\bm{x} = L^{-1}\bm{b}$ to
$\Omega(1)$, from the $O(1/\Nt)$ decay it would otherwise suffer.
The explicit form of $L$ for the smallest non-trivial case is
shown in Fig.~\ref{fig:L-matrix}; the general definition at
arbitrary $(\Nt, \NK, p)$ is given in
Ref.~\cite[Def.~1]{Berry2017-vv} under the symbol
$C_{\Nt, \NK, p}(\Deltat A)$, and the same matrix (renamed $L$) is
used in Ref.~\cite{Penuel2024-cl}.

\begin{figure}[htbp]
\centering
\[
  \setlength{\arraycolsep}{2.5pt}
  L = \left(\begin{array}{@{}cccc|cccc|ccc@{}}
    I \\[2pt]
    -\Deltat A & I \\[2pt]
    & -\tfrac{\Deltat A}{2} & I \\[2pt]
    & & -\tfrac{\Deltat A}{3} & I \\[2pt]
    \hline\\[-8pt]
    -I & -I & -I & -I & I \\[2pt]
    & & & & -\Deltat A & I \\[2pt]
    & & & & & -\tfrac{\Deltat A}{2} & I \\[2pt]
    & & & & & & -\tfrac{\Deltat A}{3} & I \\[2pt]
    \hline\\[-8pt]
    & & & & -I & -I & -I & -I & I \\[2pt]
    & & & & & & & & -I & I \\[2pt]
    & & & & & & & & & -I & I
  \end{array}\right)
\]
\caption{Time-evolution matrix $L$ for the
  smallest non-trivial case $(\Nt, \NK, p) = (2, 3, 2)$;
  each entry shown is a $Q\Nx\times Q\Nx$ block when $\Nord=1$.
}
\label{fig:L-matrix}
\end{figure}

In this work we examine Taylor truncation orders
$\NK \in \{1, 2, 3\}$ in the analysis of
Sec.~\ref{sec:results}, while the quantum circuit implementation of
Sec.~\ref{sec:circuits} is valid for any~$\NK$, using
$n_k = \lceil\log_2(\NK+1)\rceil$ qubits for the Taylor index.

\subsection{Quantum circuit implementation}
\label{sec:circuits}

The remaining task is to realize the matrices $A$ and $L$, and the QSVT solver as explicit quantum circuits.
We start from a generic sparse block-encoding construction
(Sec.~\ref{sec:be-sparse}), which is applied to the $A$-matrices at first order
(Sec.~\ref{sec:A11-block}) and second order
(Sec.~\ref{sec:a22-encoding}), and the $L$-matrix (Sec.~\ref{sec:be-L}),
completing the necessary components to build the QSVT inverter in Sec.~\ref{sec:qsvt}.

\subsubsection{Block-encoding}
\label{sec:be-sparse}

Both the rate matrix $A$ of Eq.~\eqref{eq:carleman-system} and the
$L$-matrix introduced in Sec.~\ref{sec:be-L} are implemented as a
quantum circuit through block
encodings~\cite{Low2017-qsp, Gilyen2019-qsvt} based on 
sparse-oracle access~\cite{quant-ph/0301023, ChildsPhD, quant-ph/0508139, Berry2012-bb}.
A unitary
$U_M$ acting on $n + a$ qubits is a block encoding
of a $2^n\times 2^n$ matrix $M$ if
\begin{equation}
  M = \lambda_M \,
  (\bra{0}^{\otimes a} \otimes I)\, U_M\,
  (\ket{0}^{\otimes a} \otimes I),
  \label{eq:block-encoding-def}
\end{equation}
where $\lambda_M$ is a normalization factor and $a$ is the number
of ancilla qubits.
In index form,
$M_{x'x}/\lambda_M = (\bra{x'} \otimes
\bra{0}^{\otimes a}) U_M (\ket{x} \otimes \ket{0}^{\otimes a})$,
so the $x$-th input amplitude is transported to the $x'$-th
output amplitude with coefficient $M_{x'x}/\lambda_M$
by the action of $U_M$.

We use a variant of the sparse-matrix block
encoding~\cite{Berry2012-bb} to implement our matrices, following
Ref.~\cite{Ueno2026-linear}.
Our construction acts on three registers: a primary register
$\ket{\cdot}_x$ of $n$ qubits to hold the column (input, i.e., before operation) 
or row (output, i.e., after operation) index;
a label register $\ket{\cdot}_i$ of $n_c = \lceil\log_2 s\rceil$ qubits 
to index the $s$ nonzero entries of 
the column (input) or row (output); and a target qubit $\ket{0}_t$.
The circuit structure is
\begin{equation}
  U_M = (I_x \otimes H^{\otimes n_c}_i \otimes X_t)\, O_M \,
  (I_x \otimes H^{\otimes n_c}_i \otimes I_t),
  \label{eq:UM-structure}
\end{equation}
where the Hadamard layers create and undo a uniform superposition
over the $2^{n_c}$ values of $i$, the $X$ gate on the target qubit
compensates a convention choice in the $R_Y$ rotation angle below, $I_x$ and $I_t$ denotes the identities on $x$ and $t$ registers respectively,
and the matrix-specific content is carried entirely by the oracle
$O_M$.

\begin{figure}[htbp]
\centering
\begin{quantikz}[column sep=0.35cm, row sep=0.4cm]
  \lstick{$\ket{\cdot}_x$}     & \qwbundle{n}   & \qw                    & \gate[3]{O_M} & \qw                        & \qw \\
  \lstick{$\ket{\cdot}_i$} & \qwbundle{n_c} & \gate{H^{\otimes n_c}} &               & \gate{H^{\otimes n_c}}     & \qw \\
  \lstick{$\ket{0}_t$}                   & \qw            & \qw                    &                           & \gate{X}  & \qw
\end{quantikz}
\caption{Block-encoding circuit used in this work
  (Eq.~\eqref{eq:UM-structure}).}
\label{fig:be-sparse}
\end{figure}
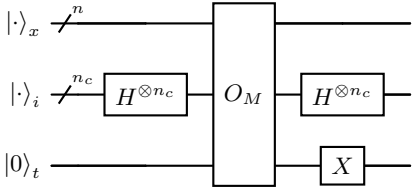

The oracle $O_M$ performs three tasks on the registers
$\ket{x}_x\ket{i}_i\ket{0}_t$ simultaneously and consistently.
First, it updates the primary register in place\footnote{The subscript labels the
register and the ket body its current contents, so $\ket{x'}_x$
denotes the primary register currently holding the value~$x'$. This convention is used throughout this paper.},
$\ket{x}_x \to \ket{x'}_x$, where $x' = \mathrm{row}(x, i)$ is the
row index of the $i$-th nonzero entry of column~$x$.
Second, it transforms the label register
$\ket{i}_i \to \ket{i'}_i$ so that the output pair $(x', i')$
uniquely determines the input pair $(x, i)$.
Third, conditioned on the values in the primary and row-label
registers (either $(x, i)$ before the update or $(x', i')$
afterwards), it applies an $R_Y$ rotation to the target qubit,
$\ket{0}_t \mapsto R_Y\!\bigl(2\arcsin(v/v_\text{max})\bigr)\ket{0}_t$,
where $v = M_{x'x}$ is the matrix element and $v_\text{max}$ is an
upper bound on the absolute entries of $M$. After the final
$X$ gate and projection onto $\ket{0}_t$ in $U_M$, the amplitude extracted
on the target is $v/v_\text{max}$, up to the Hadamard projection.
As we observe that the number of distinct values taken by $v = M_{x'x}$ is bounded
independently of $\Nt$ and $\Nx$ in our case, we hard-code the
$R_Y$ rotations with explicit angles inside $O_M$, avoiding the quantum arithmetic required in a more general construction~\cite{Berry2012-bb}. 

Combining the factor $1/2^{n_c}$ from the Hadamard projection with
the $1/v_\text{max}$ factor from the target rotation yields the
normalization
\begin{equation}
  \lambda_M = 2^{n_c}\, v_\text{max}.
  \label{eq:lambda-M}
\end{equation}
The concrete value of $v_\text{max}$ and $n_c$ depends on the matrix being
encoded.

\subsubsection{First-order block encoding ($U_{A_{11}}$)}
\label{sec:A11-block}

\begin{figure*}[htbp]
\centering
\resizebox{\textwidth}{!}{%
\begin{quantikz}[wire types={c,c,b,q,q,q}, column sep=0.35cm, row sep=0.3cm,classical gap=0.07cm]
  \lstick{$\ket{\cdot}_i$}      & \qwbundle{2}        & \gate[2]{\text{SWAP}} & \hexctrl{1}      &             & \hexctrl{1} &                                &                &  \hexctrl{1}              &\hexctrl{1} &     \gate{\text{Set $Q$}}  &  \gate[5]{\text{Uncomp.}} &  \\
  \lstick{$\ket{\cdot}_q$}      & \qwbundle{2}        &                       & \hexlctrl{4}{2}  & \hexctrl{1} & \hexctrl{1} &                \gate{X_\text{bb}} & \hexctrl{1}  &  \gate{\text{Unset $Q$}}   &\hexctrl{4} &    \hexctrl{-1}          &                         &  \\
  \lstick{$\ket{\cdot}_\alpha$} & \qwbundle{n_\alpha} &                       &                  & \hexctrl{1} & \hexctrl{2} &                                &   \gate{\pm 1}  &                          &             &                           &                      &  \\
  \lstick{$\ket{0}_{q_c}$}      &                     &                       &                  & \targ{}       &             &                \ctrl{-2}      &   \octrl{-1}   &                          &             &                           &                      &  \\
  \lstick{$\ket{0}_{q_a}$}      &                     &                       &                  &             & \targ{}       & \ctrl{1}                     &                 & \ctrl{-3}               &              &        \ctrl{-3}          &                       &  \\
  \lstick{$\ket{0}_t$}          &                     &                       & \gate{R_Y}       &             &             &  \gate{R_Y}                    &                 &                         & \gate{R_Y} &                         &                        &
\end{quantikz}%
}
\caption{Oracle circuit $O_{A_{11}}$ for the first-order Carleman
  matrix. SWAP denotes bitwise SWAP between registers, blank hexagons means that the register is a control (with some condition) for the gate, hexagons with a value in it means that the gate is applied conditioned on the register being in that value. 
  $R_Y$ gate symbols may collectively represent multiple applications of $R_Y$ with different angles, depending on the values in the registers. 
  $X_\text{bb}$ is the bounce-back velocity reversal applied on the lowest bit.
  Other details are in the main text.}
\label{fig:circuit-UA}
\end{figure*}
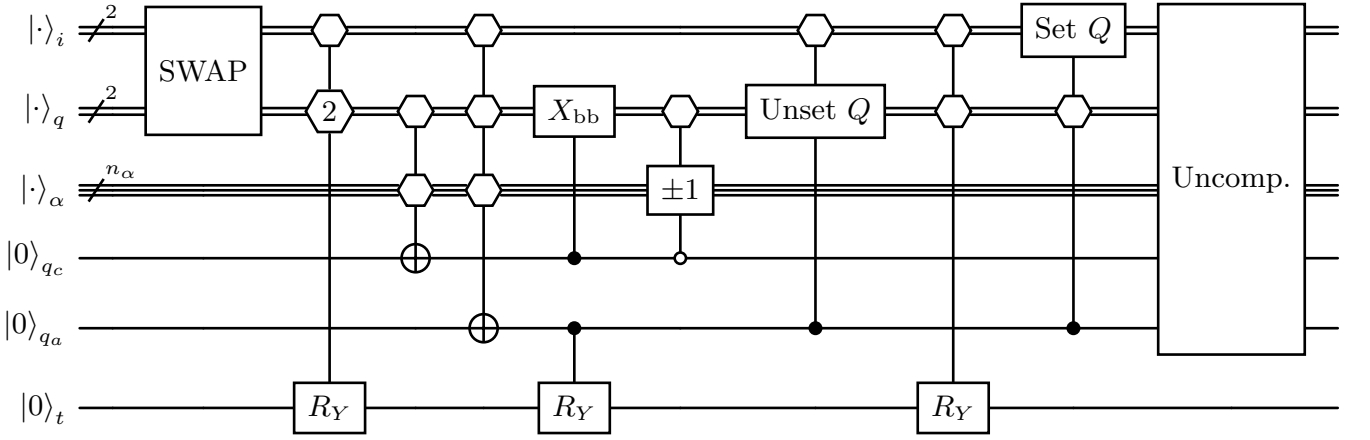

This subsection specifies the oracle $O_{A_{11}}$ 
for the first-order Carleman matrix $A_{11}$,
instantiating the generic construction of
Fig.~\ref{fig:be-sparse} with $M = A_{11}$.
Figure~\ref{fig:circuit-UA} shows the circuit structure of the
first-order block encoding oracle $O_{A_{11}}$.
The quantum register for $A_{11}$ consists of four groups:
\begin{itemize}
  \item $\ket{\cdot}_\alpha$: $n_\alpha = \lceil\log_2 \Nx\rceil$ qubits
    encoding the spatial index;
  \item $\ket{\cdot}_q$: $n_q = 2$ qubits encoding a velocity direction
    ($3$ of the $4$ values are used);
  \item $\ket{\cdot}_i$: $n_i = 2$ qubits forming the label
    register;
  \item $\ket{0}_t$: one target qubit for value encoding.
\end{itemize}
The pair $\ket{\cdot}_\alpha\ket{\cdot}_q$ plays the role of the
primary register $\ket{\cdot}_x$ in Fig.~\ref{fig:be-sparse},
mapped in place by $O_{A_{11}}$ as
$\ket{\alpha}_\alpha\ket{q}_q \mapsto
 \ket{\alpha'}_\alpha\ket{q'}_q$.
The four values of the label register $\ket{\cdot}_i$ serve as
the label $i$ for nonzero entries
in Fig.~\ref{fig:be-sparse}: $i \in \{0, 1, 2\}$ coincide with
the three D1Q3 output velocities, and $i = Q = 3$ handles the
diagonal (subtracted-identity) term.
The input/output register qubit count is $n_\alpha + n_q + n_i + 1 = n_\alpha + 5$, 
excluding ancillary qubits that are used internally and uncomputed.

The oracle $O_{A_{11}}$ encodes the matrix elements of $A_{11}$ 
(Eq.~\eqref{eq:A11-elements}) by performing the following operations.
We note that, while each step is implemented mostly in this order, the $R_Y$ value-encoding rotations of step~4
are interleaved throughout, applied as soon as the relevant register
state is available.
\begin{enumerate}
  \item \textbf{Velocity SWAP.}
    An unconditional SWAP between $\ket{\cdot}_q$ and $\ket{\cdot}_i$
    moves the row-label value $i$ into the primary velocity register
    (where it will serve as the output row velocity~$q'$);
    the original column velocity~$q$ is deposited in the
    label register, where it acts as (part of) the label of nonzero elements in the row, and is projected out by the outer Hadamard layer of
    Eq.~\eqref{eq:UM-structure}.
  \item \textbf{Case branching.}
    Two ancillae flag non-bulk branches:
    $\ket{1}_{q_c}$ on the bounce-back branch, where
    $\ket{\cdot}_\alpha$ sits at the wall aligned with the output
    velocity~$q'$ ($\alpha = \Nx - 1$ for $q' = 1$; $\alpha = 0$
    for $q' = 0$);
    $\ket{1}_{q_a}$ on the extra row-label branch
    ($i = Q = 3$ at a bulk site), which
    addresses the $-\delta_{\alpha'\alpha}\,\delta_{q' q}$ term of
    Eq.~\eqref{eq:A11-elements}.
  \item \textbf{In-place index update.}
    Three register updates produce the output row address
    $(\alpha', q')$:
    \begin{itemize}
      \item \emph{Bounce-back velocity reversal}: conditional on
        $\ket{1}_{q_c}$, a controlled-$X$ on the low bit of
        $\ket{\cdot}_q$ maps $q \to \bar{q}$.
        This works because the rest-particle value $q = 2$
        (binary~$10$) is left unchanged, while the reversal
        $0 \leftrightarrow 1$ (binary~$00 \leftrightarrow 01$) is a
        single-bit flip.
      \item \emph{Spatial shift}: conditional on
        $\ket{0}_{q_c}$ and $q \in \{0, 1\}$ (kept in the $i$-register as $\ket{q}_i$), a controlled
        increment ($q = 0$) or decrement ($q = 1$) shifts
        $\ket{\cdot}_\alpha$ by $e_q$, giving $\alpha' = \alpha + e_q$.
        For $q = 2$ and for the $i = Q$ slot, $\alpha$ is
        unchanged.
      \item \emph{Diagonal-slot velocity}: on the $\ket{1}_{q_a}$
        branch, $\ket{i}_q$ holds the known value $i=Q = 3$.
        A controlled copy (``Unset $Q$'' in the figure) overwrites it with the input velocity
        $q \in \{0, 1\}$ stored now in the row-label register as
        $\ket{q}_i$, so that the output row address
        $(\alpha', q')$ equals $(\alpha, q)$, i.e., the diagonal
        contribution leaves the velocity unchanged.
        The $i=2$ case takes care of the $q=2$ diagonal elements, and the value encoding there accounts for the identity subtraction.
        After the value encoding, a similar operation (``Set $Q$'' in the figure) 
        restores $\ket{q}_i$ to $i = Q$.
    \end{itemize}
  \item \textbf{Value encoding.}
    Multi-controlled $R_Y$ rotations on $\ket{0}_t$ encode the
    matrix element
    $(A_{11})_{(\alpha', q'),\, (\alpha, q)} / v_\text{max}$.
    Each row-label slot~$i$ addresses a distinct part of
    Eq.~\eqref{eq:A11-elements}:
    \begin{itemize}
      \item $i \in \{0, 1\}$, bulk ($\ket{0}_{q_c}$): the
        collide-then-stream term
        $(1 {-} 1/\tau)\,\delta_{q, i}
        + (w_{i}/\tau)(1 {+} 3\,e_{i}\, e_q)$,
        with output row $(\alpha + e_{q'},\, q')$ where $q' = i$.
      \item $i \in \{0, 1\}$, bounce-back ($\ket{1}_{q_c}$):
        the same collision factor, but the output velocity is
        reversed to $q' = \bar{q}$ and $\alpha' = \alpha$.
      \item $i = 2$: the rest-particle slot
        ($e_{q'} = 0$ with $q' = 2$, no spatial shift),
        encoding the on-site collision
        $(w_{q'} - \delta_{q' q})/\tau$ and other contributions such as the identity subtraction.
      \item $i = Q$ ($\ket{1}_{q_a}$): the subtracted identity
        $-\delta_{\alpha'\alpha}\,\delta_{q' q}$,
        contributing $-1$ at the diagonal entry
        $(\alpha', q') = (\alpha, q)$.
    \end{itemize}
    The distinct values that can appear are a finite set
    depending only on $\tau$, the weights $w_q$, and the
    velocities $e_q$; the encoding is a hard-coded sequence of
    multi-controlled rotations, without quantum arithmetic.
    Note that we take $v_\text{max}$ as the explicit maximum absolute value that appears in the matrix.
  \item \textbf{Uncomputation.}
    The diagonal-slot velocity copy, $q_a$, and $q_c$ are undone
    in reverse order with some adjustments to match the operations that are applied,
    restoring the ancillae to $\ket{0}$.
\end{enumerate}

\subsubsection{Second-order block encoding ($U_{A}$ for $\Nord=2$)}
\label{sec:a22-encoding}

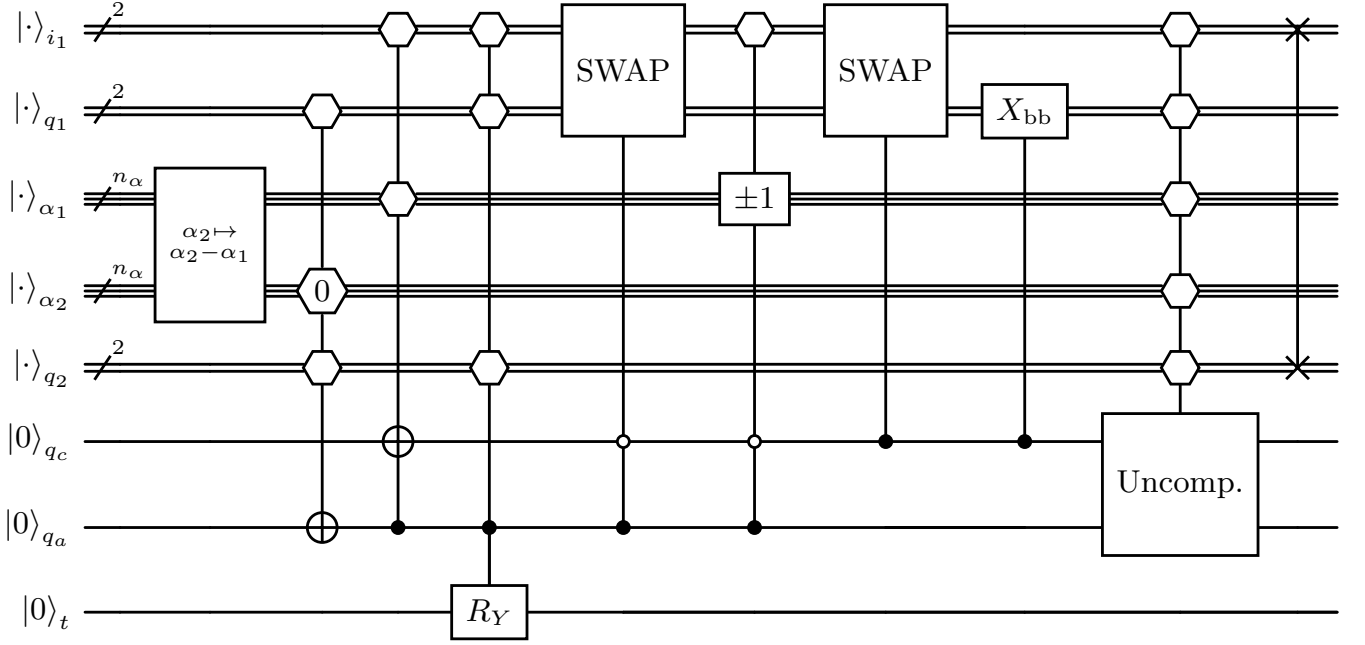
\begin{figure*}[htbp]
\centering
\resizebox{\textwidth}{!}{%
\begin{quantikz}[wire types={c,c,b,b,c,q}, column sep=0.35cm, row sep=0.3cm,classical gap=0.07cm]
  \lstick{$\ket{\cdot}_{i_1}$}      & \qwbundle{2}        &                                                             &                 &    \hexctrl{2}     &     \hexctrl{1}  & \gate[2]{\text{SWAP}} & \hexctrl{2}        & \gate[2]{\text{SWAP}}&                    & \hexctrl{1}              &   \swap{4}   &          \\
  \lstick{$\ket{\cdot}_{q_1}$}      & \qwbundle{2}        &                                                             & \hexctrl{2}     &                     &     \hexctrl{3} &                       &                    &                      & \gate{X_\text{bb}} & \hexctrl{1}              &               &          \\
  \lstick{$\ket{\cdot}_{\alpha_1}$} & \qwbundle{n_\alpha} & \gate[2]{\substack{\alpha_2 \mapsto \\ \alpha_2 - \alpha_1}}&                  &  \hexctrl{3}        &                &                       &  \gate{\pm 1}      &                      &                    & \hexctrl{1}              &               &          \\
  \lstick{$\ket{\cdot}_{\alpha_2}$} & \qwbundle{n_\alpha} &                                                             & \hexlctrl{1}{0} &                     &                &                       &                    &                       &                    &  \hexctrl{1}             &               &          \\
  \lstick{$\ket{\cdot}_{q_2}$}      & \qwbundle{2}        &                                                             & \hexctrl{2}     &                     &   \hexctrl{3}   &                       &                    &                      &                    &  \hexctrl{1}             &     \swap{0}   &          \\
  \lstick{$\ket{0}_{q_c}$}          &                     &                                                             &                 &    \targ{}          &                &     \octrl{-5}        &   \octrl{-3}       & \ctrl{-4}             & \ctrl{-4}          &  \gate[2]{\text{Uncomp.}}  &               &          \\
  \lstick{$\ket{0}_{q_a}$}          &                     &                                                             &  \targ{}        &     \ctrl{-1}       &  \ctrl{1}      &    \ctrl{-1}          &   \ctrl{-1}        &                       &           \qw      &                          &               &          \\
  \lstick{$\ket{0}_t$}              &                     &                                                             &                 &                     &   \gate{R_Y}  &                       &       \qw           &        \qw            &    \qw             &       \qw               &     \qw           &\qw 
\end{quantikz}%
}
\caption{Oracle circuit $O_{A_{12}}$. SWAP denotes bitwise SWAP between registers, blank hexagons means that the register is a control (with some condition) for the gate, hexagons with a value in it means that the gate is applied conditioned on the register being in that value. 
The $R_Y$ gate symbol collectively represent multiple applications of $R_Y$ with different angles, depending on the values in the registers. 
$X_\text{bb}$ is the bounce-back velocity reversal applied on the lowest bit. The final SWAP gate is 
a one-bit swap between bit~1 of $\ket{\cdot}_{i_1}$
and bit~0 of $\ket{\cdot}_{q_2}$.
Other details are in the main text.
}
\label{fig:circuit-OA12}
\end{figure*}

At second-order Carleman truncation, the block encoding must
implement the full rate matrix~\eqref{eq:carleman-system}
with three nontrivial blocks:
the diagonal blocks $A_{11}$ and
$A_{22} = A_{11}\otimes I + I\otimes A_{11}$,
and the off-diagonal block $A_{12}$ encoding the nonlinear
collision coupling from $\bm{f}^{(2)}$ to~$\bm{f}^{(1)}$.
Figure~\ref{fig:circuit-OA12} shows the circuit structure of the
$A_{12}$ oracle, and Fig.~\ref{fig:circuit-ord2} illustrates the
oracle case-branching structure for the second-order block encoding.

\begin{figure}[htbp]
\centering
\resizebox{\columnwidth}{!}{%
\begin{quantikz}[wire types={c,q,b,c,q,b,c}, column sep=0.35cm, row sep=0.3cm,classical gap=0.07cm]
  \lstick{$\ket{\cdot}_\text{case}$}                       & \qwbundle{2}          && \hexlctrl{1}{0}      & \hexlctrl{1}{3} & \hexlctrl{1}{3}         & \hexlctrl{1}{1}          & \hexlctrl{1}{2}                    & \\
  \lstick{$\ket{\cdot}_{\text{ord}}$}                      &                       && \octrl{2}            & \ctrl{2}       & \gate{X}         & \ctrl{2}              & \ctrl{4}                    & \\
  \lstick{$\ket{\cdot}_{\alpha_1}\!\ket{\cdot}_{q_1}$}     & \qwbundle{n_\alpha+2} &&  \gate[3]{O_{A_{11}}}    &  \gate[4]{O_{A_{12}}}      & & \gate[3]{O_{A_{11}}}  &                              & \\
    \lstick{$\ket{\cdot}_{i_1}$}                             & \qwbundle{2}          && &   &  & &       \gate[3]{O_{A_{11}}}        &  \\
  \lstick{$\ket{0}_t$}                                     &                       &&          &            &                         &                       &  &  \\
  \lstick{$\ket{\cdot}_{\alpha_2}\!\ket{\cdot}_{q_2}$}     & \qwbundle{n_\alpha+2} &&                      &                         &                     &  &                       &
\end{quantikz}%
}
\vspace{0.2cm}\\
\small
\begin{tabular}{lll}
  \hline
  Condition & Action & Block \\
  \hline
  $\text{ord}{=}0$, $\text{case}{=}0$
    & $O_{A_{11}}(\alpha_1, q_1)$
    & $A_{11}$ \\
  $\text{ord}{=}1$, $\text{case}{=}1$
    & $O_{A_{11}}(\alpha_1, q_1)$
    & $A_{11} \otimes I$ \\
  $\text{ord}{=}1$, $\text{case}{=}2$
    & $O_{A_{11}}(\alpha_2, q_2)$
    & $I \otimes A_{11}$ \\
  $\text{ord}{=}1$, $\text{case}{=}3$
    & $O_{A_{12}}$
    & $A_{12}$ \\
  \hline
\end{tabular}
\caption{Oracle circuit $O_A$ for the second-order Carleman block
  encoding.
  Hexagons with a value in it means that the gate is applied conditioned on the register being in that value.}
\label{fig:circuit-ord2}
\end{figure}

The quantum register duplicates the first-order registers
$(\ket{\cdot}_{\alpha_1}, \ket{\cdot}_{q_1})$ and
$(\ket{\cdot}_{\alpha_2}, \ket{\cdot}_{q_2})$
for the two copies in the tensor-product space, and adds three
control registers:
a single-qubit \emph{order register} $\ket{\cdot}_{\text{ord}}$
that selects the Carleman subspace ($\ket{0}_\text{ord}$ for
$\bm{f}^{(1)}$, $\ket{1}_\text{ord}$ for $\bm{f}^{(2)}$),
$n_\text{case}=2$ \emph{case-selection} qubits
$\ket{\cdot}_{\text{case}}$,
and $n_i=2$ additional row-label qubits
$\ket{\cdot}_{i_1}$.
The additional row-label register acts similarly to its role in
the first-order encoding: $i_1$ encodes the output velocity label,
and the active block ($A_{11}$, $A_{12}$, or one of the two LCU
slots of $A_{22}$) is selected by $(\text{ord}, \text{case})$ so
that a single $i$-register suffices for all four cases.
In the notation of the general construction in Fig.~\ref{fig:be-sparse}, the primary
register $\ket{\cdot}_x$ is
$\ket{\cdot}_{\alpha_1}\ket{\cdot}_{q_1}
 \ket{\cdot}_{\alpha_2}\ket{\cdot}_{q_2}\ket{\cdot}_{\text{ord}}$,
and the row-label register $\ket{\cdot}_i$ is
$\ket{\cdot}_{\text{case}}\ket{\cdot}_{i_1}$.
The Hadamard layer in Eq.~\eqref{eq:UM-structure} is applied to
the $n_i + n_\text{case} = 4$ row-label qubits.
Including one target qubit, the input/output register count
is $2\lceil\log_2 \Nx\rceil + 10$;
additional ancilla qubits required by the sub-oracles
(case flags, subtractor workspace) are reported together
with the compiled gate counts in Sec.~\ref{sec:results}.

The oracle $O_A$ branches on the order and case registers
as follows.

\paragraph{Diagonal blocks ($A_{11}$ and $A_{22}$).}
When $\text{ord}=0$ and $\text{case}=0$, the oracle applies the
first-order subroutine $O_{A_{11}}$ to the first-copy registers
$\ket{\cdot}_{\alpha_1}\ket{\cdot}_{q_1}$, encoding the $A_{11}$
block.

For $A_{22}$, we exploit the decomposition
$A_{22} = A_{11}\otimes I + I\otimes A_{11}$ as a linear
combination of unitaries (LCU).
The case register serves directly as the LCU selection qubit:
when $\text{ord}=1$ and $\text{case}=1$, the oracle applies
$O_{A_{11}}$ to the first-copy registers
$\ket{\cdot}_{\alpha_1}\ket{\cdot}_{q_1}$, implementing
$A_{11}\otimes I$;
when $\text{case}=2$, it applies $O_{A_{11}}$ to the
second-copy registers $\ket{\cdot}_{\alpha_2}\ket{\cdot}_{q_2}$,
implementing $I\otimes A_{11}$.
The Hadamard preparation on case creates the equal
superposition, and projection onto $\ket{0}_\text{case}$ in the
final Hadamard layer extracts the sum, yielding
$A_{22}/\lambda_A$ in the block encoding.
This construction reuses the first-order oracle as a controlled
subroutine without additional ancilla qubits.

\paragraph{Off-diagonal block ($A_{12}$).}
When $\text{ord}=1$ and $\text{case}=3$, the oracle $O_{A_{12}}$ encodes the nonlinear collision
coupling~$A_{12}$.
Since $A_{12}$ maps the second-order subspace ($\text{ord}=1$) to
the first-order subspace ($\text{ord}=0$), a controlled-$X$ on
the order register, controlled by $\text{case}=3$, flips
$\ket{1}_\text{ord} \to \ket{0}_\text{ord}$
(see Fig.~\ref{fig:circuit-ord2}).

The $A_{12}$ oracle acts on the first-copy registers
$\ket{\cdot}_{\alpha_1}\ket{\cdot}_{q_1}$, the second-copy registers
$\ket{\cdot}_{\alpha_2}\ket{\cdot}_{q_2}$, the row-label register
$\ket{\cdot}_{i_1}$ (where $i_1$ encodes the output
row velocity $q' \in \{0, 1, 2\}$), and a target qubit $\ket{0}_t$.
The input state represents a column
$(\alpha_1, q_1, \alpha_2, q_2)$ of~$A_{12}$; the oracle updates
$\ket{\cdot}_{\alpha_1}\ket{\cdot}_{q_1}$ in place to hold the
output row $(\alpha', q')$.
The structure of Eq.~\eqref{eq:A12-elements} is realized by the
following operations:
\begin{enumerate}
  \item \textbf{Spatial matching.}
    A Majority-based adder circuit~\cite{Cuccaro2004-majority}
    computes $\ket{\alpha_1 {-} \alpha_2}_{\alpha_2}$ in place,
    overwriting $\ket{\cdot}_{\alpha_2}$.
    An ancilla $q_a$ is then set to $\ket{1}_{q_a}$ when
    $\alpha_1 = \alpha_2$ (i.e.\ $\ket{0}_{\alpha_2}$) and both
    velocity registers are valid ($q_1, q_2 < Q$), enforcing the
    $\delta_{\alpha_1\alpha_2}$ factor.
  \item \textbf{Boundary detection.}
    A second ancilla $q_c$ is set to $\ket{1}_{q_c}$ on the
    bounce-back branch, exactly as in the first-order oracle: the
    output row velocity stored in the first two qubits of
    $\ket{\cdot}_{i_1}$ is checked against the boundary of
    $\ket{\cdot}_{\alpha_1}$.
  \item \textbf{Value encoding.}
    Multi-controlled $R_Y$ rotations on $\ket{0}_t$ encode the
    collision kernel $K^{(12)}_{q',\,q_1,\,q_2}$ of
    Eq.~\eqref{eq:A12-kernel}, conditioned on
    $\ket{\cdot}_{q_1}$ and $\ket{\cdot}_{q_2}$ and on the output
    row velocity $q'$ held in the first row-label register
    $\ket{\cdot}_{i_1}$.
    Because $K^{(12)}_{0,\,q_1,\,q_2} = K^{(12)}_{1,\,q_1,\,q_2}$
    for all $q_1, q_2$ in D1Q3, the $q' = 0$ and $q' = 1$ branches
    share the same rotation angles and are controlled jointly.
    This encoding is performed \emph{before} the in-place index
    update, while the velocity registers still hold their input
    values.\footnote{In our numerical experiments, the $q_a$ control on the
    $R_Y$ rotations is omitted, as it has no effect on the output in our setup.}
  \item \textbf{In-place index update.}
    As in the first-order oracle, a bitwise SWAP between
    $\ket{\cdot}_{q_1}$ and $\ket{\cdot}_{i_1}$ moves the output row
    velocity (originally in $\ket{\cdot}_{i_1}$) into the $q_1$
    register to be $\ket{i_1}_{q_1}$. A subsequent
    controlled spatial shift $\alpha_1 \to \alpha_1 + e_{i_1}$ on
    the bulk branch completes the streaming step.
    On the bounce-back branch ($\ket{1}_{q_c}$) the same SWAP is
    applied, followed by a low-bit flip
    $i_1 \to \bar{i_1}$ on $\ket{i_1}_{q_1}$;
    $\ket{\cdot}_{\alpha_1}$ is left unchanged.
  \item \textbf{Uncomputation.}
    The flags $q_c$ and $q_a$ are undone (with control registers
    adjusted for the post-SWAP state).
    Finally, a one-bit swap between bit~1 of $\ket{q_1}_{i_1}$
    and bit~0 of $\ket{\cdot}_{q_2}$ is applied. Using that 
    $q_1, q_2 \in \{0, 1\}$ for a nonzero matrix value Eq.~(\ref{eq:A12-kernel}) (so
    their bit~1 is zero), this swap overwrites
    $\ket{q_1}_{i_1}$ with the two-bit value $q_1 + 2 q_2$ and
    resets $\ket{q_2}_{q_2}$ to $\ket{0}_{q_2}$. This operation
    achieves (i)~the row-label register including
    $\ket{\cdot}_{i_1}$ uniquely labels the nonzero entries of the
    column, (ii)~the second-copy velocity register is cleared to $\ket{0}_{q_2}$,
    matching the $\ket{0}_{\alpha_2}\ket{0}_{q_2}$ convention of
    the $\ket{0}_\text{ord}$ subspace, while maintaining the
    invertibility.
    The subtractor of step~1 is not inverted, and more generally
    the $q_a = \ket{0}$ branch is left uncleaned: any residue there
    lies outside the physical $\ket{0}_\text{ord}$ row subspace
    (with $\alpha_2 = q_2 = 0$) that the block-encoding projection
    picks out, so it does not contribute to physical matrix
    elements.
\end{enumerate}

\paragraph{Verification.}
We verified both block encodings entry by entry.  Each compiled
circuit's encoded matrix is extracted by sweeping computational-basis
inputs $\ket{j}_{\mathrm{sys}}\ket{0}_{\mathrm{anc}}$, applying the
circuit, reading the $\ket{0}_{\mathrm{anc}}$ projection of the
output, and multiplying by the encoder normalisation $\lambda_A$.
Block-by-block comparison against the classical
$A_{11}$~\eqref{eq:A11-elements}, $A_{12}$~\eqref{eq:A12-elements},
and $A_{22} = A_{11}\otimes I + I \otimes A_{11}$ at $\nu = 2.0$
agrees to machine precision: maximum entry-wise differences are
$4\times 10^{-16}$ for $\Nord = 1$ at $\Nx = 128$, and below
$1.5 \times 10^{-14}$ for each of $A_{11}$, $A_{12}$, $A_{22}$
at $\Nord = 2$, $\Nx = 8$.  Details are given in
Appendix~\ref{app:verification}.

\subsubsection{Block encoding of $L$}
\label{sec:be-L}

The block encoding $U_L$ of the time-evolution matrix $L$
(Fig.~\ref{fig:L-matrix}) is implemented using the $A$-matrix
oracles of Sec.~\ref{sec:A11-block} and~\ref{sec:a22-encoding} as
controlled subroutines.

\begin{figure*}[htbp]
\centering
\resizebox{\textwidth}{!}{%
\begin{quantikz}[wire types={b,c,q,b,b,q}, row sep=0.4cm, column sep=0.35cm]
  \lstick{$\ket{\cdot}_m$}                & \qwbundle{n_m}   &                            &                       & \hexctrl{3}          & \hexctrl{3}          & \hexctrl{3}          & \hexctrl{1}      & \hexctrl{5}          & \gate{+1}        &                                                  &  \hexctrl{2}     & \gate[2]{(k,m){+}{+}} &              &  \\
  \lstick{$\ket{\cdot}_k$}                & \qwbundle{n_k=2} &                            &                       & \hexlctrl{2}{0}      & \hexlctrl{2}{1}      & \hexlctrl{2}{2}      & \gate{+1}        &  \hexctrl{4}         &  \hexctrl{-1}    &  \gate[3]{\substack{k\mapsto 0 \\ i \mapsto k}}  &  \hexctrl{1}     &                       &              &  \\
  \lstick{$\ket{0}_{i_{L,0}}$}                &                  & \gate{H}                   & \ctrl{3}              & \octrl{1}            & \octrl{1}            & \octrl{1}            & \octrl{-1}       & \ctrl{3}             & \ctrl{-2}        &                                                  & \octrl{3}        & \octrl{-2}            & \gate{H}     &  \\
  \lstick{$\ket{0}_{i_A}$}                & \qwbundle{n_i^A} & \gate{H^{\otimes {n_i^A}}} & \hexlctrl{2}{\bm{1}}  & \gate[3]{U_A^{(0)}}  & \gate[3]{U_A^{(1)}}  & \gate[3]{U_A^{(2)}}  &                  & \hexlctrl{2}{0}      &  \hexlctrl{-2}{0}&                                                  & \hexlctrl{2}{0}  &  \hexlctrl{-1}{0}     & \gate{H^{\otimes {n_i^A}}}     &  \\
  \lstick{spatial + vel.}                 & \qwbundle{}      &                            &                       &                      &                      &                      &                  &                      &                  &                                                  &                  &                       &              &  \\
  \lstick{$\ket{0}_t$}                    &                  &                            & \gate{R_Y}            &                      &                      &                      &                  & \gate{R_Y}           &                  &                                                  & \gate{R_Y}       &                       & \gate{X}     &
\end{quantikz}%
}
\caption{Block encoding circuit $U_L$ for the $L$-matrix
  ($\NK = 3$, $n_k = 2$).
  Blank hexagons means that the register is a control (with some condition) for the gate, hexagons with a value in it means that the gate is applied conditioned on the register being in that value, while bold $\bm{1}$ means all-1's in the register, i.e., $\bm{1}=2^{n_i^A}-1$. 
  $R_Y$ gate symbols may collectively represent multiple applications of $R_Y$ with different angles, depending on the values in the registers. 
  Other details are in the main text.}
\label{fig:circuit-UL}
\end{figure*}
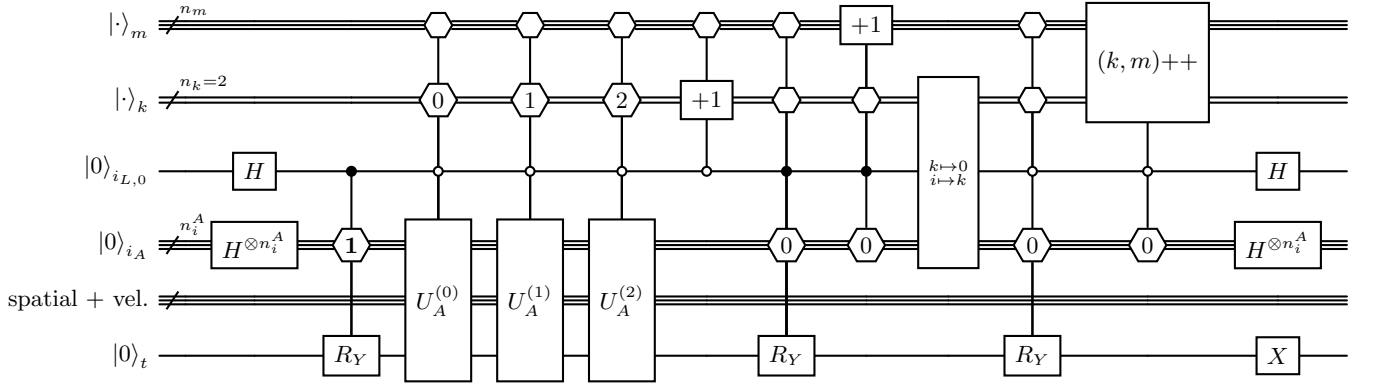

The block encoding of~$L$ requires a Taylor-index register
$\ket{\cdot}_k$ of $n_k=\lceil\log_2 (\NK+1) \rceil$ qubits,
a time-step register $\ket{\cdot}_m$ of
$n_m = \lceil\log_2 (\Nt)+1 \rceil$ qubits, and a composite
row-label register whose total size is
$n_i^L = \max(1 + n_i^A,\; 2 + n_k)$ qubits,
where $n_i^A$ is the row-label size of $U_A$.
The block-encoding normalization is then
$\lambda_L = 2^{n_i^L}\,L_\text{max}$ with
$L_\text{max} \equiv \max\bigl(1,\, \Deltat\,\max_{i,j}|A_{ij}|\bigr)$.
The time register $\ket{\cdot}_m$ with $m<\Nt$ tracks the time evolution phase,
while $m\geq\Nt$ is used for the idling operation.
It means that the idling number $p$ is automatically $\Nt\times \NK$ in this implementation.
For the row-label register $\ket{\cdot}_{i_L}=:\ket{\cdot}_{i_{L,0}} \ket{\cdot}_{i_A}$, the first $n_i^A$
qubits are passed to $U_A$ as $\ket{\cdot}_{i_A}$, while the
$L$-matrix block encoding exploits unused values of $i_L$ to label
other contributions.
The oracle $O_L$ branches on the row label~$i_L$ and the
current $(m, k)$ indices:
\begin{enumerate}
  \item \textbf{Diagonal ($i_L = 2^{n_i^L}{-}1$):} identity
    contribution ($R_Y$ encodes $+1$).
  \item \textbf{Taylor off-diagonal ($i_L < 2^{n_i^A}$,
    $k < \NK$, $m < \Nt$):} a controlled call to $U_A$
    encoding $-\Deltat\,A/(k{+}1)$, followed by incrementing
    $k \to k{+}1$.
  \item \textbf{Time-step coupling
    ($i_L = 2^{n_i^L - 1}$, $m < \Nt$):}
    encodes $-1$, increments $m \to m{+}1$, and resets $k \to 0$.
  \item \textbf{Terminal copy ($m \geq \Nt$):} encodes $-1$
    with increment of the block row index.
\end{enumerate}
Figure~\ref{fig:circuit-UL} shows the circuit structure of $U_L$.

A na\"ive implementation of $U_L$ would place $\NK$ independent
copies of the $A_{11}$ oracle inside $U_L$, one per subdiagonal
block as in Fig.~\ref{fig:circuit-UL}. 
The index-manipulation part (streaming, bounce-back,
bit flips) is, however, identical for every $k$, and only the
$R_Y$ rotation angles depend on~$k$ through the Taylor coefficient
$-\Deltat/(k{+}1)$.  A single $O_A$ structure can therefore be
shared across all $\NK$ subdiagonal blocks, with the $R_Y$ rotations
applied by a $k$-conditional control. We adopt this shared-oracle
variant of $U_L$ throughout this work; the resulting savings in
gate count are quantified in Sec.~\ref{sec:gate-scaling}.

\subsubsection{\texorpdfstring{Solving $L\bm{x}=\bm{b}$ via QSVT}{Solving Lx=b via QSVT}}
\label{sec:qsvt}

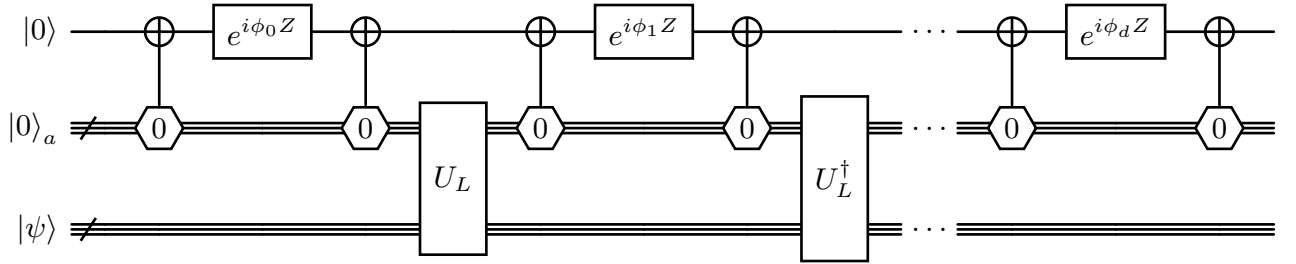
\begin{figure*}[htbp]
\centering
\resizebox{0.95\textwidth}{!}{%
\begin{quantikz}[wire types={q,b,b}, row sep=0.4cm, column sep=0.35cm]
  \lstick{$\ket{0}$}           &             & \targ{}          & \gate{e^{i\phi_0 Z}} & \targ{}          &                & \targ{}          & \gate{e^{i\phi_1 Z}} & \targ{}          &                       & \ \ldots\  & \targ{}          & \gate{e^{i\phi_d Z}} & \targ{}          &  \\
  \lstick{$\ket{0}_a$}         & \qwbundle{} & \hexlctrl{-1}{0} &                      & \hexlctrl{-1}{0} & \gate[2]{U_L}  & \hexlctrl{-1}{0} &                      & \hexlctrl{-1}{0} & \gate[2]{U_L^\dagger} & \ \ldots\  & \hexlctrl{-1}{0} &                      & \hexlctrl{-1}{0} &  \\
  \lstick{$\ket{\psi}$}        & \qwbundle{} &                  &                      &                  &                &                  &                      &                  &                       & \ \ldots\  &                  &                      &                  &
\end{quantikz}%
}
\caption{QSVT circuit for polynomial transformation of the
  block-encoded $L$-matrix.
  Hexagons with 0 in it means that the gate is applied conditioned on the register being all-0's.
  The signal-processing qubit (top wire) is acted upon by $d_\text{QSVT}{+}1$
  phase rotations $e^{i\phi_j Z}$, interleaved with alternating
  applications of $U_L$ and $U_L^\dagger$.
  The ancilla register $\ket{0}_a$ includes all ancilla qubits
  of the $L$-matrix block encoding (row labels $\ket{\cdot}_{i_A}$
  and $\ket{\cdot}_{i_L}$, and target $\ket{0}_t$).
  The system register $\ket{\psi}$ encodes the time-step,
  spatial, and velocity indices.}
\label{fig:circuit-qsvt}
\end{figure*}

The linear system $L \bm{x} = \bm{b}$ is solved using the
quantum singular value transformation
(QSVT)~\cite{Gilyen2019-qsvt}.
QSVT applies a polynomial transformation $p(\sigma)$ to each
singular value $\sigma$ of the block-encoded matrix.
By choosing $p(\sigma) \approx 1/\sigma$ (a polynomial approximation
to the inverse function), we implement $L^{-1}$ as a quantum circuit.
The QSVT circuit alternates applications of $U_L$ and $U_L^\dagger$
with single-qubit phase rotations on a dedicated QSVT ancilla qubit,
for a total of $d_\text{QSVT} + 1$ oracle calls, where $d_\text{QSVT}$ is the polynomial degree.
The QSP phase factors $\{\phi_0, \ldots, \phi_d\}$ are computed
classically using the optimization-based algorithm~\cite{Dong2021-qsp}.
The degree~$d_\text{QSVT}$ required for a given precision $\varepsilon$
scales as $d_\text{QSVT} = O(\kappa \log(\kappa/\varepsilon))$, where
$\kappa = \lambda_L / \sigma_\text{min}(L)$ is the \emph{effective
condition number}, defined as the ratio of the block-encoding
normalization $\lambda_L$ of $U_L$ to the smallest relevant singular
value of~$L$, following Ref.~\cite{Ueno2026-linear};
see also Ref.~\cite{Gilyen2019-qsvt} for the QSVT cost bound.
The QSVT circuit structure is shown in Fig.~\ref{fig:circuit-qsvt}.

The right-hand side $\bm{b}$ encodes the initial condition: $\bm{b}$
is zero everywhere except in the $m{=}0$ block, where it
contains the initial distribution function $\bm{f}(0)$.
In the quantum circuit, $\bm{b}$ is prepared by initializing the
time-step register to $\ket{0}$ and loading
$\bm{f}(0)$ into the state register.
For the step-function initial condition used in this work, $\bm{f}(0)$
is a local equilibrium state that can be prepared efficiently.

\section{Results}
\label{sec:results}

In this section we demonstrate the circuits and algorithms introduced
above on a concrete one-dimensional flow problem and characterize
their numerical accuracy and asymptotic cost.
The presentation is organized as follows.
Section~\ref{sec:problem-setup} defines the test problem and shows the
classically solved reference flow (Fig.~\ref{fig:bgk-ode-flow}).
Section~\ref{sec:carleman-error} quantifies the Carleman truncation
error as a function of simulation time
$T = \Deltat \Nt$ (time-step size $\Deltat$, number of time steps $\Nt$;
Fig.~\ref{fig:carleman-l2norm}) and of the initial density step
$\Delta\rho$ (Fig.~\ref{fig:carleman-C123}), together with the
corresponding density profile and residual
(Fig.~\ref{fig:carleman-profile-error}).
Section~\ref{sec:taylor-order} evaluates the Taylor truncation
error as a function of the Taylor order $\NK$ at three time-step
sizes (Fig.~\ref{fig:taylor-K}).
Section~\ref{sec:kappa-scaling} characterizes the effective
condition number $\kappa(L)$ of the Taylor-ODE linear system as a
function of the spatial grid size $\Nx$ and $\Nt$ (Fig.~\ref{fig:kappa-baseline}), of $\NK$
(Fig.~\ref{fig:kappa-vs-K}), and of the Carleman truncation order
$\Nord$ (Fig.~\ref{fig:kappa-C1-vs-C2}), and analyses the
optimal choice of $\NK$ (Fig.~\ref{fig:K-vs-h}).
Section~\ref{sec:qsvt-results} demonstrates the QSVT-based time
evolution for $\Nord = 1$, including a sweep of the polynomial
degree $d_\text{QSVT}$ (Fig.~\ref{fig:qsvt-d-comparison}) and a comparison of
$\NK = 1$ and $\NK = 3$ under the same QSVT cost (Fig.~\ref{fig:qsvt-K-comparison});
the $\Nord = 2$ QSVT simulation is presented in
Fig.~\ref{fig:qsvt-c2-verify}.
Section~\ref{sec:gate-scaling} reports gate-count and qubit-count
scaling for both $\Nord = 1$ and $\Nord = 2$
(Figs.~\ref{fig:gatecount} and~\ref{fig:qubitcount}).

All quantum circuit simulations were performed through the
QURI SDK~\cite{QURISDK} framework, using
Qulacs~\cite{Suzuki2021-qulacs} as the CPU state-vector simulator
and the cuQuantum GPU backend via
QURI Parts cuQuantum~\cite{QuriPartsCuquantum}.
All quantum circuits are compiled so that they only contain
the gates in an elementary gate set
and their multi-controlled versions, before simulation.

\subsection{Problem setup and classical reference}
\label{sec:problem-setup}

We consider density wave propagation in a one-dimensional channel
with bounce-back walls at both ends.
The initial condition is a step-function density profile:
$\rho(\alpha) = 1 + \Delta\rho/2$ for $\alpha < \Nx/2$ and
$\rho(\alpha) =1 - \Delta\rho/2$ for $\alpha \geq \Nx/2$.
We use $\Delta\rho = 0.4$ as the
default case, and study the dependence on the initial density step
$\Delta\rho$ in Sec.~\ref{sec:carleman-error}.
The initial distribution is set to the local
equilibrium~\eqref{eq:feq} at zero velocity.

We parametrize the flow by a grid-independent dimensionless
viscosity~$\nu$ that is related to the BGK relaxation time through
\begin{equation}
  \tau = \left(\frac{\Nx}{512}\right)\frac{\nu}{c_s^2},
  \label{eq:tau-nu}
\end{equation}
so that $\tau$ scales linearly with the grid resolution and $\nu$
labels a family of physically equivalent runs\footnote{The
classical LBM reference used for comparison is implemented in its
standard fully discrete form with collision operator
$f \leftarrow f - 2\beta\,(f - f^{\mathrm{eq}})$ and
$\beta = 1/(2\tau + 1)$~\cite{Kruger2016-lj}, with $\tau$ taken
from Eq.~\eqref{eq:tau-nu}; Chapman--Enskog analysis of that
scheme recovers the kinematic viscosity $c_s^2\,\tau$, so runs
at the same~$\nu$ are directly comparable between the
continuous-time formulation of Sec.~\ref{sec:lbm} and the
discrete-time reference.}.
Unless otherwise noted, all results in the rest of the paper use
$\nu = 2.0$.

\begin{figure}[htbp]
  \centering
  \includegraphics[width=1.0\columnwidth]{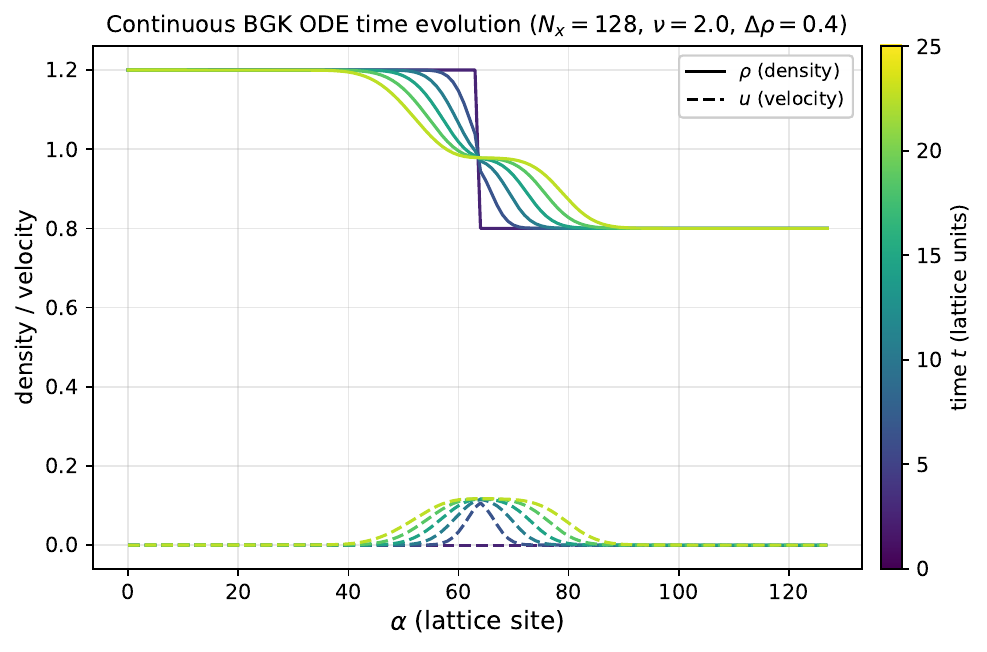}
  \caption{Continuous BGK reference flow at $T = 25$
    $\Nx = 128$, $\nu = 2.0$, $\Delta\rho = 0.4$:
    density~$\rho$ and velocity~$u$.}
  \label{fig:bgk-ode-flow}
\end{figure}

Figure~\ref{fig:bgk-ode-flow} shows a classical solution of Eq.~\eqref{eq:lbm},
obtained by a
fourth-order Runge--Kutta scheme with step size $\Deltat = 0.01$,
at $\Nx = 128$ with the simulation time $T = 25$.
The initial step evolves into a pair of density waves propagating
in opposite directions; the nonlinear collision term in the BGK
equation~\eqref{eq:dvbe} (via its quadratic dependence on the local
velocity in $f^\mathrm{eq}$) produces a left--right asymmetry in the
velocity profile, and this is the nonlinear effect targeted throughout this paper.

We adopt this \emph{continuous BGK} model,
opposed to the standard \emph{discrete-time} LBM,
as the reference solution for
the rest of this paper, since it is the $\Nord \to \infty$ limit of
the Carleman ODE that the quantum algorithm of
Sec.~\ref{sec:circuits} solves.  At $T = 25$ the
discrete LBM and the continuous BGK agree pointwise at
$L^\infty \approx 6 \times 10^{-3}$, i.e., the largest density difference
along the spatial direction was at most $\sim 0.6\%$
(Appendix~\ref{app:lbm-bgk-residual}).
We therefore use the continuous BGK as the reference solution for
measuring algorithmic errors in the following sections.

\subsection{Carleman truncation error}
\label{sec:carleman-error}

\begin{figure*}[htbp]
  \centering
  \includegraphics[width=0.95\textwidth]{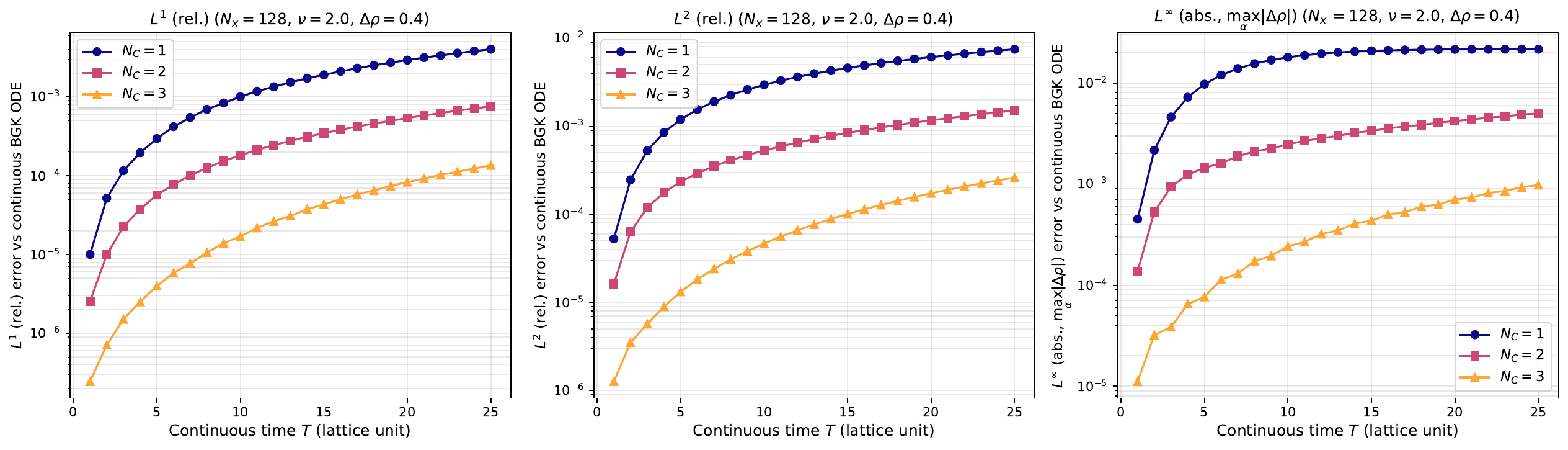}
  \caption{Relative $L^1$ (left), $L^2$ (middle), and pointwise
    $L^\infty$ (right) error of the Carleman-linearized density
    with respect to the continuous BGK reference as a function of
    simulation time $T = \Deltat\Nt$
    ($\Nx = 128$, $\nu = 2.0$, symmetric initial condition
    $\rho \in [1-\Delta\rho/2,\,1+\Delta\rho/2]$ with
    $\Delta\rho = 0.4$) for $\Nord = 1, 2, 3$.}
  \label{fig:carleman-l2norm}
\end{figure*}

\begin{figure}[htbp]
  \centering
  \includegraphics[width=0.95\linewidth]%
    {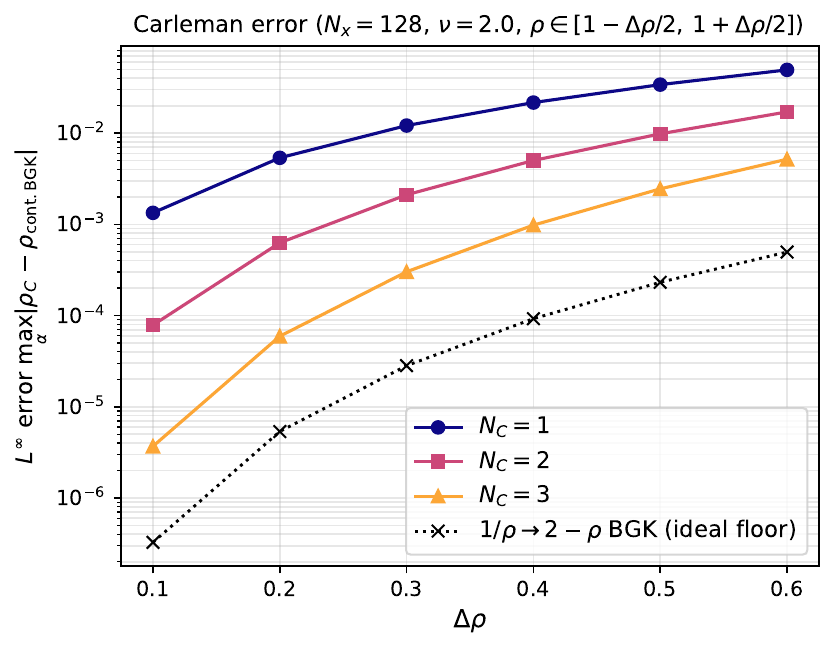}
  \caption{Final-time pointwise $L^\infty$ error
    $\max_\alpha|\rho_C(\alpha) - \rho_\text{BGK}(\alpha)|$ of the
    Carleman density against the continuous BGK reference as a
    function of the initial density step~$\Delta\rho$ for the
    symmetric setup
    $\rho \in [1-\Delta\rho/2,\,1+\Delta\rho/2]$
    ($\Nx = 128$, $\nu = 2.0$) and $\Nord = 1, 2, 3$.
    The dotted black curve is obtained by replacing the exact
    $1/\rho$ in the BGK collision with its first-order expansion
    $2 - \rho$.}
  \label{fig:carleman-C123}
\end{figure}

\begin{figure*}[htbp]
  \centering
  \includegraphics[width=0.92\textwidth]%
    {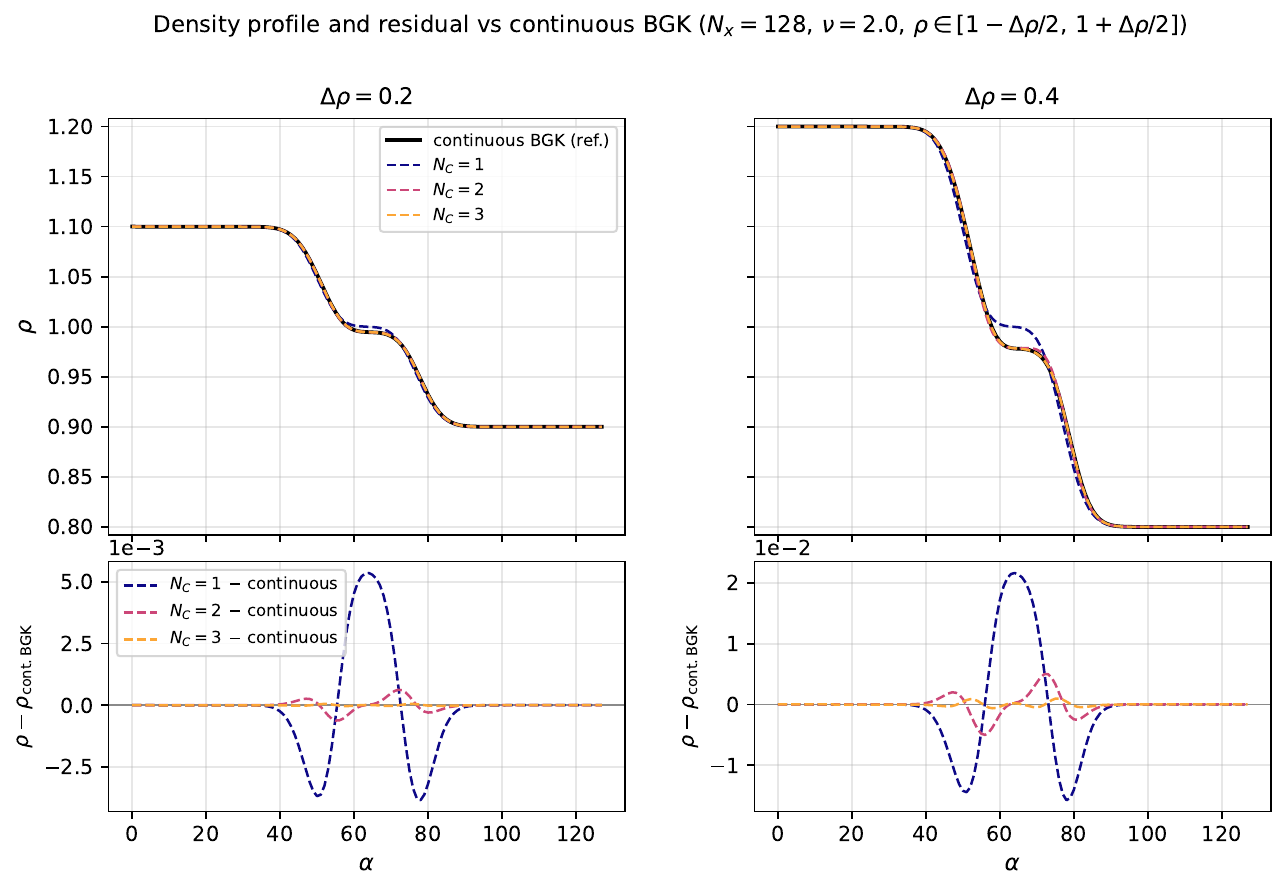}
  \caption{Final-time density profile (top row) and residual
    relative to the continuous BGK reference (bottom row) at
    $\Delta\rho = 0.2$ (left) and $\Delta\rho = 0.4$ (right).
    Parameters as in Fig.~\ref{fig:carleman-C123}. The top panels
    share the same $\rho$-axis range; residual panels have
    $\Delta\rho$-dependent scales.}
  \label{fig:carleman-profile-error}
\end{figure*}

This subsection studies the Carleman truncation error against the
continuous BGK reference at
$\Nx = 128$.  Three figures are presented:
Fig.~\ref{fig:carleman-l2norm} shows the time evolution of the
relative $L^1$, $L^2$, and pointwise $L^\infty$ errors at
$\Delta\rho = 0.4$;
Fig.~\ref{fig:carleman-C123} reports the final-time $L^\infty$ error
as a function of the initial density step $\Delta\rho$ for
$\Nord = 1, 2, 3$;
Fig.~\ref{fig:carleman-profile-error} compares the density profile
and residual at $\Delta\rho = 0.2$ and $0.4$.

Classical Carleman results are obtained by fourth-order
Runge--Kutta integration of the system
$\dot{\mathbf{f}}^{(c)} = \sum_{d=1}^{\Nord} A_{cd}\,\mathbf{f}^{(d)}$
($c = 1, \ldots, \Nord$), with each Carleman block
$\mathbf{f}^{(c)}$ stored as a separate state vector and the
couplings $A_{cd}$ applied directly, using the halving/quartering convention of
Appendix~\ref{app:scales}.
Errors are reported against the continuous BGK reference; for
comparison we additionally integrate the same equation with the
collision $1/\rho$ replaced by its first-order expansion
$2 - \rho$, which serves as the irreducible floor that any
finite-$\Nord$ Carleman trajectory inherits from the $1/\rho$
truncation alone.
Figure~\ref{fig:carleman-l2norm} shows the relative $L^1$, $L^2$,
and pointwise $L^\infty$ errors of
$\rho_{(\Nord)}(t) - \rho_\text{BGK}(t)$ as a function of
simulation time $T$ at the default $\Delta\rho = 0.4$.
The three norms are qualitatively similar; we focus on the
pointwise $L^\infty$ error in the discussion below,
which picks out the largest deviation at any spatial point.
The error grows with $T$ and decreases monotonically with
$\Nord$ at every time, indicating convergence in the Carleman
truncation order.

Figure~\ref{fig:carleman-C123} reports the final-time pointwise
$L^\infty$ error against the continuous BGK reference as a
function of the initial density step $\Delta\rho \in [0.1, 0.6]$.
Monotone improvement with $\Nord$ is maintained across the range:
at $\Delta\rho = 0.4$, $\Nord = 3$ already reaches below $0.1\%$.
The $1/\rho \to 2 - \rho$ 
continuous BGK reference (dashed) sits below $0.01\%$,
about an order of magnitude below at the
same $\Delta\rho$.  The dominant residual at $\Nord = 3$ is
therefore the Carleman truncation itself, not the
$1/\rho \to 2 - \rho$ approximation of the BGK collision.

Figure~\ref{fig:carleman-profile-error} shows final-time density
profiles and residuals at $\Delta\rho = 0.2$ and $0.4$;
$\Nord = 3$ collapses onto the continuous BGK reference at
sub-percent level at both density gaps.
The residual panels make visible how the spatial structure of the
deviation is progressively flattened as $\Nord$ increases.

\subsection{Effect of Taylor truncation order}
\label{sec:taylor-order}

To study the error introduced by the Taylor truncation
in the Taylor ODE solver, we solve the $L$-matrix
linear system classically by sparse LU factorization and
compare the resulting density profiles against the exact
matrix exponential $e^{TA}\bm{f}(0)$.

\begin{figure*}[htbp]
  \centering
  \includegraphics[width=0.95\textwidth]{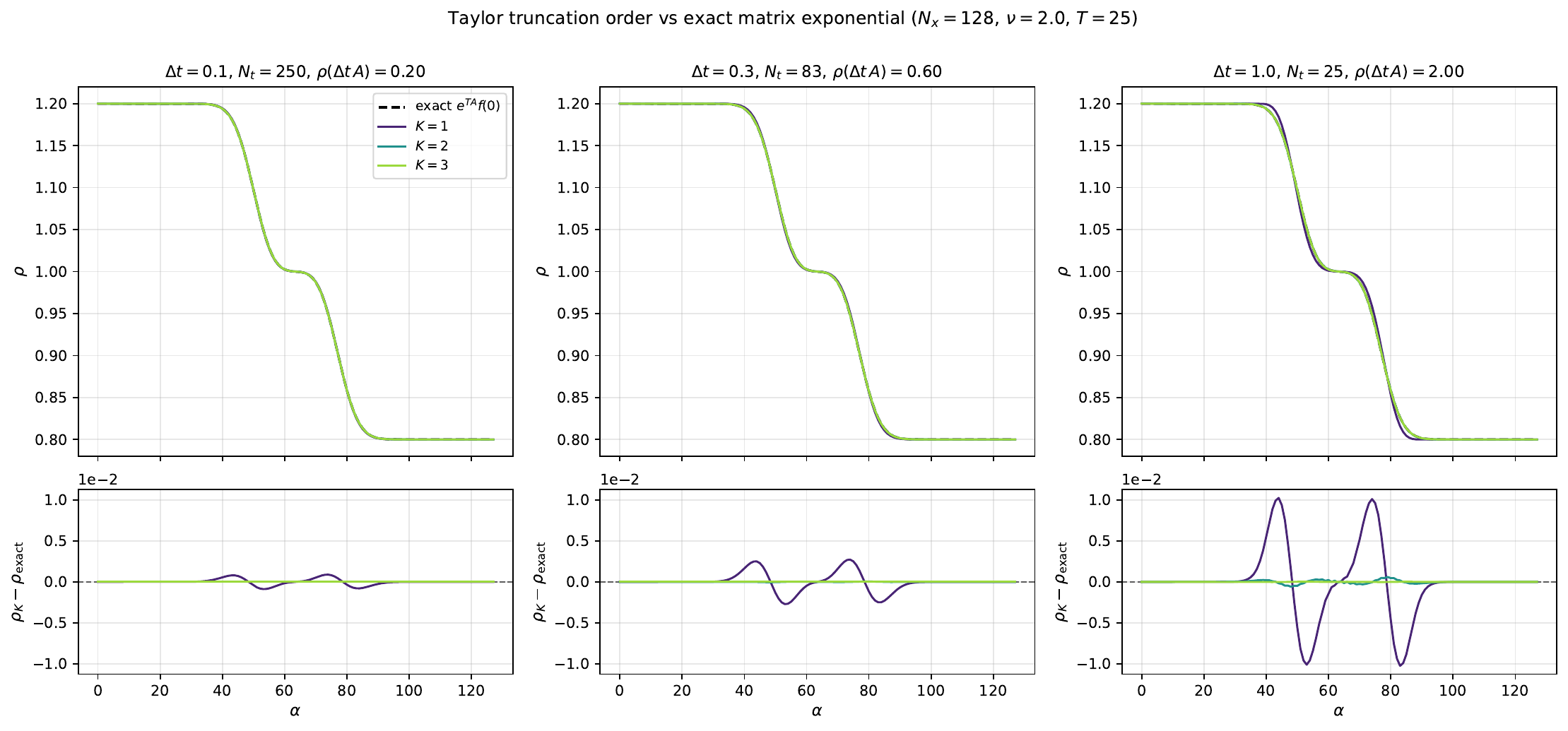}
  \caption{Effect of Taylor truncation order~$\NK$ on the density
    profile at time $T = 25$ ($\Nx = 128$, $\nu = 2.0$,
    first-order Carleman).
    Dashed black: exact matrix exponential $e^{TA}\bm{f}(0)$.
    Solid: sparse $L$-matrix solve for $\NK = 1, 2, 3$.
    Left: $\Deltat = 0.1$ ($\Nt = 250$,
    $\mu(\Deltat A) = 0.2$).
    Middle: $\Deltat = 0.3$ ($\Nt = 83$,
    $\mu(\Deltat A) = 0.6$).
    Right: $\Deltat = 1.0$ ($\Nt = 25$,
    $\mu(\Deltat A) = 2.0$).}
  \label{fig:taylor-K}
\end{figure*}

Figure~\ref{fig:taylor-K} shows the density at time $T = 25$
for $\Nx = 128$ with the standard $\Delta\rho = 0.4$,
at three time-step sizes $\Deltat = 0.1$ ($\Nt = 250$),
$\Deltat = 0.3$ ($\Nt = 83$), and $\Deltat = 1.0$ ($\Nt = 25$).
At $\Deltat = 0.1$, all three Taylor orders $\NK = 1, 2, 3$ match the
exact matrix exponential to the $10^{-3}$ level.
At $\Deltat = 1.0$ ($\rho = 2$) the $\NK = 1$ truncation visibly
underestimates the density gradient, while $\NK = 2$ and $\NK = 3$
remain close to the exact solution. It is notable that
the error in $\Deltat=1.0$ for $\NK=2$ is below that of $\NK=1$
in $\Deltat=0.1$, and the $\NK=3$ error is not visible even in $\Deltat=1.0$.
This shows that raising $\NK$ relaxes the $\Deltat$ constraint set by
accuracy; whether the resulting $\Nt$ reduction outweighs the
per-oracle gate count increase is discussed in Sec.~\ref{sec:disc-K}.
We note that, while the Taylor expansion itself has infinite convergence radius and converges for arbitrary $\Deltat$, the effective condition number $\kappa(L)$ that controls the QSVT cost (Sec.~\ref{sec:qsvt}) diverges once $\Deltat$ exceeds certain range (Sec.~\ref{sec:kappa-scaling}). Raising $\NK$ alone therefore cannot accommodate arbitrarily large $\Deltat$.

\subsection{Effective condition number scaling of the $L$-matrix}
\label{sec:kappa-scaling}

This subsection presents the effective condition number
$\kappa(L)$ that is a dominant factor in the QSVT cost in various settings.
Recall from Sec.~\ref{sec:qsvt} that the QSVT polynomial degree $d_\text{QSVT}$
required for accurate matrix inversion scales as
$d_\text{QSVT} = O(\kappa(L) \log(\kappa/\varepsilon))$ in the
\emph{effective condition number}
$\kappa(L) = \lambda_L/\sigma_\text{min}(L)$,
the ratio of the block-encoding normalization $\lambda_L$ of $U_L$
to the smallest relevant singular value $\sigma_\text{min}(L)$ of~$L$.
Understanding how $\kappa(L)$ depends on $\Nx$, $\Nt$, and the
Carleman order~$\Nord$ is therefore essential for estimating the
total computational cost.
We study the $\kappa(L)$ scaling in $\Nt$
and $\Nx$ (Fig.~\ref{fig:kappa-baseline}),
in the Taylor truncation
order $\NK$ (Fig.~\ref{fig:kappa-vs-K}), 
and in the Carleman order $\Nord$ (Fig.~\ref{fig:kappa-C1-vs-C2}).
Optimal choices of $\NK$ and $\Deltat$ for a given simulation time $T$ and accuracy
are discussed based on Fig.~\ref{fig:K-vs-h}.

\subsubsection{Dependence on $\Nx$ and $\Nt$}
\begin{figure*}[htbp]
  \centering
  \includegraphics[width=0.95\textwidth]{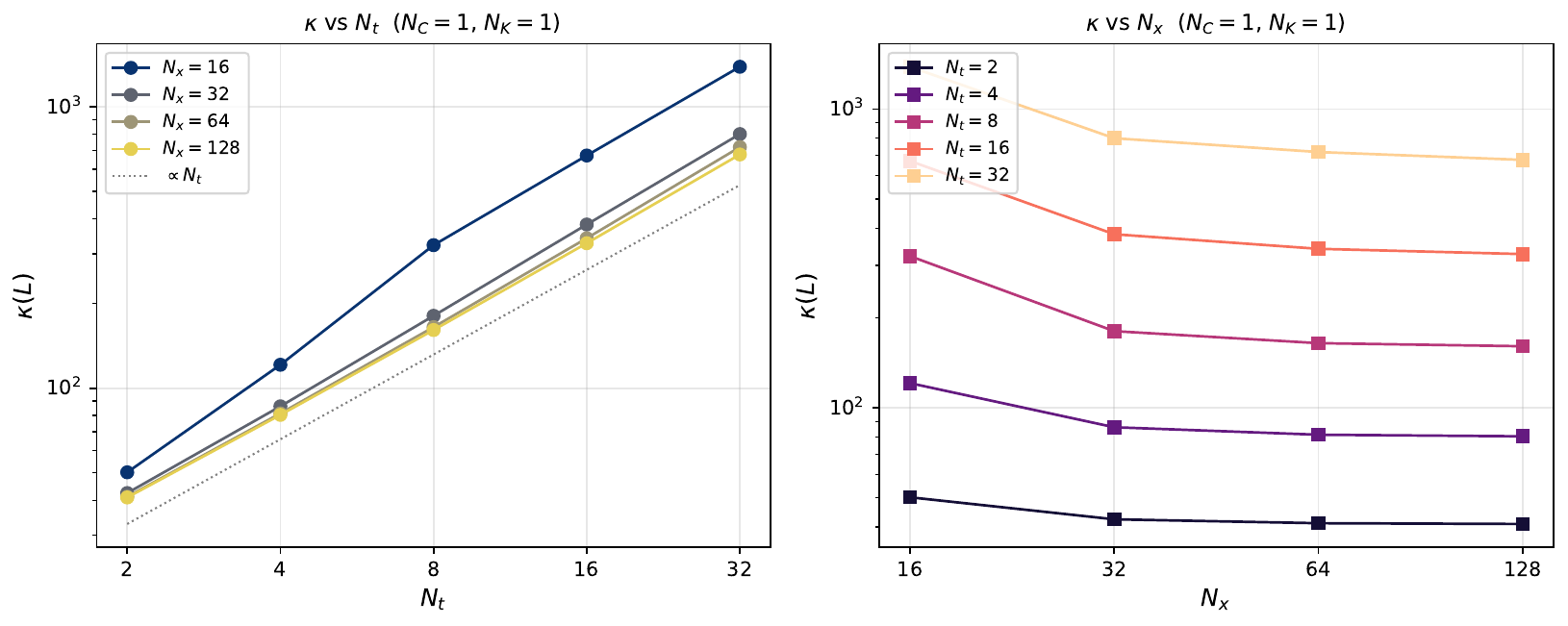}
  \caption{Baseline effective condition number $\kappa(L)
    = \lambda_L/\sigma_\text{min}(L)$ at $\Nord = 1$, $\NK = 1$,
    $\Deltat = 0.1$, $\nu = 2.0$. Left: $\kappa$ vs $\Nt$
    for $\Nx \in \{16, 32, 64, 128\}$; the dotted line is a
    $\propto \Nt$ guide. Right: $\kappa$ vs $\Nx$ for
    $\Nt \in \{2, 4, 8, 16, 32\}$.}
  \label{fig:kappa-baseline}
\end{figure*}

Figure~\ref{fig:kappa-baseline} shows $\kappa(L)$ at $\Nord = 1$,
$\NK = 1$ as a function of $\Nt$ and $\Nx$.
At fixed $\Nt$ (right panel), $\kappa(L)$ decays with increasing $\Nx$ to a
constant value.
This saturation reflects the $\Nx$ dependence of the spectral
radius $\mu(A)$ of the rate matrix
(Appendix~\ref{app:spectral-radius}): $\mu(A) \propto 1/\Nx$ when the
collision rate $1/\tau$ dominates (small $\Nx$, via
Eq.~\eqref{eq:tau-nu}), and $\mu(A) \to 2$ for $\Nord = 1$
($4$ for $\Nord = 2$) when streaming dominates (large $\Nx$).
We refer to the streaming-dominant phase $\mu(\Deltat A) \lesssim 1$ as
the \emph{stable regime}.
Within the stable regime ($\Nx \geq 32$ at $\Deltat = 0.1$, where
$\mu(\Deltat A_{11}) \leq 0.27$), $\kappa(L)$ grows linearly with
$\Nt$ (left panel) at slope $\kappa/\Nt \approx 80$--$100$, saturating the
structural $\Omega(T)$ lower
bound~\cite{Berry2017-vv, Jennings2025-theory} and matching
Ref.~\cite{Ueno2026-linear}; the $\Nx = 16$ line lies outside the
stable regime ($\mu(\Deltat A_{11}) = 0.53$) and sits visibly above
this asymptote.

\subsubsection{Dependence on the Taylor truncation order $\NK$}
\label{sec:kappa-vs-K}
\begin{figure}[htbp]
  \centering
  \includegraphics[width=\columnwidth]{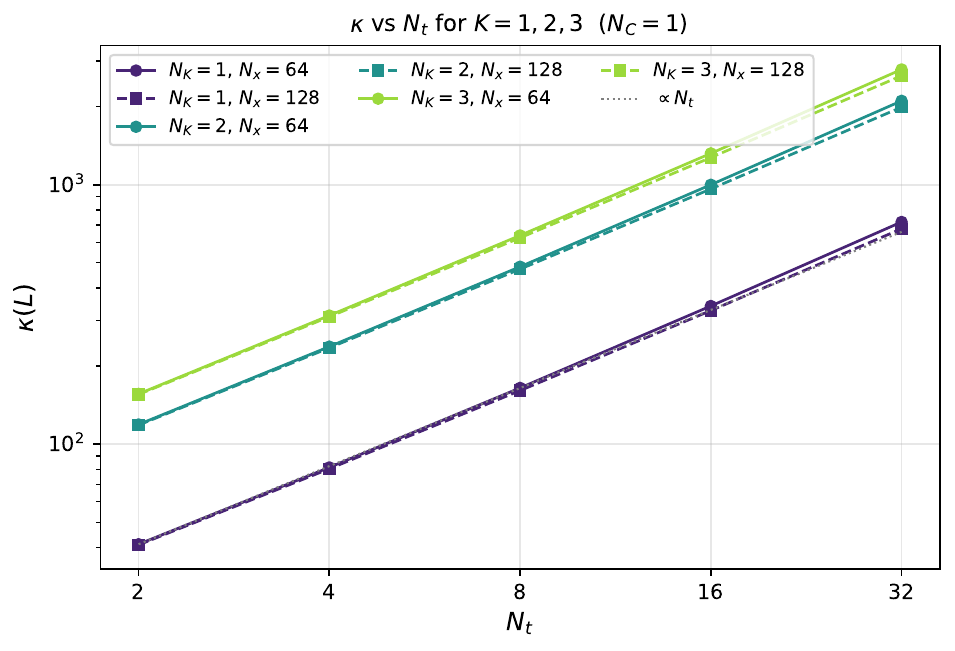}
  \caption{Effective condition number $\kappa(L)$ as a function
    of~$\Nt$ for Taylor orders $\NK = 1, 2, 3$ ($\Nord = 1$, $\Deltat = 0.1$, $\nu = 2.0$).}
  \label{fig:kappa-vs-K}
\end{figure}

Figure~\ref{fig:kappa-vs-K} shows $\kappa(L)$ for
$\NK = 1, 2, 3$ across $\Nx \in \{64, 128\}$ and
$\Nt \in \{2, 4, 8, 16\}$: the linear
$\kappa(L) \propto \Nt$ scaling persists for each $\NK$, and
$\kappa_{\NK=3}/\kappa_{\NK=1} \sim 4$ throughout the stable regime.
Half\footnote{The rest is attributed to the decrease in the smallest singular value.} of this comes from the block-encoding normalization
$\lambda_L = 2^{n_i^L}\,L_\text{max}$ doubling: at $\Nord = 1$ we have
$2 + n_k > 1 + n_i^A$ for $n_k\geq 2$, so
$n_i^L = \max(1 + n_i^A,\, 2 + n_k) = 2 + n_k$ jumps from $3$
to $4$ when $\NK$ goes from $1$ to $3$ (i.e.~$n_k$ from $1$ to $2$).
At $\Nord = 2$ (with $n_i^A = 4$), $2 + n_k > 1 + n_i^A$ holds only
for $n_k\geq 4$, so this contribution vanishes up to $\NK = 15$.
Considering that the Taylor truncation error shrinks rapidly
with $\NK$ (Sec.~\ref{sec:taylor-order}), $\NK = 3$ can be a natural
operating point on the $\kappa$--$\Deltat$ trade-off, discussed
in detail in Sec.~\ref{sec:K-vs-h}.

\subsubsection{Dependence on the Carleman order $\Nord$}
\label{sec:kappa-C2}

\begin{figure*}[htbp]
  \centering
  \includegraphics[width=0.95\textwidth]{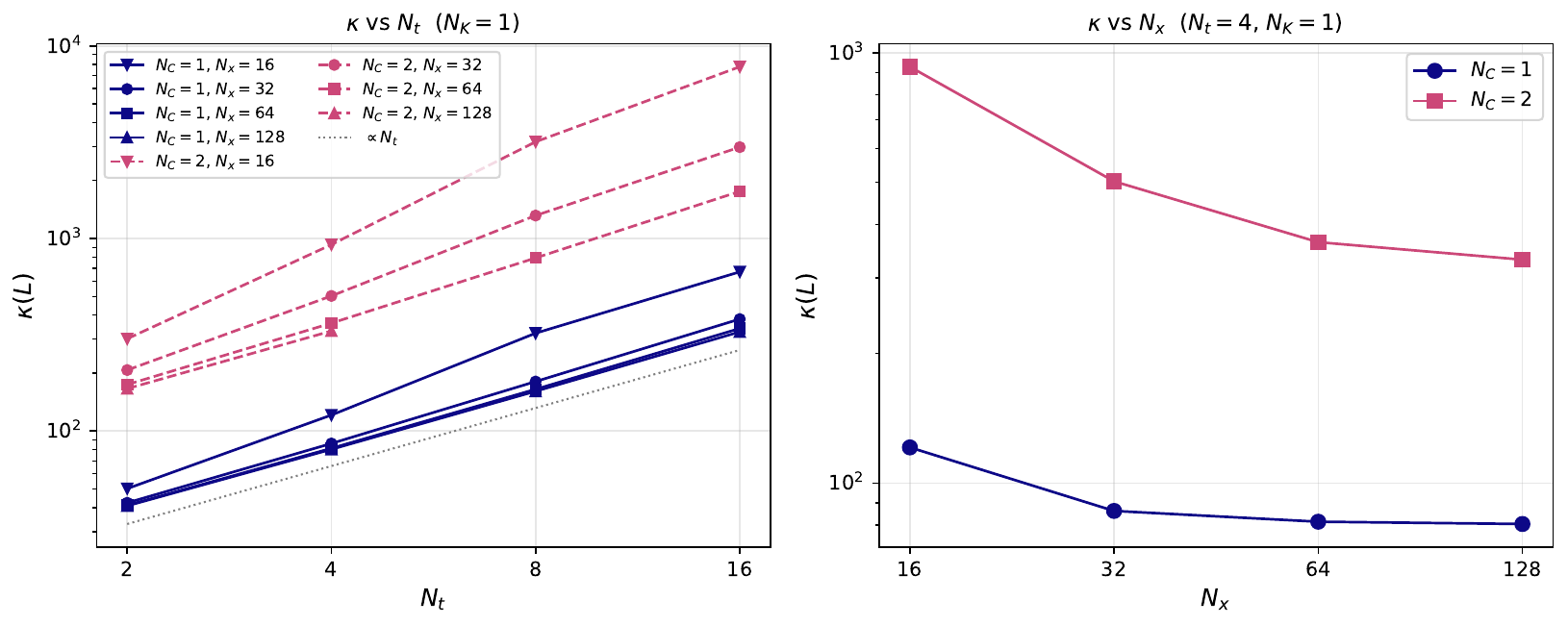}
  \caption{Effective condition number $\kappa(L)$ for
    $\Nord = 1$ (blue) and $\Nord = 2$ (red) at $\NK = 1$,
    $\Deltat = 0.1$, $\nu = 2.0$.
    Left: $\kappa$ vs $\Nt$ for
    $\Nx \in \{16, 32, 64, 128\}$; the gray dotted line is a
    $\propto \Nt$ guide and some of $\Nx=128$ data is missing due to computational limitations.
    Right: $\kappa$ vs $\Nx$ at $\Nt = 4$.}
  \label{fig:kappa-C1-vs-C2}
\end{figure*}

Extending the analysis to the second-order Carleman ($\Nord = 2$)
requires solving an enlarged linear system whose state-space dimension
is $Q\Nx + Q^2\Nx^2$ instead of $Q\Nx$.
The spectral radius $\mu(A)$ is exactly $2\times$ its
$\Nord = 1$ value, which shifts the stable regime to larger~$\Nx$.
Figure~\ref{fig:kappa-C1-vs-C2} compares $\kappa(L)$ for $\Nord = 1$
and $\Nord = 2$ at $\NK = 1$.
In the stable regime ($\Nx \geq 64$ for $\Nord = 2$),
$\kappa_{\Nord=2}(L)/\kappa_{\Nord=1}(L)$ stays close to~$\approx 4$:
at $\Nx = 64$ the ratio ranges from $\approx 4.2$ at $\Nt = 2$ to
$\approx 5.2$ at $\Nt = 16$, and at $\Nx = 128$ it is $\approx 4.1$
at $\Nt \in \{2, 4\}$.
This is dominated by the expansion of the block-encoding
normalization $\lambda_L = 2^{n_i^L} \cdot L_\text{max}$.
At $\NK = 1$ we have $n_i^A = 2$ for $\Nord = 1$ (the velocity index
alone) and $n_i^A = 4$ for $\Nord = 2$ (velocity index plus the
$2$-qubit case register, Sec.~\ref{sec:a22-encoding}), giving
$n_i^L = \max(1{+}n_i^A,\,2{+}n_k) = 3$ and $5$ respectively;
$L_\text{max} = 1$ in both cases since
$\Deltat \cdot \max_{ij}|A_{ij}| \leq 1$.
The resulting $\lambda_L$ thus grows from $8$ to $32$, a factor of~$4$,
while the singular-value-based spectral condition number contributes
only a $\sim 1.05$--$1.3$ factor on top.
At $\Nx = 32$ the ratio is larger ($\approx 5$--$7$ for
$\Nt \in \{2, 4, 8\}$) because 
the spectral radius 
$\mu(\Deltat A_{22}) \approx 1.07$
in this case
sits on the stability boundary, the point before the divergence in
the right panel of Fig.~\ref{fig:kappa-C1-vs-C2},
lifting $\sigma_{\min}$ above the asymptotic value; the linear $\kappa(L) \propto \Nt$ scaling becomes
clean only for $\Nx \geq 64$.

\subsubsection{$\NK$-vs-$\Deltat$ trade-off and the QSVT polynomial degree}
\label{sec:K-vs-h}

As we have seen in Sec.~\ref{sec:taylor-order}, a higher Taylor
truncation order $\NK$ allows a larger time step $\Deltat$ to be taken
at a fixed accuracy target.  This sets up a non-trivial
trade-off between $\NK$ and $\Deltat$ at fixed simulation time
$T = \Nt \Deltat$: increasing $\NK$ raises the per-block contribution
to $\kappa(L)$ (Fig.~\ref{fig:kappa-vs-K}), while the larger
$\Deltat$ that higher $\NK$ admits reduces $\Nt$ and therefore
$\kappa(L)$, which is observed to be proportional to $\Nt$
(Fig.~\ref{fig:kappa-baseline}).

\begin{figure*}[htbp]
  \centering
  \includegraphics[width=0.95\textwidth]{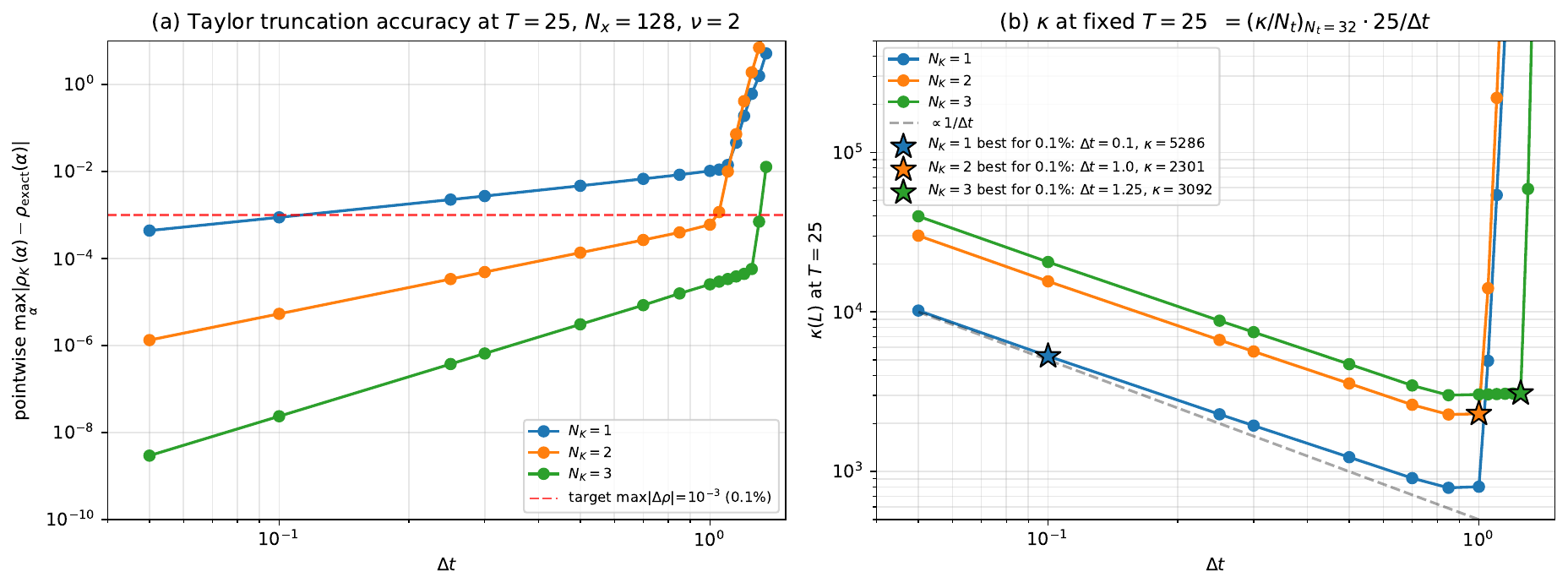}
  \caption{$\NK$-vs-$\Deltat$ trade-off at $T = 25$, $\Nx = 128$,
    $\nu = 2.0$.  Left: pointwise $L^\infty$ Taylor truncation error
    against the exact matrix exponential as a function of $\Deltat$
    for $\NK = 1, 2, 3$.  Red dashed: $\max_\alpha|\Delta\rho|=10^{-3}$
    target.
    Right: $\kappa(L) = \lambda_L/\sigma_\text{min}(L)$ at
    $T = 25$ as a function of $\Deltat$ ($\propto 1/\Deltat$ guide
    in black dashed); stars mark the largest $\Deltat$ at each $\NK$
    that meets the $10^{-3}$ pointwise accuracy target.}
  \label{fig:K-vs-h}
\end{figure*}

This trade-off is evident in Fig.~\ref{fig:K-vs-h}. 
The analysis in this subsection is performed at first-order
Carleman ($\Nord = 1$) throughout.
The left panel
plots the pointwise Taylor truncation error of the classical
$L$-matrix solve against the exact matrix exponential as a function
of $\Deltat$, at $T = 25$, $\Nx = 128$ and
$\NK = 1, 2, 3$; the right panel plots the corresponding effective
condition number $\kappa(L)$ at the same operating points.
For example, let us take the target error of $10^{-3}$,
so that the Carleman truncation error is the dominant source of inaccuracy.
In that case, the choice of $\Deltat$ in
$\NK = 1$ is \emph{accuracy-limited}: the pointwise residual already
reaches the $10^{-3}$ target at $\Deltat \approx 0.1$, so
$\Nt \gtrsim 250$ steps are needed and the polynomial-design
parameter $\kappa_\text{QSVT}$ must accommodate
$\kappa \approx 5300$.
$\NK = 3$ is \emph{stability-limited}: the Taylor residual stays
several orders of magnitude below $10^{-3}$ all the way up to the
stability boundary at $\Deltat \approx 1.25$, where $\Nt = 20$ steps
suffice and the corresponding $\kappa \approx 3100$ is
about $1.7\times$ smaller.  In particular, $\Deltat > 1$ is
practical for $\NK = 3$, which hints that our approach
of continuous time evolution may be advantageous over the discrete-time LBM
approach in terms of the time step, where $\Deltat$ is fixed to be one by the formulation.

An increasing per-oracle gate cost to implement larger $\NK$ may spoil this advantage, though.
The Taylor-index register grows only as
$\lceil \log_2(\NK+1) \rceil$ qubits, so the qubit-count overhead of
raising $\NK$ is negligible. 
At the gate level, the shared-oracle
implementation of $U_L$ (Sec.~\ref{sec:be-L}) keeps the dominant
Toffoli contribution nearly $\NK$-independent, but the $R_Y$ count
grows roughly linearly in $\NK$ and, given the large T-count of an
$R_Y$ rotation (roughly $100$ and $200$ T gates for accuracies
$10^{-10}$ and $10^{-20}$, respectively~\cite{Ross2016-rs}),
dominates the per-oracle T cost (Fig.~\ref{fig:UL-gatecount}).
For $\NK = 3$ to be advantageous in this particular case, the
$R_Y$ count must therefore be reduced by an order of magnitude to be sub-dominant in T-count, 
which may be achievable by, for instance, grouping the
cases that encode the same rotation angle, possibly with the aid
of additional ancilla qubits,
while it is true that other optimizations could also improve the Toffoli count.

A similar analysis at $\Nx = 32$, $T = 5$, the size at which the
$\Nord = 1$ QSVT runs of Sec.~\ref{sec:qsvt-results} are performed,
is collected in Appendix~\ref{app:K-vs-h-Nx32}.  The qualitative
picture is unchanged at the smaller size, while $\Deltat>1$ 
is no longer available here.

\subsection{QSVT time evolution}
\label{sec:qsvt-results}

This subsection collects the QSVT-circuit results of the paper.
Three figures are presented.
Fig.~\ref{fig:qsvt-d-comparison} demonstrates the QSVT
time evolution by a minimal setup of $\Nord=1$, $\NK=1$,
and also studies the effect of the QSVT polynomial degree $d_\text{QSVT}$ 
by sweeping $d_\text{QSVT}/\kappa_\text{QSVT}$
for two polynomial design parameters $\kappa_\text{QSVT}$ that
bracket $\kappa(L)$.
Fig.~\ref{fig:qsvt-c2-verify} extends the demonstration to
$\Nord = 2$ and $\NK=3$ with small but nonzero nonlinear corrections,
providing the end-to-end verification of the
Carleman-linearized Boltzmann QSVT circuit with leading nonlinear effects.
Fig.~\ref{fig:qsvt-K-comparison} compares the
$\NK = 1$ accuracy-limited setup against the $\NK = 3$
stability-limited setup at the \emph{same} polynomial design
parameter $\kappa_\text{QSVT}$, illustrating the
$\NK$-vs-$\Deltat$ trade-off in $d_\text{QSVT}$; the
operating points are set based on
Fig.~\ref{fig:K-vs-h-Nx32} in
Appendix~\ref{app:K-vs-h-Nx32}, following the discussion of
Sec.~\ref{sec:K-vs-h}.

\subsubsection{QSVT simulation with various $d_\text{QSVT}$ and $\kappa_\text{QSVT}$}
\label{sec:qsvt-degree-saturation}

We perform a state-vector simulation
of the full QSVT circuit at $\Nx = 32$, $\Nt = 32$
(first-order Carleman $\Nord=1$, $\NK = 1$, $\Deltat = 0.1$), using the
$L$-matrix QSVT circuit of Sec.~\ref{sec:qsvt}.
The effective condition number at this operating point is
$\kappa(L) = \lambda_L/\sigma_\text{min}(L) \approx 800$
with $\lambda_L = 2^{n_{Li}} \cdot L_\text{max} = 8$; the QSP polynomial
is constructed to approximate $1/x$ on
$[1/\kappa_\text{QSVT}, 1]$, and we compare two settings:
$\kappa_\text{QSVT} = 400$ (an undershoot case at
$\kappa_\text{QSVT} \approx \tfrac{1}{2} \kappa$) and
$\kappa_\text{QSVT} = 1000$ (an overshoot case).

\begin{figure*}[htbp]
  \centering
  \includegraphics[width=0.85\textwidth]{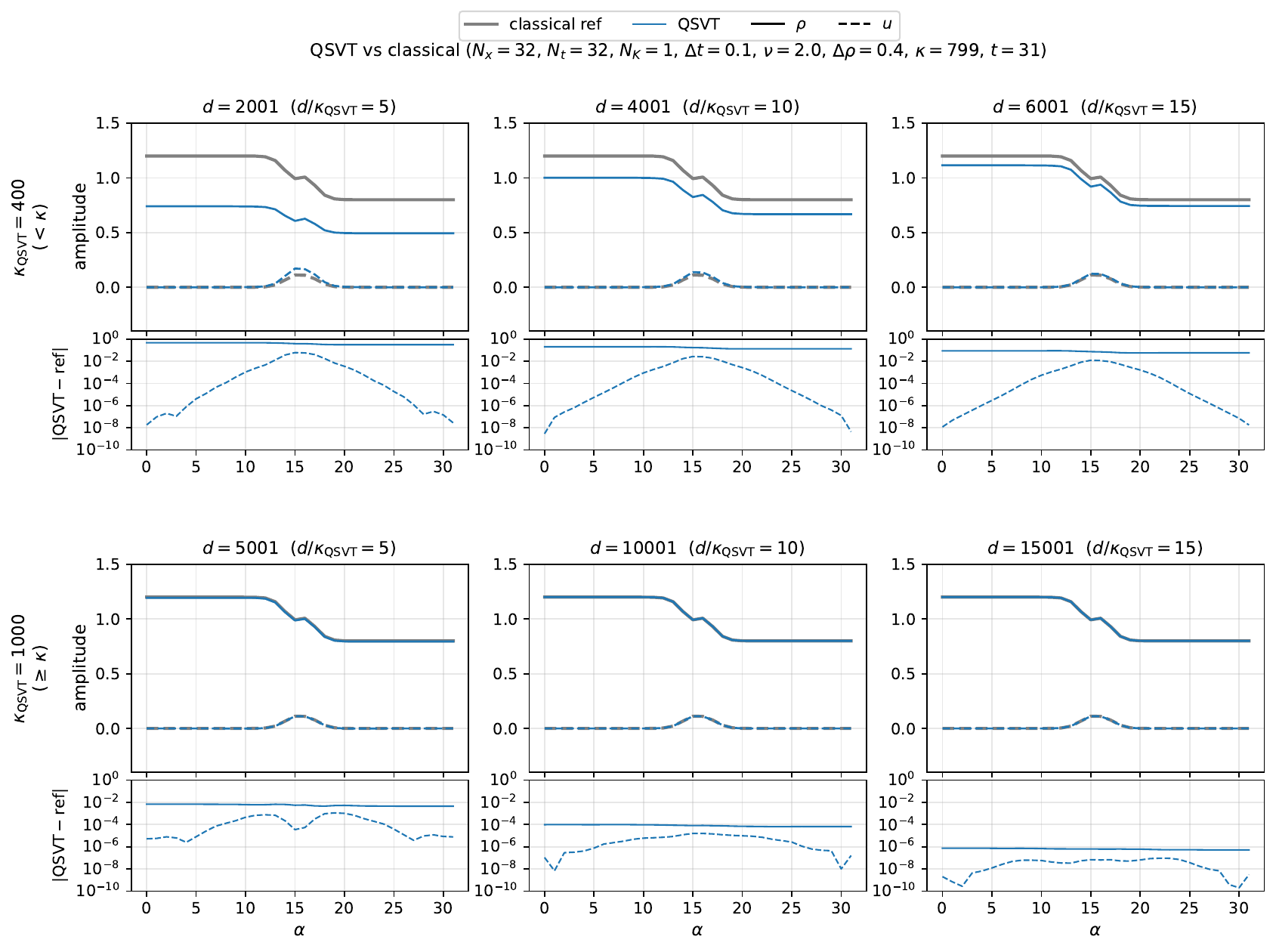}
  \caption{QSVT-based multi-step time evolution for $\Nx = 32$,
    $\Nt = 32$, $\Nord = 1$, $\NK = 1$, $\Deltat = 0.1$, $\nu = 2.0$,
    symmetric $\rho \in [0.8, 1.2]$ ($\Delta\rho = 0.4$), at the
    final time step $t = 31$.
    Two polynomial design targets span the measured effective
    condition number $\kappa(L) \approx 800$:
    $\kappa_\text{QSVT} = 400$ (top row, undershoot,
    $\kappa_\text{QSVT} \approx \tfrac{1}{2} \kappa$)
    and $\kappa_\text{QSVT} = 1000$ (bottom row, overshoot).
    Each row scans three polynomial degrees $d_\text{QSVT} = 5\kappa_\text{QSVT}
    {+}1$, $10\kappa_\text{QSVT}{+}1$, $15\kappa_\text{QSVT}{+}1$
    (i.e. $d_\text{QSVT}/\kappa_\text{QSVT} \in \{5, 10, 15\}$).
    Top sub-panel of each cell: classical reference (gray) and
    QSVT output (blue); density~$\rho$ and velocity~$u$ are
    distinguished by line style.
    Bottom sub-panel: absolute residual $|\text{QSVT} - \text{ref}|$
    on a logarithmic axis.}
  \label{fig:qsvt-d-comparison}
\end{figure*}

Figure~\ref{fig:qsvt-d-comparison} shows the resulting final-time
density, velocity, and pressure profiles for these two
$\kappa_\text{QSVT}$ settings, each scanned at three polynomial
degrees $d_\text{QSVT}/\kappa_\text{QSVT} \in \{5, 10, 15\}$.
As $d_\text{QSVT}$ grows, the QSVT output converges to the classical reference
both in shape and in amplitude.
When $\kappa_\text{QSVT} \geq \kappa$, the polynomial
approximates $1/x$ accurately on the full singular-value spectrum
of $L/\lambda_L$ and the convergence is essentially complete at
$d_\text{QSVT}/\kappa_\text{QSVT} = 10$: the QSVT output matches the classical reference
to a $\lesssim 10^{-5}$ relative error, as visible in the residual
sub-panel of each cell.
When $\kappa_\text{QSVT} < \kappa$, singular values
below $\lambda_L/\kappa_\text{QSVT}$ lie outside the polynomial's
design domain and are under-inverted, visible as the reduced
amplitude at $d_\text{QSVT}/\kappa_\text{QSVT} = 5$ in the top row; the residual deficit
shrinks with increasing $d_\text{QSVT}$ but saturates at a non-zero value
controlled by the mismatch $\kappa/\kappa_\text{QSVT}$.
In the undershoot row, it may seem like the QSVT density retains the overall
profile shape with a suppressed amplitude, but the velocity
profile, which is independent of the overall amplitude, is also visibly distorted;
the deviation therefore cannot be absorbed into a uniform amplitude rescaling.

This result confirms that $\sigma_\text{min}(L)$ is the
appropriate indicator for the QSVT polynomial-degree cost: even at
$\kappa_\text{QSVT} = \kappa/2 =
\lambda_L/(2\sigma_\text{min})$, the result is visibly inaccurate,
implying that the singular values in
$[\sigma_\text{min}, 2\sigma_\text{min}]$ that fall outside the
polynomial design domain contribute non-trivially to the
physically relevant state space, so $\sigma_\text{min}$
cannot be replaced by a larger threshold without sacrificing
accuracy.  The $\sigma_\text{min}$-based cost estimate of
Sec.~\ref{sec:gate-scaling} is therefore appropriate in this sense.

\subsubsection{QSVT simulation of $\Nord = 2$ case}
\label{sec:qsvt-c2-verification}

\begin{figure}[htbp]
  \centering
  \includegraphics[width=\columnwidth]{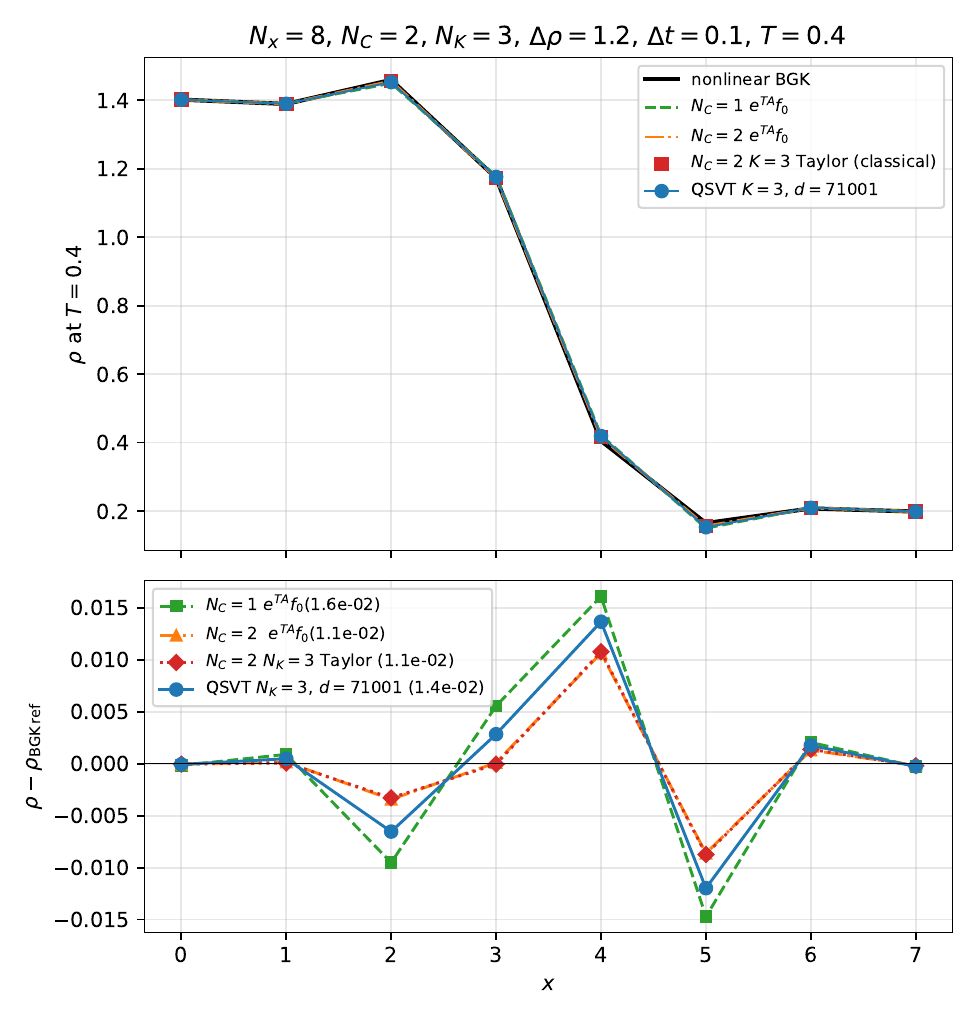}
  \caption{$\Nord = 2$ QSVT simulation at
  $\Nx = 8$, $\Nt = 4$, $\NK = 3$, $\Deltat = 0.1$, $T = 0.4$,
  $\Delta\rho = 1.2$ ($\kappa_\text{eff} \approx 7{,}100$,
  $27$~qubits).
  Top: density profile $\rho(x)$ at $T = 0.4$, overlaying the
  nonlinear BGK reference (black), the linear $\Nord = 1$
  exponential $e^{T A_{11}} \mathbf{f}_0$ (green dashed), the
  exact $\Nord = 2$ exponential $e^{T A}$ (orange dash-dot), the
  classical $\NK = 3$ Taylor evaluation of $\Nord = 2$ (red squares,
  the target of the QSVT circuit), and the QSVT output
  ($d = 71{,}001$, blue circles).
  Bottom: residual of each method against the BGK reference, on a
  shared axis.}
  \label{fig:qsvt-c2-verify}
\end{figure}

To verify the $\Nord = 2$ implementation end-to-end, we run the
QSVT circuit at
$\Nx = 8$, $\Nt = 4$, $\NK = 3$, $\nu = 2$, $\Deltat = 0.1$,
$\Delta\rho = 1.2$, giving $T = \Nt \Deltat = 0.4$ and the
classical nonlinear correction
$|\rho_{\Nord = 1} - \rho_{\Nord = 2}|_\infty \approx 6.1 \times
10^{-3}$ (i.e.\ $\approx 0.5\%$ of the shock amplitude).
The effective condition number is
$\kappa_{\rm eff} = \lambda_L / \sigma_{\min}(L) \approx 7{,}100$
and the QSVT polynomial degree is chosen at
$d = 10\,\kappa_{\rm eff} = 71{,}001$ with
$\kappa_\text{QSVT} = 7{,}100$.
The circuit uses $27$ qubits and was simulated on a single H100 GPU in $\approx 53$~hours.
Figure~\ref{fig:qsvt-c2-verify} shows the result.
The top panel shows the density profile at $T = 0.4$; on this
absolute scale all classical and QSVT curves overlap to within
the line width.
The residual plot on the bottom shows the following structure:
the $\Nord = 1$ residual is
largest, which counts the entire nonlinear effect that the problem
originally have;
the exact
$\Nord = 2$ exponential shows the amount of nonlinear effects
that could be addressed by the $\Nord = 2$ Carleman linearization
in principle, and 
the classical $\NK = 3$ Taylor 
sits essentially on top
of $\Nord = 2$ expm,
implying that the Taylor truncation error is negligible, i.e., $\Deltat$ is sufficiently small
in this setup.
The QSVT output lies in between the $\Nord = 1$ and $\Nord = 2$ curves, confirming
that part of the nonlinear correction is successfully captured.
The difference to the $\Nord = 2$ is consistent to be considered as a QSVT error:
as the error for $d = 10\,\kappa_\text{QSVT}$ is $\sim 10^{-4}$ in Fig.~\ref{fig:qsvt-d-comparison} at $\kappa_\text{QSVT} = 1000$ and the error is expected to grow with $\kappa_\text{QSVT}$, so the observed $\sim 10^{-3}$ residual is reasonable, considering that the $\Delta \rho$ is different in the two cases as well.

\subsubsection{$\NK$-vs-$\Deltat$ trade-off in QSVT}
\label{sec:qsvt-K-trade-off}

\begin{figure*}[htbp]
  \centering
  \includegraphics[width=0.95\textwidth]{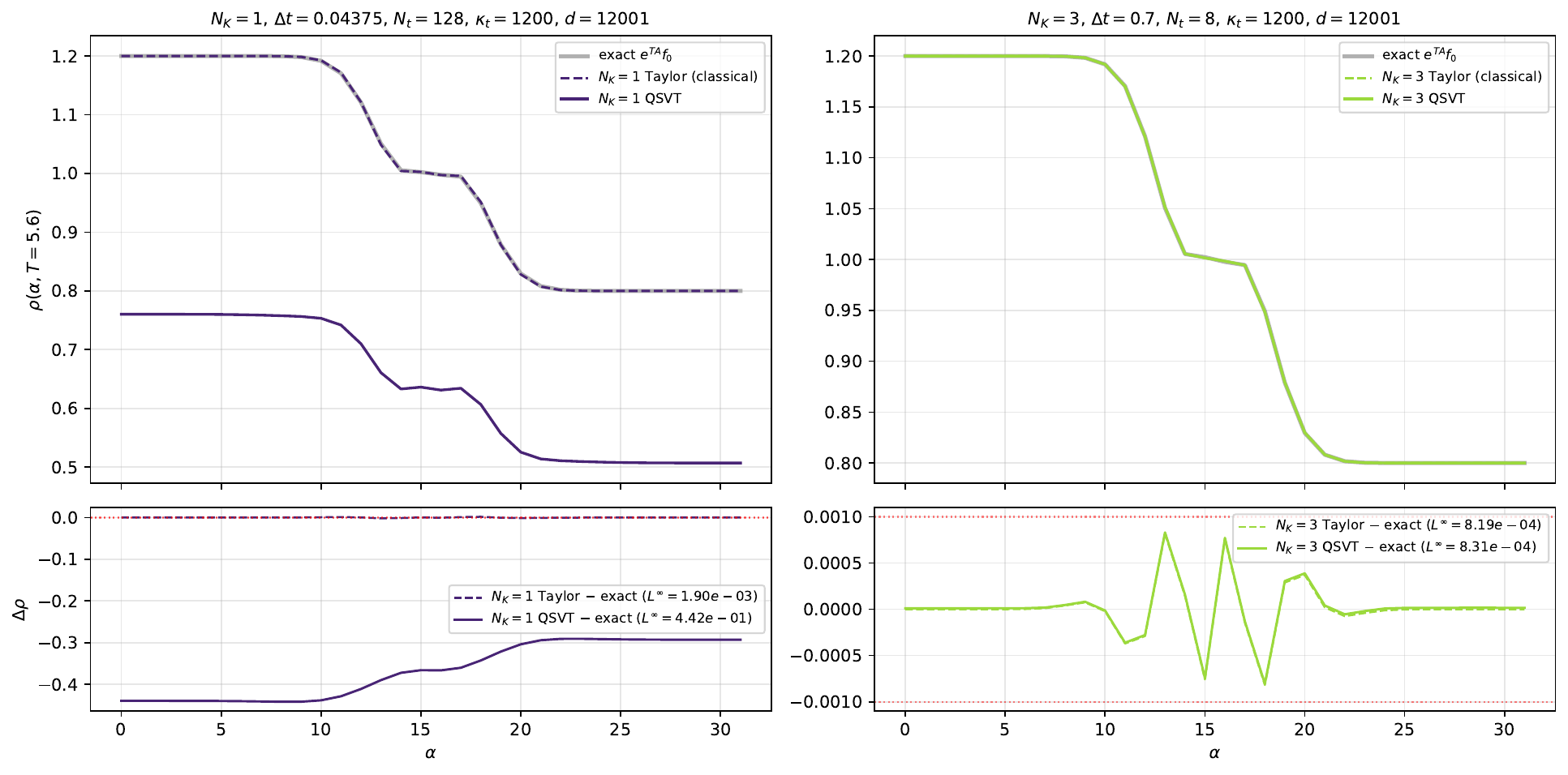}
  \caption{QSVT $\NK = 1$ vs $\NK = 3$ comparison at fixed
    $T = 5.6$, $\Nx = 32$, $\nu = 2.0$, $\Delta\rho = 0.4$, both
    runs at the same QSVT polynomial design parameter
    $\kappa_\text{QSVT} = 1200$, $d_\text{QSVT} = 12\,001$.
    Top panels: density profiles (QSVT output, solid; classical
    Taylor solve, dashed).
    Bottom panels: residual relative to the exact
    $e^{TA} \mathbf{f}_0$ with the $\pm 10^{-3}$ target (red dotted).
    Left: $\NK = 1$ accuracy-limited setup ($\Deltat = 0.04375$,
    $\Nt = 128$, $\kappa \approx 3283$).
    Right: $\NK = 3$ stability-limited setup ($\Deltat = 0.7$,
    $\Nt = 8$, $\kappa \approx 1058$).}
  \label{fig:qsvt-K-comparison}
\end{figure*}

We illustrate the $\NK$-vs-$\Deltat$ trade-off
(Sec.~\ref{sec:K-vs-h}) in the QSVT circuit at the same
physical time $T = 5.6$, $\Nx = 32$, $\Delta\rho = 0.4$,
running both the accuracy-limited $\NK = 1$ setting
($\Deltat = 5.6/128 \approx 0.044$, $\Nt = 128$) and the
stability-limited $\NK = 3$ setting
($\Deltat = 0.7$, $\Nt = 8$) at the same QSVT polynomial design
parameter $\kappa_\text{QSVT} = 1200$, $d_\text{QSVT} = 12\,001$
(so $d_\text{QSVT}/\kappa_\text{QSVT} = 10$, well into the saturation regime).
At this $\kappa_\text{QSVT}$, $\NK = 3$ lies inside the polynomial design domain
($\kappa^{(\NK=3)} \approx 1058 < 1200$) and the QSVT output
matches the classical Taylor solve to $L^\infty \!\approx\!
8 \times 10^{-4}$, well within the $10^{-3}$ target;
$\NK = 1$ lies outside ($\kappa^{(\NK=1)} \approx 3283 > 1200$)
and the singular values below $\lambda_L/\kappa_\text{QSVT}$ are
under-inverted, producing a $\sim 0.5$ amplitude collapse of the
density profile (Fig.~\ref{fig:qsvt-K-comparison}).
This confirms that the $\NK=3$ is a more efficient choice
in terms of the number of oracle calls,
at least at a target accuracy of $10^{-3}$ (or below).

\subsection{Gate count and qubit count scaling}
\label{sec:gate-scaling}

Having verified the QSVT workflow end-to-end and characterised
the effective condition number $\kappa(L)$ that determines the
QSVT polynomial degree (Sec.~\ref{sec:kappa-scaling},
\ref{sec:qsvt-results}), we now report the qubit and gate counts
of the compiled circuits and project them to the asymptotic
$\Nx \to \infty$ regime.

\subsubsection{Qubit count}

\begin{figure*}[htbp]
  \centering
  \includegraphics[width=0.95\textwidth]{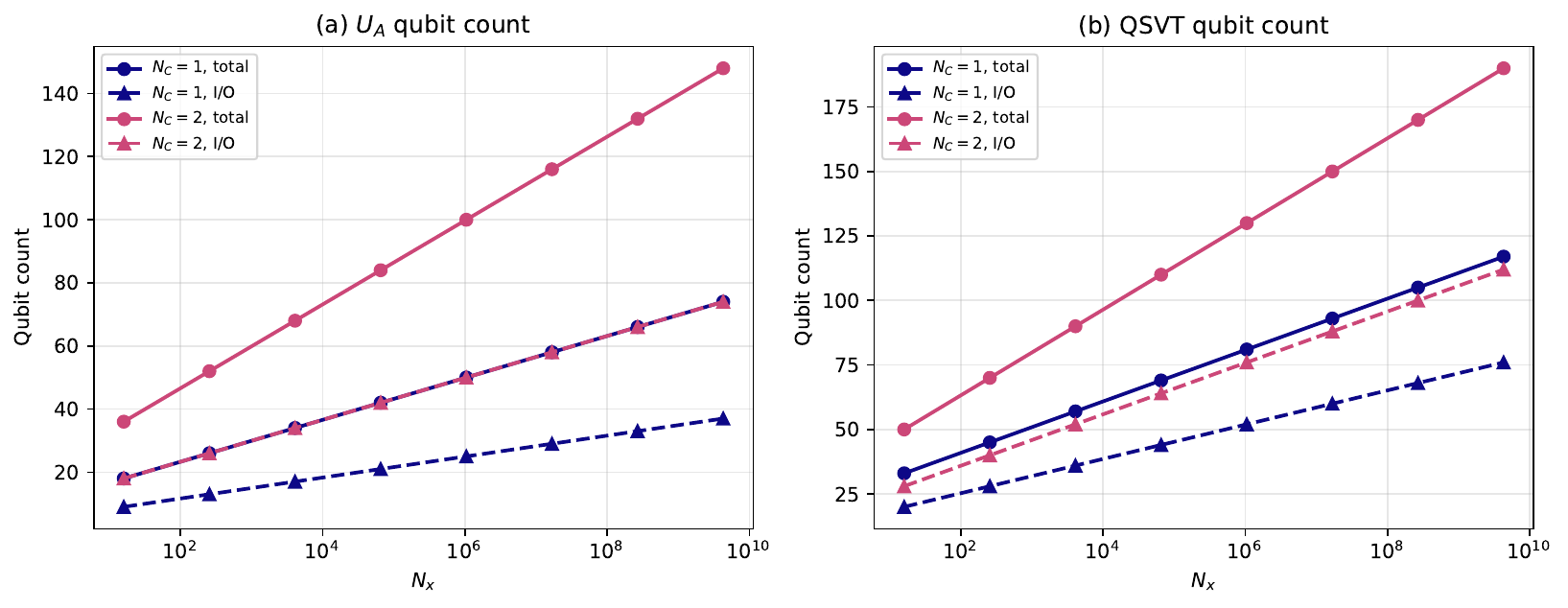}
  \caption{Qubit count scaling with~$\Nx$ for $\Nord = 1$ (blue) and
    $\Nord = 2$ (red). Dashed: I/O register qubits. Solid lines: total qubit count including I/O and ancilla qubits.
    Left panel: $U_A$ block-encoding.
    Right panel: full QSVT-compiled circuit.}
  \label{fig:qubitcount}
\end{figure*}

Table~\ref{tab:qubit-count} summarizes the input/output (I/O)
qubit count of each level of the algorithm; 
the total qubit counts including ancillary qubits are visualised in
Fig.~\ref{fig:qubitcount}.  Both panels confirm the expected
logarithmic growth with the grid size~$\Nx$, with the first-order
Carleman implementation requiring half as many spatial qubits as
the second-order one due to the second register copy.
\begin{table}[h]
\centering
\begin{tabular}{lc}
\hline
Implementation & Qubit count \\
\hline
$A$-matrix, $\Nord=1$ & $\log_2 \Nx + 5$ \\
$A$-matrix, $\Nord=2$ & $2\log_2 \Nx + 10$ \\
$L$-matrix, $\Nord=1$, $\NK=1$ & $\log_2 \Nx + \lceil\log_2 2\Nt\rceil + 7$ \\
$L$-matrix, $\Nord=1$, $\NK=3$ & $\log_2 \Nx + \lceil\log_2 2\Nt\rceil + 9$ \\
$L$-matrix, $\Nord=2$, $\NK=1$ & $2\log_2 \Nx + \lceil\log_2 2\Nt\rceil + 12$ \\
$L$-matrix, $\Nord=2$, $\NK=3$ & $2\log_2 \Nx + \lceil\log_2 2\Nt\rceil + 13$ \\
QSVT & $L$-matrix $+ 1$ \\
\hline
\end{tabular}
\caption{I/O qubit count for each level of the algorithm.}
\label{tab:qubit-count}
\end{table}

\subsubsection{Gate count}

\begin{figure*}[htbp]
  \centering
  \includegraphics[width=0.48\textwidth]{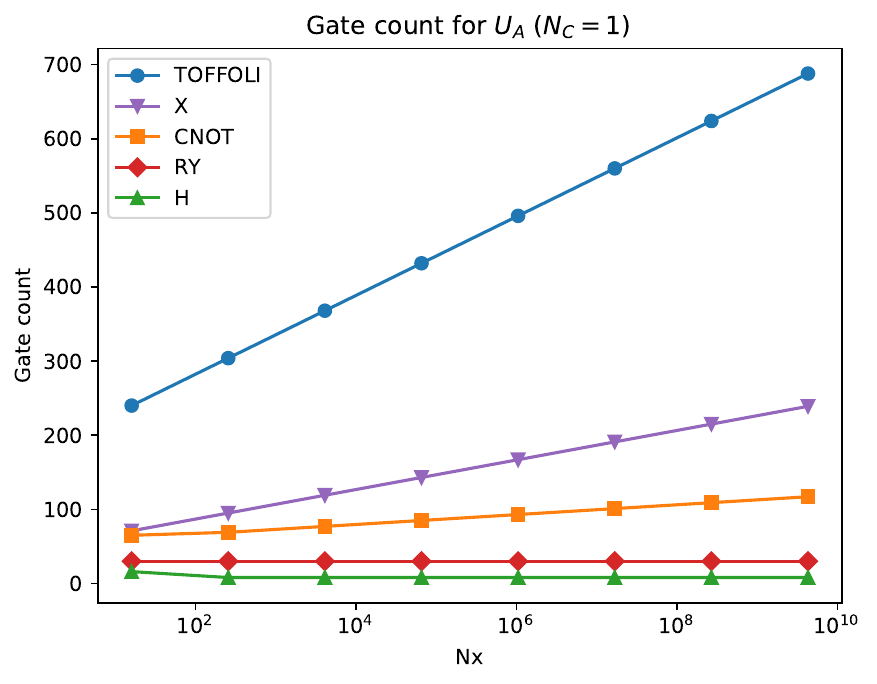}
  \hfill
  \includegraphics[width=0.48\textwidth]{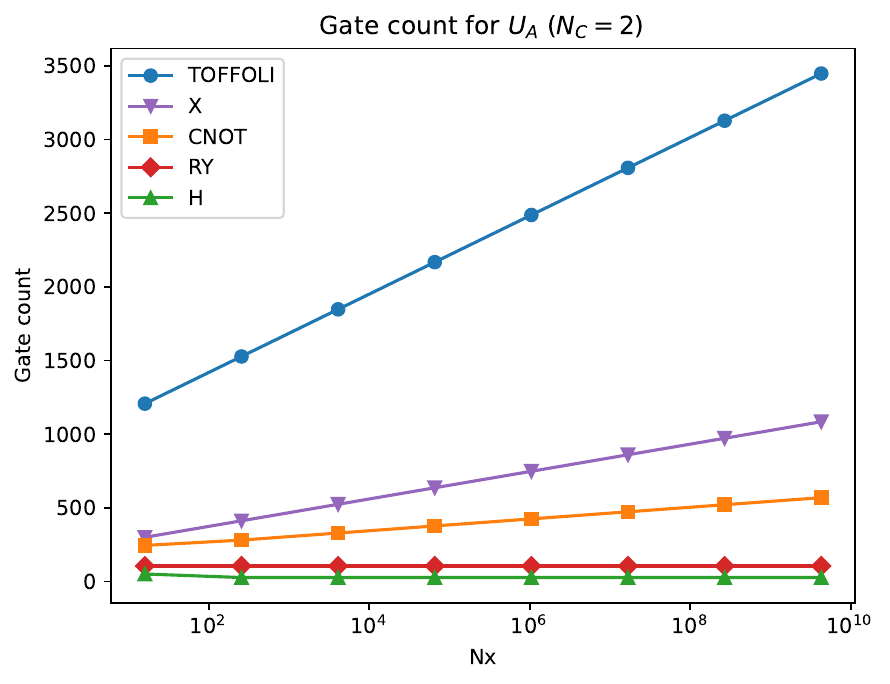}
  \caption{Gate count of the block-encoding circuit
    $U_A$ as a function of grid size~$\Nx$ (after optimization).
    Left: first-order Carleman ($\Nord = 1$).
    Right: second-order Carleman ($\Nord = 2$).}
  \label{fig:gatecount}
\end{figure*}

Figure~\ref{fig:gatecount} reports the per-call gate count of the
$U_A$ block encoding for both Carleman orders, broken down by
gate type.
The $U_A$ quantum circuit is compiled to the gate set
$\{\text{TOFFOLI},\, \text{CNOT},\, H,\, X,\, R_Y,\,  S,\, S^\dagger\}$
by QURI SDK~\cite{QURISDK}.
The compiled circuits are then post-processed by a peephole pass
that cancels adjacent identical Toffoli pairs and $X$--$X$ pairs.
A clear $O(\log \Nx)$ scaling is visible for both $\Nord = 1$ and
$\Nord = 2$.

\begin{figure*}[htbp]
  \centering
  \includegraphics[width=0.48\textwidth]{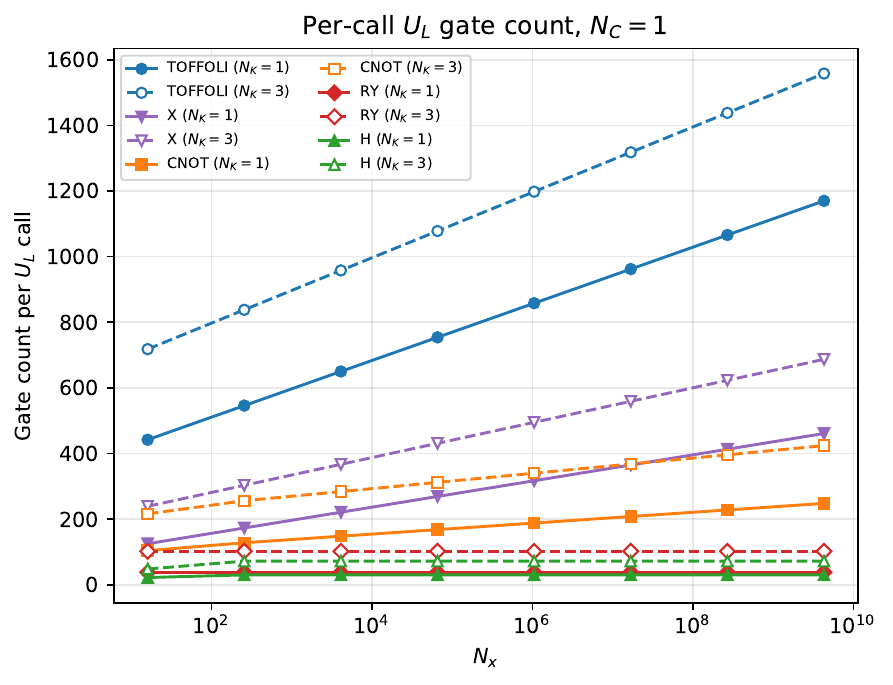}
  \hfill
  \includegraphics[width=0.48\textwidth]{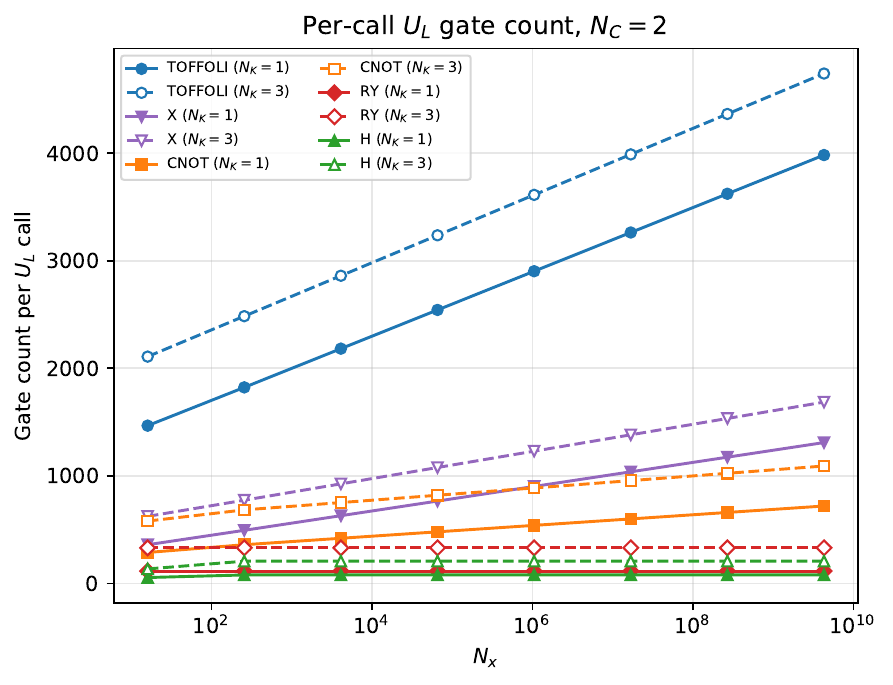}
  \caption{Gate count of the propagator $U_L$ as a
    function of grid size~$\Nx$ (after optimization), at
    $\Deltat = 0.1$ under the fixed-physical-time convention
    $\Nt = \lceil t_{\max}\,\Nx/(N_\mathrm{ref}\,\Deltat)\rceil$
    (${t_{\max}} = 100$, $N_\mathrm{ref} = 512$).
    Solid lines: $\NK = 1$. Dashed lines: $\NK = 3$.
    Left: first-order Carleman ($\Nord = 1$).
    Right: second-order Carleman ($\Nord = 2$).
    Each $U_L$ call invokes $U_A$ once via the shared-oracle
    construction of Sec.~\ref{sec:be-L}.}
  \label{fig:UL-gatecount}
\end{figure*}

Figure~\ref{fig:UL-gatecount} reports the per-call gate count of
the propagator $U_L$ at $\Deltat = 0.1$ under the
fixed-physical-time convention $\Nt(\Nx) = \lceil t_{\max}\,\Nx /
(N_\mathrm{ref}\,\Deltat) \rceil$ (${t_{\max}} = 100$,
$N_\mathrm{ref} = 512$), with Taylor orders $\NK = 1$ (solid) and
$\NK = 3$ (dashed) overlaid. We always employ the shared-oracle
implementation of $U_L$ introduced in Sec.~\ref{sec:be-L}, so each
$U_L$ call invokes (an extended version of) $U_A$ once rather than $\NK$ times; the
remaining per-oracle overhead is the transition-row block and the
time-step counter increment, both with $\NK$- and $\Nt$-dependent
contributions.
The growth from $\NK = 1$ to $\NK = 3$ is mild for the Toffoli
contribution ($\sim 1.2$--$1.4\times$ in the asymptotic
large-$\Nx$ regime), and roughly linear in~$\NK$ for the $R_Y$
component.  In terms of count alone, Toffoli therefore dominates
the per-oracle gate budget.  In T-count, however, an $R_Y$ rotation
compiles to $\sim 100$--$200$ T gates against the $\sim 7$ T gates
of a Toffoli~\cite{Ross2016-rs}, so the $R_Y$ contribution
dominates the per-oracle T cost in our implementation, and the
per-oracle T count grows roughly as $\sim 3\times$ from $\NK = 1$ to
$\NK = 3$.  Whether the $1.7\times$ reduction in $d_\text{QSVT}$
achievable with $\NK = 3$ (Sec.~\ref{sec:K-vs-h}) translates into a
net advantage therefore depends on reducing this $R_Y$ overhead
by an optimized compilation, which we leave to future work.

Each QSVT call applies $U_L$ (and $U_L^\dagger$) $d_\text{QSVT}$
times together with single-qubit phase rotations and a fixed
number of multi-controlled $X$ gates~\cite{Camps2024-sparse-be},
so the elementary gate count is linear in $d_\text{QSVT}$.  We
therefore measure the gate count at $d_\text{QSVT} = 3$
and $d_\text{QSVT} = 13$ and interpolate linearly in
$d_\text{QSVT}$ to the target degree.

For the target degree we adopt the polynomial-degree bound~\cite{Gribling2024-qls} 
for the $1/x$ approximation on $[1/\kappa, 1]$ at precision~$\varepsilon$,
\begin{equation}
  d_\text{QSVT} \;=\;
    \frac{\kappa(L)\,\log\!\bigl(2\kappa(L)/\varepsilon\bigr)}{2},
  \label{eq:d-star}
\end{equation}
rounded up to the next odd integer.  We use the \emph{measured}
effective condition number $\kappa(L)$ from
Sec.~\ref{sec:kappa-scaling} and set $\varepsilon = 10^{-3}$ for the
figures below.  We note that this choice
corresponds to $d_\text{QSVT} /\kappa \approx 7$ in the setup of
Fig.~\ref{fig:qsvt-d-comparison}.
For $\Nx$ beyond the largest measured grid ($\Nx = 128$),
we exploit the fact that $\kappa(L)$ saturates with $\Nx$ in the
stable regime (Sec.~\ref{sec:kappa-scaling}) and grows linearly
with $\Nt$: we take the $\Nx = 128$ row as representative of the
$\Nx$-saturated value and extrapolate linearly in $\Nt$ to the
target $\Nx > 128$.  The time axis itself is set by the fixed
physical-time convention $\Nt(\Nx) = \lceil t_{\max}\,\Nx /
(N_{\mathrm{ref}}\,\Deltat_L) \rceil$ (${t_{\max}} = 100$,
$N_{\mathrm{ref}} = 512$, $\Deltat_L = 0.1$ in the figures) so
that the simulated physical time is held constant as the grid is
refined.  Combined with $d_\text{QSVT} =\tilde{O}(\kappa)$, $\kappa \propto
\Nt \propto \Nx$, and logarithmic per-oracle cost, the total gate count scales as
$\tilde{O}(\Nx)$, where $\tilde{O}$ denotes asymptotic scaling up to logarithmic factors.

\begin{figure*}[htbp]
  \centering
  \includegraphics[width=0.95\textwidth]{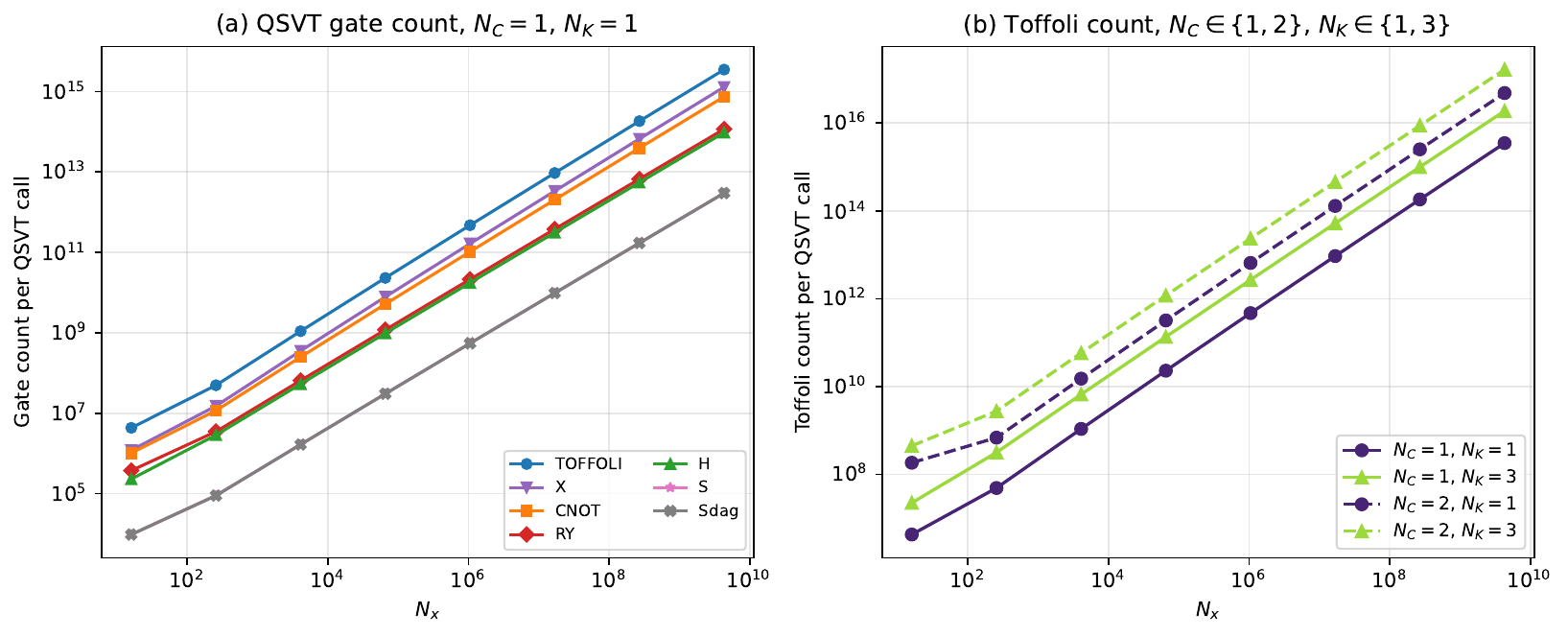}
  \caption{Total gate count per QSVT call as a function of~$\Nx$,
    obtained by linear interpolation in $d_\text{QSVT}$ from
    measurements at $d_\text{QSVT} = 3$ and $d_\text{QSVT} = 13$,
    evaluated at the target degree $d_\text{QSVT}(\Nx)$ set by
    the measured $\kappa(L)$ via Eq.~(\ref{eq:d-star}) with
    $\varepsilon = 10^{-3}$.  All curves are at the common time
    step $\Deltat = 0.1$ at which both $\kappa(L)$
    (Sec.~\ref{sec:kappa-scaling}) and the per-call gate count
    (Fig.~\ref{fig:UL-gatecount}) are measured.
    Left: full breakdown by gate type for $\Nord = 1$, $\NK = 1$.
    Right: Toffoli count only, comparing all four
    $(\Nord, \NK) \in \{1, 2\} \times \{1, 3\}$ combinations.
    $\NK$ is encoded by colour (matching the convention of
    Fig.~\ref{fig:kappa-vs-K}); $\Nord$ by linestyle
    (solid: $\Nord = 1$, dashed: $\Nord = 2$).}
  \label{fig:qsvt-gatecount}
\end{figure*}

\section{Discussion}
\label{sec:discussion}

The workflow verified above solves a one-dimensional CFD problem 
whose size is set by the spatial grid~$\Nx$ and the number of time
steps~$\Nt$ (set by the physical time~$T$ and the time
step~$\Deltat$). 
To solve that problem at a prescribed accuracy, one must further
specify
the Taylor truncation order~$\NK$, the Carleman
truncation order~$\Nord$, and the QSVT polynomial-design
parameters~$\kappa_\text{QSVT}$ and $d$. 
Two quantities determine the total
quantum cost: the per-call cost of the block encoding (qubit
count and gate count, both $O(\log\Nx)$ in our implementation)
and the effective condition number $\kappa(L)$ that sets the QSVT
polynomial degree~$d_\text{QSVT}$.  Below we discuss how these costs
respond to those parameter choices in turn, in the same
order as Sec.~\ref{sec:kappa-scaling}: $\Nx$ and $T$ first, then
$\NK$, then $\Nord$, before turning to topics that fall outside
the present scope.

\subsection{Asymptotic scaling in $\Nx$ and $T$}
\label{sec:disc-NxNt}

Two observations from Sec.~\ref{sec:kappa-scaling} and
Sec.~\ref{sec:gate-scaling} fix the asymptotic cost of the
algorithm in $\Nx$ and $T$. 
First, in the stable regime $\mu(\Deltat A) \lesssim 1$, the
effective condition number satisfies $\kappa(L) \propto \Nt$ with
$\kappa/\Nt \approx 80$--$100$, and the prefactor flattens with
$\Nx$ once the streaming term dominates the rate matrix~$A$
(Fig.~\ref{fig:kappa-baseline}); the saturation tracks the
$\Nx$-dependence of $\mu(A)$, which transitions from
collision-dominated ($\mu(A) \propto 1/\Nx$) to streaming-dominated
($\mu(A)\to 2$ for $\Nord=1$, $\to 4$ for $\Nord=2$;
Table~\ref{tab:spectral-radius}).  Second, the per-call $U_A$ gate
count is $O(\log\Nx)$ (Fig.~\ref{fig:gatecount}), and the $U_L$
overhead added on top (Fig.~\ref{fig:UL-gatecount}) stays
sub-logarithmic in~$\Nx$ in the asymptotic regime, so each QSVT
call costs $O(d_\text{QSVT}\log\Nx) = O(\Nt\log\Nt\log\Nx)$
elementary gates with the polynomial-degree bound
$d_\text{QSVT} = \kappa(L)\log(2\kappa(L)/\varepsilon)/2$ of
Eq.~\eqref{eq:d-star}.  Assuming a constant physical time, i.e., $\Nt
\propto \Nx$, this yields a total of $O(\Nx\,\polylog\Nx)$ gates,
which should be compared to $O(\Nx^{D+1})$ for a classical CFD solver in $D$
spatial dimensions after properly considering the readout overhead.

\subsection{Cost vs.~accuracy at Taylor order~$\NK$}
\label{sec:disc-K}

A larger Taylor order~$\NK$ allows a larger time-step $\Deltat$
for a given accuracy. At a $10^{-3}$ accuracy target
($\Nx=128$, $T=25$), $\NK = 3$ admits $\Deltat \approx 1.25$
($\Nt = 20$, $\kappa \approx 3100$), giving a $\sim 1.7\times$
reduction in $d_\text{QSVT}$ over the accuracy-limited $\NK = 1$
setting (Sec.~\ref{sec:K-vs-h}). This gain is offset by
the linear-in-$\NK$ growth of the $R_Y$ component, which dominates
the T count at $\sim 100$--$200$ T per rotation~\cite{Ross2016-rs}
over the $\times 7$ T gate contribution from Toffoli gates:
the per-oracle T budget grows by $\sim 3\times$ from $\NK = 1$ to
$\NK = 3$ (Fig.~\ref{fig:UL-gatecount}), so the total T cost shows
no net advantage in this preliminary implementation.
While $\NK = 3$ remains a promising choice, it is thus
an important open question if a sufficient $R_Y$ reduction
can be achieved to realize a practical advantage over $\NK = 1$,
which we leave to future work.

\subsection{Cost vs.~accuracy at Carleman order $\Nord$}
\label{sec:disc-Nord}

Extending the Carleman order from $\Nord = 1$ to $\Nord = 2$ 
incorporates the leading nonlinear effect of the BGK collision, 
at the price of the following cost increases.
The
qubit count grows from $\log_2 \Nx + 5$ to $2\log_2 \Nx + 10$
(Table~\ref{tab:qubit-count}), reflecting one extra copy of the
spatial register together with a small case register that
selects among the three blocks $A_{11}$, $A_{12}$, $A_{22}$
(Sec.~\ref{sec:a22-encoding}); the per-oracle Toffoli count
remains $O(\log\Nx)$, with a modest prefactor multiplier between
the two orders visible in Fig.~\ref{fig:gatecount}.  The
condition-number cost is concentrated in the block-encoding
normalization: in the stable regime ($\Nx \geq 64$ at
$\Deltat = 0.1$) we measure $\kappa_{\Nord=2}/\kappa_{\Nord=1}
\approx 4$ (Fig.~\ref{fig:kappa-C1-vs-C2}), and almost all of
this arises from the $4\times$ growth of $\lambda_L = 2^{n_i^L}\,
L_\text{max}$ ($8 \to 32$); the singular-value contribution
$\sigma_\text{min}^{(\Nord=1)}/\sigma_\text{min}^{(\Nord=2)}$
adds only a $1.05$--$1.3$ factor on top.

On the accuracy side (Fig.~\ref{fig:carleman-C123}), at the
default $\Delta\rho = 0.4$ the pointwise $L^\infty$ residual
against the continuous BGK reference shrinks from
$2.2\times 10^{-2}$ at $\Nord = 1$ to $5.0\times 10^{-3}$ at
$\Nord = 2$ (a factor of $\sim 4$), and to $9.8\times 10^{-4}$
at $\Nord = 3$ (a further factor of $\sim 5$).  The dominant
residual at $\Nord = 3$ is still the Carleman truncation itself, not
the $1/\rho \to 2 - \rho$ approximation of the BGK collision,
whose floor sits at $\sim 9 \times 10^{-5}$, an order of
magnitude below.
Note that this analysis assumes the Carleman expansion converges,
which is known to break down at higher Reynolds
numbers~\cite{Jennings2025-theory}.

Although our quantum implementation is specific to $\Nord \leq 2$, 
with $\Nord = 3$ explored only classically (Sec.~\ref{sec:carleman-error}),
the construction extends to $\Nord \geq 3$ without conceptual
obstacle: each Carleman block $A_{cd}$ has tensor-product structure
across the duplicated spatial registers, and the case-register LCU
branching of Sec.~\ref{sec:a22-encoding} accommodates additional blocks
by enlarging the case register, giving a qubit count of
$\Nord\,\log_2\Nx + O(\log\Nord)$.  The number of nonzero blocks per
Carleman row stays linear in $\Nord$ under the $1/\rho \approx 2 - \rho$
closure ($A_{c,c}, A_{c,c+1}, A_{c,c+2}$), while a higher-order
expansion of $1/\rho$ widens the block-bandwidth and trades sparsity
for closure accuracy.  The number of tensor-product summands within
each block, however, grows combinatorially with~$\Nord$~\cite{Jennings2025-theory}, and this can affect both the
per-oracle gate count and the block-encoding normalization
$\lambda_L$.
A quantitative study of whether the $\sim 4\times$ $\kappa$ inflation
observed between $\Nord = 1$ and $\Nord = 2$ persists at higher
Carleman orders, and to what extent an optimized implementation can mitigate it, is left to future work.

\subsection{Outlook}
\label{sec:outlook}

\subsubsection{Observable extraction and amplitude estimation}
\label{sec:disc-observable}

The QSVT circuit constructed in this work prepares a state
$\ket{\psi} \propto \ket{f(T)}$ on the relevant subspace of the
solution register, with a success amplitude that scales as
$1/\kappa$ set by the normalization of the QSP polynomial used
to approximate $1/x$.
Extracting a physical observable such as the drag on a body
therefore requires two further ingredients on top of the present
construction: amplitude amplification, which boosts the success
amplitude to $O(1)$ at the cost of $O(\kappa)$ additional oracle
calls, and amplitude
estimation~\cite{Brassard2002-amplitude, Penuel2024-cl}, which
estimates the resulting amplitude to relative precision
$\tilde\varepsilon$ at the cost of $O(1/\tilde\varepsilon)$
repetitions.
Implementing either of these on top of the present construction
is left for follow-up work.

The workflow is favourable only for a small number of
scalar quantities such as drag, lift, or local velocity at a specific point;
reading out the full distribution at every lattice site requires
$O(\Nx^D / \tilde\varepsilon)$ repetitions of the final state preparation and spoils the
underlying quantum advantage. 
For a single scalar observable,
standard amplitude amplification yields an end-to-end cost of
$O(\kappa^2 / \tilde\varepsilon)$ oracle calls in our setup based on QSVT.
As we have observed that $\kappa \propto \Nt$, this means that
$O(\Nt^2/ \tilde\varepsilon)$ oracle calls required to extract the observable.
Taking a standard simulation time of $\Nt \propto \Nx$, this in turn means
that the per-scalar-observable cost is $O(\Nx^2 \log(\Nx) / \tilde\varepsilon)$, which yields no quantum speedup in one dimension and a quadratic speedup in three dimensions against an $O(\Nx^{D+1})$ classical solve.

The situation can be improved by replacing the QSVT-based solver with a more efficient linear-system solver~\cite{Ambainis2012-vtaa, Costa2022-optimal-qls,
Costa2026-optimal-practice, Lin2020-eigenstate-filtering,
Morales2024-qls-survey} that runs with $\tilde{O}(\kappa)$ rather than $\tilde{O}(\kappa^2)$ oracle calls, 
which improves the quantum advantage to quadratic and quartic, respectively, in one and three dimensions.

As a back-of-the-envelope estimate, consider a representative
industrial target with $\Nx \sim 10^4$ per axis, $\Deltat \sim 1$,
and $\Nt \sim 10^4$ time steps (e.g., a trillion-cell
three-dimensional CFD calculation, which is beyond the current
capabilities of classical high-performance computing). With
$\kappa(L) \sim 100\,\Nt \sim 10^6$ and $d_\text{QSVT} \sim
10\,\kappa \sim 10^7$, combined with the per-oracle Toffoli
count of $\sim 10^3$ at $\Nord = 2$ (Fig.~\ref{fig:UL-gatecount}),
a single QSVT-prepared state costs of order $10^{10}$ Toffoli
gates. Extracting one scalar observable at $\tilde\varepsilon
= 1\%$ accuracy via amplitude amplification and estimation
multiplies this by another $O(\kappa/\tilde\varepsilon) \sim
10^8$ oracle invocations, for an order-of-magnitude estimate of
$\sim 10^{18}$ gates, which would take tens of years even under
an extremely optimistic $1\,\text{GHz}$ Toffoli rate.

Replacing the QSVT solver with an $\tilde O(\kappa)$
linear-system algorithm would shave $\sim \kappa \sim 10^6$
off this count, bringing the runtime to of order ten days even at
a less optimistic $1\,\text{MHz}$ Toffoli rate and making a polynomial
quantum advantage plausible. Although the QSVT-specific constant-factor
analysis used above is no longer valid, it is reported~\cite{Costa2026-optimal-practice}
that the cost per $\kappa$ is 3--8 for $\tilde O(\kappa)$ solvers, similar to the QSVT polynomial degree per $\kappa$ in our setup.
Implementing such an efficient linear-system solver, together with amplitude
amplification and amplitude estimation on top of the final state
preparation, is thus a prerequisite for industrially useful
quantum CFD.
Further gate-count reductions are also required to
leave room for extensions to higher spatial dimensions,
non-trivial geometries, and higher Carleman orders.

\subsubsection{Higher spatial dimensions and non-trivial geometry}
\label{sec:disc-higher-d}

The block-encoding construction generalizes to higher spatial
dimensions without a qualitative change.
A $D$-dimensional LBM with $Q$ velocity channels and $\Nx$ grid
points per axis requires $D \lceil \log_{2} \Nx \rceil$ spatial
qubits and $\lceil \log_{2} Q \rceil$ velocity qubits per
Carleman slot; at order~$\Nord$ the total qubit count is
$\Nord\,(D \log_{2} \Nx + \lceil \log_{2} Q \rceil)
+ O(\log \Nt) + O(1)$, and the per-step gate count is likely be
still $O(\log \Nx)$ if boundary detection does not require
complex arithmetic operations.
The streaming oracle decomposes as a product of
increment/decrement operators on each spatial axis, and
non-trivial geometries (bounce-back boundaries, interior
obstacles) can be incorporated by augmenting the oracle with a
binary mask register~\cite{Penuel2024-cl, Ueno2026-linear}.

\section{Conclusion}
\label{sec:conclusion}

We have constructed and verified, at the elementary-gate
level, block encodings of the rate matrix $A$ and the
Taylor-ODE matrix $L$ for the one-dimensional D1Q3
Boltzmann equation with bounce-back walls, at Carleman
orders $\Nord = 1, 2$ and Taylor orders $\NK = 1, 3$.
$U_A$ and $U_L$ were checked against their classical
counterparts to machine precision, and the QSVT output of
the $\Nord = 2$ circuit successfully captured the leading 
nonlinear effects verified against the classical reference.

The total final-state preparation cost is the product of the
per-oracle gate count of $U_L$ and the number $d_\text{QSVT}$
of oracle calls per QSVT run. The former was verified to
scale as $O(\log\Nx)$ in our implementation, while the
latter is set by the effective condition number
$\kappa(L) = \lambda_L/\sigma_\text{min}(L)$, which scales
as $\Nt$ in the stable regime $\mu(\Deltat A) \lesssim 1$ while independent
of $\Nx$ for sufficiently large $\Nx$. Together they give
$O(\Nx \polylog \Nx)$ gates per state preparation under the
fixed-physical-time convention $\Nt \propto \Nx$.
Extending the Carleman truncation from $\Nord = 1$ to $\Nord = 2$
incorporates the leading nonlinear effect of the BGK
collision, at the
price of an additional $\log_2 \Nx + 5$ qubits, a
several-fold increase in the per-oracle gate count, and a
$\sim 4\times$ inflation of $\kappa(L)$ that is concentrated
almost entirely in $\lambda_L$ rather than in
$\sigma_\text{min}(L)$.
Raising the Taylor order to $\NK = 3$ in the Taylor ODE solver
admits larger $\Deltat$ and a
correspondingly shorter QSVT polynomial for a given accuracy, which makes $\NK = 3$
a promising choice in terms of $d_\text{QSVT}$; we reveal that
the net advantage in T-count, however, requires reducing
the $R_Y$ counts by a more efficient implementation.

Moving from this one-dimensional concrete baseline to industrially
relevant quantum CFD requires various steps of development,
from implementing amplitude amplification, amplitude estimation for the readout
of a small number of scalar observables,
and a linear-system solver with $\tilde O(\kappa)$
rather than $\tilde O(\kappa^2)$ scaling of QSVT,
to an extension of the block-encoding construction to higher spatial
dimensions and Carleman orders with more complex geometries,
and to the development of a better linearization scheme that is capable
of handling much higher Reynolds numbers.
The present work provides
concrete elementary-gate-level numbers for the
one-dimensional case as a starting point on which such
developments can build.

\acknowledgements
The authors thank Yasunori Lee for valuable discussion.

\appendix

\section{Carleman blocks at third order and coupling-scale
  conventions}
\label{app:nc3}

Section~\ref{sec:carleman} worked at $\Nord = 2$ and introduced
$A_{11}$, $A_{12}$, and $A_{22} = A_{11}\otimes I + I\otimes A_{11}$.
Extending to $\Nord = 3$ requires two additional blocks,
$A_{13}$ (cubic correction to $\bm{f}^{(1)}$) and $A_{23}$
(quadratic contribution to $\bm{f}^{(2)}$ by $\bm{f}^{(3)}$), as well as
$A_{33}$. This appendix collects their
derivations and then addresses a scale convention needed to cancel
the overshoot produced by finite truncation.

\subsection{Derivation of $A_{13}$}

The nonlinearity of the BGK collision enters only through
$\rho u^2 = m_1^2/\rho$ in the equilibrium~\eqref{eq:feq}, with
$m_1 = \sum_j e_j f_j$. Taylor-expanding $1/\rho$ around $\rho_0 = 1$,
\begin{equation}
  \frac{1}{\rho} \;=\; 1 - \Delta + \Delta^2 - \cdots,
  \qquad \Delta := \rho - 1,
  \label{eq:rho-binomial}
\end{equation}
and truncating after the linear term in~$\Delta$ gives
$1/\rho \approx 2 - \rho$. Substituting this into $\rho u^2$ yields
\begin{equation}
  \frac{m_1^2}{\rho} \;\approx\; 2\,m_1^2 \;-\; m_1^2\,\rho,
  \label{eq:2term}
\end{equation}
a sum of a quadratic-in-$\bm{f}$ leading term and a cubic-in-$\bm{f}$
correction. The quadratic piece is the source of $A_{12}$
(Eq.~\eqref{eq:A12-kernel}); the cubic piece yields $A_{13}$, with
matrix elements
\begin{equation}
  (A_{13})_{\alpha i,\, \alpha_1 i_1\, \alpha_2 i_2\, \alpha_3 i_3}
  = \delta_{\alpha,\,\alpha_1 + e_i}\,
    \delta_{\alpha_1\alpha_2}\,
    \delta_{\alpha_1\alpha_3}\,
    K^{(13)}_{i\,i_1\,i_2\,i_3},
  \label{eq:A13-elements}
\end{equation}
where
\begin{equation}
  K^{(13)}_{i\,i_1\,i_2\,i_3}
  = -\frac{w_i}{\tau}\left[
    \frac{9}{2}\,(e_{i_1}\!\cdot e_i)(e_{i_2}\!\cdot e_i)
    - \frac{3}{2}\,(e_{i_1}\!\cdot e_{i_2})
    \right],
  \label{eq:A13-kernel}
\end{equation}
with the third velocity index $i_3$ carrying only the density sum
$\sum_{i_3} = \rho$ at the shared site $\alpha_1$.

\subsection{Derivation of $A_{23}$}

Differentiating $\bm{f}^{(2)} = \bm{f}^{(1)} \otimes \bm{f}^{(1)}$
under the full nonlinear dynamics gives
\begin{equation}
  \frac{d \bm{f}^{(2)}}{dt}
  = \bigl(\dot{\bm f}^{(1)}\bigr)\otimes \bm{f}^{(1)}
  + \bm{f}^{(1)} \otimes \bigl(\dot{\bm f}^{(1)}\bigr).
  \label{eq:leibniz-f2}
\end{equation}
Substituting $\dot{\bm{f}}^{(1)} = A_{11}\bm{f}^{(1)}
+ F(\bm{f}^{(1)}, \bm{f}^{(1)}) + G(\bm{f}^{(1)}, \bm{f}^{(1)}, \bm{f}^{(1)})
+ \ldots$, where $F$ and $G$ are the bi- and trilinear maps defined
by Eqs.~\eqref{eq:A12-kernel} and~\eqref{eq:A13-kernel}, the
right-hand side of Eq.~\eqref{eq:leibniz-f2} separates by
polynomial degree:
\begin{align}
  \text{degree 2 in } \bm{f}^{(1)}:\quad
    & A_{22} \bm{f}^{(2)},
    \label{eq:A22-from-leibniz} \\
  \text{degree 3:}\quad
    & A_{23} \bm{f}^{(3)},
    \label{eq:A23-from-leibniz}\\
  \text{degree 4:}\quad
    & A_{24} \bm{f}^{(4)} \text{ (dropped at } \Nord = 3\text{).}
    \label{eq:A24-dropped}
\end{align}
Evaluated on the product state $\bm{f}^{(3)} = \bm{f}^{(1)\otimes 3}$,
\begin{equation}
  A_{23}\bigl(\bm{f}^{(1)\otimes 3}\bigr)
  = F\bigl(\bm{f}^{(1)}\!,\bm{f}^{(1)}\bigr) \otimes \bm{f}^{(1)}
  + \bm{f}^{(1)} \otimes F\bigl(\bm{f}^{(1)}\!,\bm{f}^{(1)}\bigr),
  \label{eq:A23-product}
\end{equation}
a Kronecker-sum-like structure that embeds the bilinear $F$ into
one slot of the third-rank tensor $\bm{f}^{(3)}$ at a time.

\subsection{Derivation of $A_{33}$}

Repeating the Leibniz argument at rank~3,
\begin{equation}
  A_{33}
  \;=\; A_{11} \otimes I \otimes I
  \;+\; I \otimes A_{11} \otimes I
  \;+\; I \otimes I \otimes A_{11},
  \label{eq:A33}
\end{equation}
a Kronecker sum over the three factors. Under truncation at
$\Nord = 3$, the couplings $A_{34} \bm{f}^{(4)}$ and higher are
dropped. The Kronecker-sum structure preserves tensor
products, so given $\bm{f}^{(3)}(0) = \bm{f}_0^{\otimes 3}$ the
solution is
\begin{equation}
  \bm{f}^{(3)}(t)
  = \bigl(e^{tA_{11}} \bm{f}_0\bigr)^{\otimes 3}
  = \bm{f}^{(1)}_{\text{lin}}(t)^{\otimes 3},
  \label{eq:f3-proxy}
\end{equation}
i.e., $\bm{f}^{(3)}$ factorizes into three copies of the pure linear
trajectory $\bm{f}^{(1)}_{\text{lin}}$. We use this factorization to efficiently
simulate the $\Nord = 3$ case,
avoiding the need to explicitly store a rank-3 tensor
of size $(Q\Nx)^3$.

\subsection{Coupling-scale convention and numerical verification}
\label{app:scales}

The $1/\rho \approx 2 - \rho$ expansion that generates
$A_{12}$ and $A_{13}$ is the two-term truncation of a convergent
series (Eq.~\eqref{eq:rho-binomial}). At finite Carleman
truncation, dropping $A_{1,\Nord+1}$ breaks the cancellation
between the leading~$2\,m_1^2$ in Eq.~\eqref{eq:2term} and the
cubic correction~$-\rho\,m_1^2$, and the nonlinear source
overshoots. To denote these two pieces we write
\begin{equation}
  F \equiv \frac{2\,w_i}{\tau}\,K_i\,m_1^2, \qquad
  G \equiv -\frac{w_i}{\tau}\,K_i\,m_1^2\,\rho,
  \label{eq:FG-def}
\end{equation}
so that~$F$ embeds $A_{12}$ and~$G$ embeds $A_{13}$ in the
rank-1 output of the 2-term collision. At $\rho = 1$ they cancel
down to a single unit-magnitude piece, $F + G \to F/2$.

In the following, we argue that $A_{23}$ should be multiplied by $1/4$,
in the same sprit of halving the $A_{12}$ term in the $\Nord=2$ case discussed in the main text.

\paragraph{$\Nord=2$ case: halve $A_{12}$.}
First, let us recall the discussion in $\Nord=2$.
In this case, dropping $A_{13}$ leaves the leading $F$ piece uncompensated. Using
$1/\rho \approx 1$ instead of $1/\rho \approx 2 - \rho$ at this
coupling is equivalent to halving $A_{12}$.

\paragraph{$\Nord=3$ case: quarter $A_{23}$.}
In the case of third order Carleman linearization,
the nonlinear effect contribution from $A_{23}$ propagates to $\bm{f}^{(1)}$
via two steps. The first is the direct contribution from $\bm{f}^{(3)}$
to $\bm{f}^{(2)}$, where the $A_{23}$ contribution should be halved
by the same logic as for the $A_{12}$ in $\Nord=2$.
Then the nonlinear effect propagated from $\bm{f}^{(3)}$ to $\bm{f}^{(2)}$
contributes to $\bm{f}^{(1)}$ via $\bm{f}^{(2)}$.
Here, another factor two appears: The nonlinear effect from $\bm{f}^{(3)}$
should contribute to $\bm{f}^{(1)}$ via $F+G$ for consistency,
while such $G$ contribution is missing as $\bm{f}^{(3)}$ is evolving only linearly, i.e., by $A_{11}$. Note that here $A_{12}$ should \emph{not} be halved, as now $A_{13}$ is present and $A_{12}$ should compensate the $\rho\sim1$ contribution from $A_{13}$.
The solution here is to further halve the $A_{23}$ term.

\begin{figure}[h]
  \centering
  \includegraphics[width=0.95\linewidth]%
    {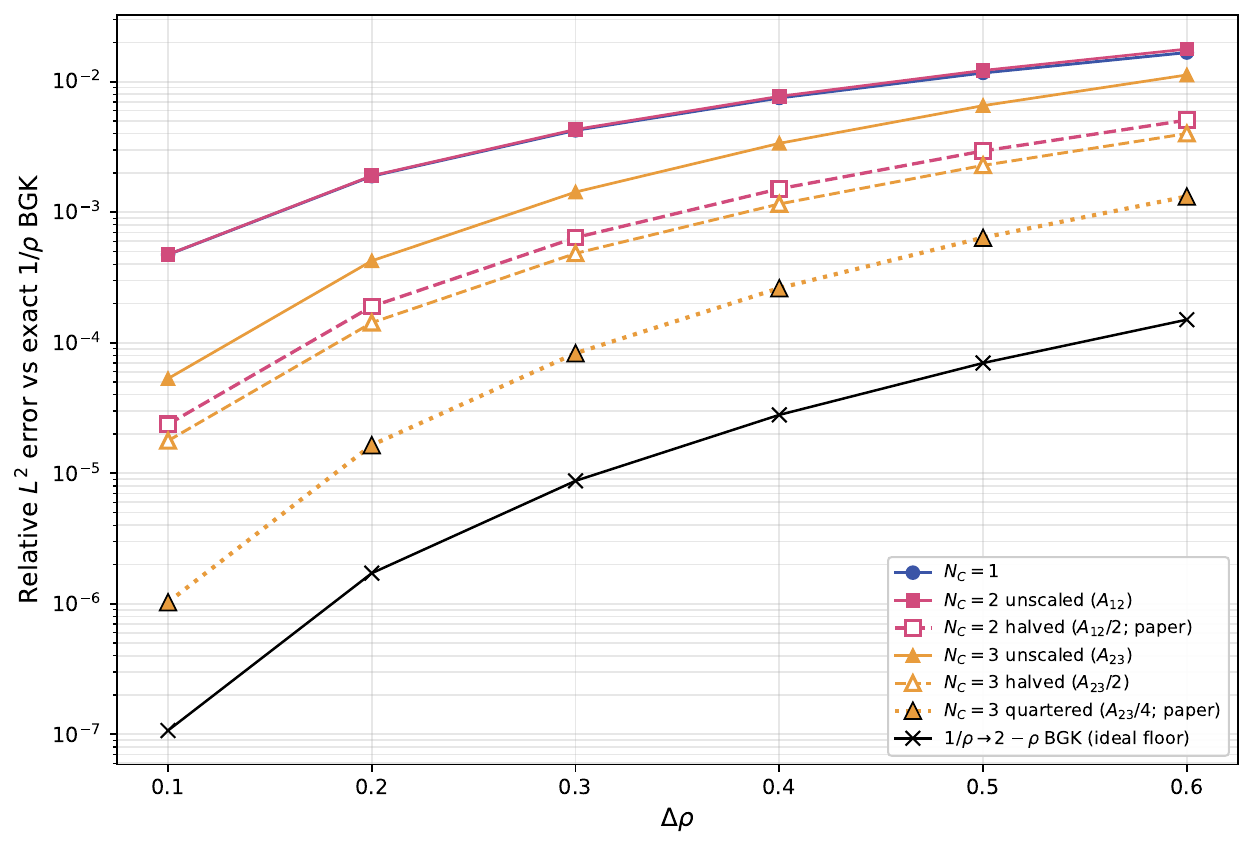}
  \caption{Coupling-scale verification
    $N_x = 128$, $\nu = 2.0$, $T = 25$, symmetric initial condition.
    Final-time relative $L^2$ density error against the exact-$1/\rho$
    continuous BGK reference for $\Nord = 1,\,2,\,3$, with and
    without the halving/quartering convention of
    Appendix~\ref{app:scales}.}
  \label{fig:carleman-halve}
\end{figure}

\paragraph{Numerical evidence.}
We perform a numerical verification of this seemingly ad-hoc trick.
Figure ~\ref{fig:carleman-halve} shows the errors of Carleman-linearized classical
simulations with various coefficients compared to
the exact-$1/\rho$ continuous BGK reference. 
In particular, the naive
halving~($s = 1/2$) at $\Nord = 3$ sits essentially on top of the
unscaled $\Nord = 2$ curve, whereas the quartering~($s = 1/4$)
reduces the final-time $L^2$ error by a factor of~$17$ over the
naive $1/2$ at $\Delta\rho = 0.1$, narrowing to a factor of~$3$
at $\Delta\rho = 0.6$.

\section{Block encoding verification}
\label{app:verification}

We verify $U_A$ entry by entry by extracting the encoded matrix from
the compiled circuit.  For each computational-basis input
$\ket{j}_{\mathrm{sys}}\ket{0}_{\mathrm{anc}}$ on the system register,
we apply the circuit, read the $\ket{0}_{\mathrm{anc}}$ projection of
the output, and multiply by the encoder normalisation
$\lambda_A = \max_A \cdot 2^{n_{\mathrm{i}}}$ for $\Nord = 1$ and
$\lambda_A = \max_A \cdot 2^{n_{\mathrm{i_1}} + n_{\mathrm{case}}}$
for $\Nord = 2$.  For this test, we pick $\max_A = 64$, comfortably above
$\max_{ij}|A_{ij}|$ for the smallest tested~$\Nx$.  Comparing block
by block against the classical $A_{11}$~\eqref{eq:A11-elements},
$A_{12}$~\eqref{eq:A12-elements}, and $A_{22} = A_{11}\otimes I + I \otimes A_{11}$
at $\nu = 2.0$ gives the following maximum entry-wise differences:

\begin{center}
\begin{tabular}{ccccc}
\hline
$\Nord$ & $\Nx$ & $A_{11}$ & $A_{12}$ & $A_{22}$ \\
\hline
1 & 8   & $2.7\times 10^{-15}$ & --- & --- \\
1 & 128 & $4.4\times 10^{-16}$ & --- & --- \\
2 & 4   & $1.1\times 10^{-14}$ & $2.8\times 10^{-14}$ & $1.4\times 10^{-14}$ \\
2 & 8   & $5.3\times 10^{-15}$ & $1.4\times 10^{-14}$ & $7.1\times 10^{-15}$ \\
\hline
\end{tabular}
\end{center}

\noindent confirming machine-precision agreement across all blocks
and the lower-triangular zero structure required by the Carleman
linearisation.

The implementation of optimized $U_L$ used in Sec.~\ref{sec:gate-scaling} is verified
by extracting
representative columns of both $U_L$ implementations on a small
$(\Nx, \Nt)$ instance and comparing them as matrix entries.  For
$\Nord = 1$ we obtain agreement to machine precision ($\leq 10^{-14}$)
at $\Nx = 4$, $\Nt = 2$, $\NK \in \{1, 3\}$; the corresponding test for
$\Nord = 2$ likewise agrees to
machine precision at the same parameters.

\section{Discrete LBM vs continuous BGK residual}
\label{app:lbm-bgk-residual}

\begin{figure}[htbp]
  \centering
  \includegraphics[width=1.0\columnwidth]{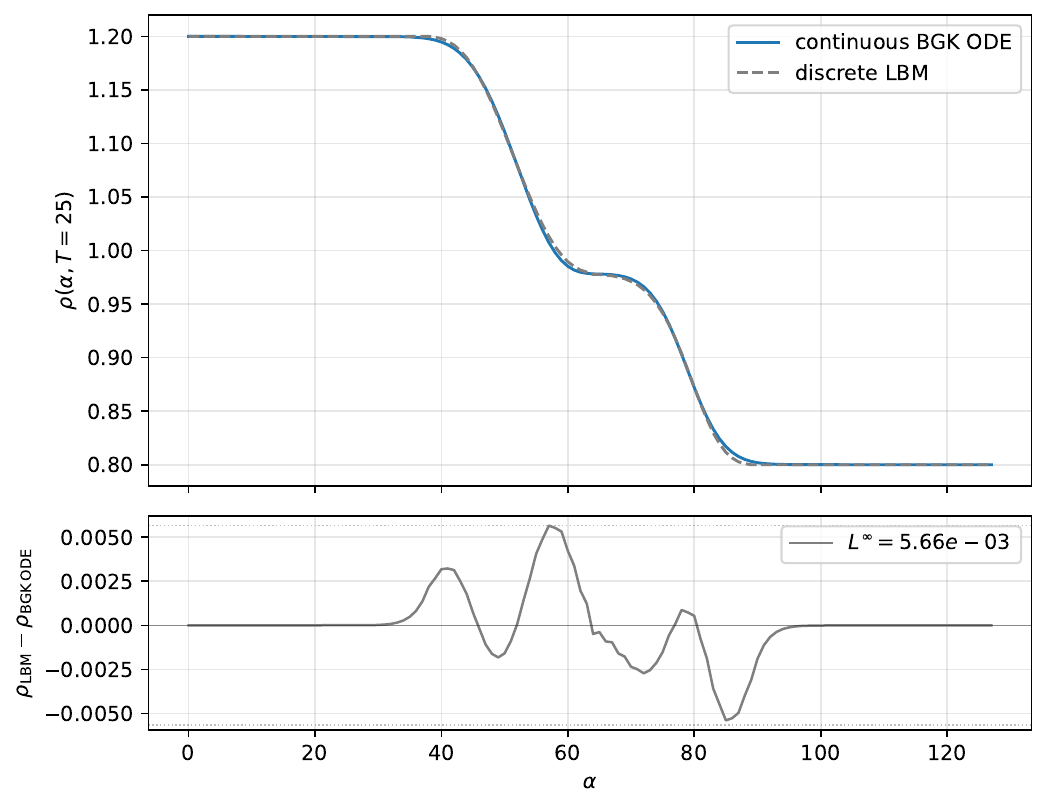}
  \caption{Density profile (top) and pointwise residual (bottom)
    for the discrete LBM and the continuous BGK at $T = 25$,
    $\Nx = 128$, $\nu = 2.0$, $\Delta\rho = 0.4$.}
  \label{fig:lbm-bgk-residual}
\end{figure}

Figure~\ref{fig:lbm-bgk-residual} compares the discrete LBM
($\beta = 1/(2\tau + 1)$) with the continuous BGK reference used
throughout the paper, at $\Nx = 128$, $\nu = 2.0$, $\Delta\rho = 0.4$,
$T = 25$.  The pointwise residual is
$L^\infty(\rho_{\mathrm{LBM}} - \rho_{\mathrm{BGK}})
\approx 5.7 \times 10^{-3}$, i.e.\ $\sim 1.4\%$ of $\Delta\rho$,
and shrinks as $O(\Delta x)$ on grid refinement
($9.4 \times 10^{-4}$ at $\Nx = 1024$), 
suggesting that the discrepancy is only from
the difference in scheme, and that our convention of omitting $+1/2$ in $\tau$ is correct.

\section{Spectral radius of the rate matrix}
\label{app:spectral-radius}
Table~\ref{tab:spectral-radius} lists the spectral radius
$\mu(A)$, i.e., the largest absolute value of eigenvalues of $A$, of the rate matrix and the corresponding critical
step $\Deltat_c = 1/\mu(A)$ at which $\mu(\Deltat_c\,A) = 1$,
for both Carleman orders.
For $\Nord = 1$, $\mu(A_{11})$ decreases as $1/\Nx$ for
small $\Nx$ (where the collision rate $1/\tau \propto 1/\Nx$
dominates) and saturates at $\mu(A_{11}) \approx 2$ for
large $\Nx$ (where streaming dominates).
For $\Nord = 2$, the Kronecker-sum structure
$A_{22} = A_{11}\otimes I + I\otimes A_{11}$ makes the spectrum
the pairwise sum of $A_{11}$ eigenvalues, doubling the spectral
radius: $\mu(A_{22}) = 2\,\mu(A_{11})$.
Stability ($\mu(\Deltat A) \lesssim  1$) is therefore achieved by
choosing $\Deltat < \Deltat_c$, with $\Deltat_c$ halved at
second order.
We note that the optimal $\Deltat$ observed in Fig.~\ref{fig:K-vs-h} is larger than $\Deltat_c$, which
suggests that optimal $\Deltat$ for $\Nord=2$ may be larger than $\Deltat_c\to 0.25$.

\begin{table}[h]
\centering
\begin{tabular}{rccccc}
\hline
$\Nx$ & $\tau$ & $\mu(A)$, $\Nord\!=\!1$ & $\mu(A)$, $\Nord\!=\!2$ & $\Deltat_c$, $\Nord\!=\!1$ & $\Deltat_c$, $\Nord\!=\!2$ \\
\hline
4   & 0.047 & 21.3 & 42.7 & 0.047 & 0.023 \\
8   & 0.094 & 10.7 & 21.3 & 0.094 & 0.047 \\
16  & 0.188 &  5.3 & 10.7 & 0.188 & 0.094 \\
32  & 0.375 &  2.7 &  5.3 & 0.375 & 0.188 \\
64  & 0.750 &  2.0 &  4.0 & 0.500 & 0.250 \\
128 & 1.500 &  2.0 &  4.0 & 0.500 & 0.250 \\
256 & 3.000 &  2.0 &  4.0 & 0.500 & 0.250 \\
\hline
\end{tabular}
\caption{Spectral radius $\mu(A)$ of the rate matrix and the
  corresponding critical time step
  $\Deltat_c = 1/\mu(A)$ at which $\mu(\Deltat_c A) = 1$
  ($\nu = 2.0$, matching the condition-number analysis of
  Sec.~\ref{sec:kappa-scaling}).}
\label{tab:spectral-radius}
\end{table}

\section{$\NK$-vs-$\Deltat$ trade-off at the QSVT operating size}
\label{app:K-vs-h-Nx32}

\begin{figure*}[htbp]
  \centering
  \includegraphics[width=0.95\textwidth]{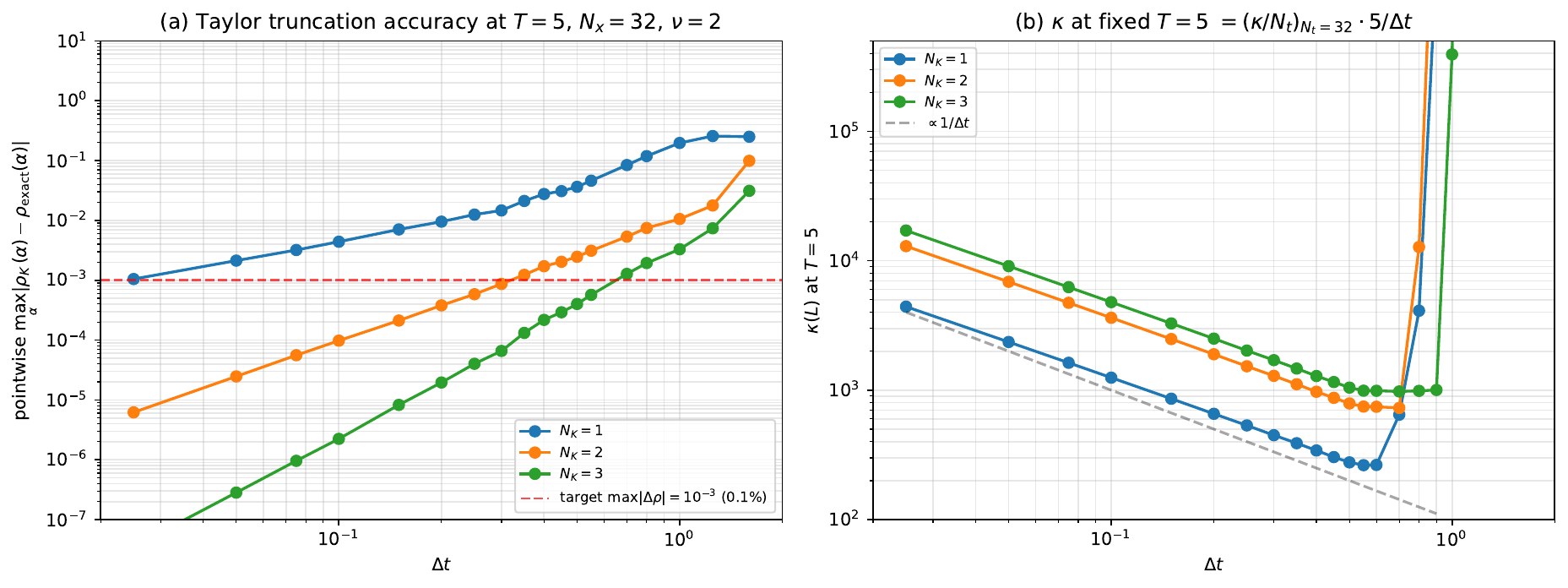}
  \caption{$\NK$-vs-$\Deltat$ trade-off at $T = 5$, $\Nx = 32$,
    $\nu = 2.0$ (the QSVT operating size).
    Layout and axis conventions are
    identical to Fig.~\ref{fig:K-vs-h}.}
  \label{fig:K-vs-h-Nx32}
\end{figure*}

The body of the paper presents the $\NK$-vs-$\Deltat$ trade-off at
$\Nx = 128$, $T = 25$ (Fig.~\ref{fig:K-vs-h}) to make
the analysis as practical as possible, while the QSVT runs of
Sec.~\ref{sec:qsvt-results} are performed at the smaller
size $\Nx = 32$, $T = 5.6$ that fits on the GPU
state-vector simulator.  Figure~\ref{fig:K-vs-h-Nx32} reports the
same trade-off at the QSVT operating size.

\bibliography{bib}
\end{document}